\documentclass[a4paper,11pt]{article}
\usepackage{jheppub}

\usepackage[english]{babel}
\usepackage{newlfont}
\usepackage{float}
\usepackage{tikz}
\usetikzlibrary{matrix, positioning, decorations.pathmorphing, calc, 3d, arrows.meta, decorations.markings}

\def\KK{{\scriptscriptstyle \mathrm{KK}}}
\def\W{{\scriptscriptstyle \mathrm W}}
\def\I{{\scriptstyle \mathrm L}}
\def\LVS{{\scriptscriptstyle \mathrm{LVS}}}
\def\CW{{\scriptscriptstyle \mathrm{CW}}}
\def\K3{{\scriptscriptstyle \mathrm{K3}}}
\def\dP{{\scriptscriptstyle \mathrm{dP}}}
\def\FI{{\scriptscriptstyle \mathrm{FI}}}
\def\P1{{\mathbb{P}^1}}
\newcommand{\vo}{\mathcal{V}}

\title{Moduli Stabilisation for ADD and the Dark Dimension Scenario}

\author[a]{Andreas P. Braun,}
\author[b,c]{Michele Cicoli,}
\author[b,c]{Riccardo Milioli,}
\author[d,e]{Roberto Valandro}

\affiliation[a]{Department of Mathematical Sciences, Durham University, Upper Mountjoy Campus,
Stockton Rd, Durham DH1 3LE, UK}
\affiliation[b]{\footnotesize Dipartimento di Fisica e Astronomia, Università di Bologna, via Irnerio 46, 40126 Bologna, Italy}
\affiliation[c]{\footnotesize INFN, Sezione di Bologna, viale Berti Pichat 6/2, 40127 Bologna, Italy}
\affiliation[d]{\footnotesize Dipartimento di Fisica, Università di Trieste, Strada Costiera 11, I-34151 Trieste, Italy}
\affiliation[e]{INFN, Sezione di Trieste, Via Valerio 2, I-34127 Trieste, Italy}

\emailAdd{andreas.braun@durham.ac.uk}
\emailAdd{michele.cicoli@unibo.it}
\emailAdd{riccardo.milioli2@unibo.it}
\emailAdd{roberto.valandro@ts.infn.it}

\abstract{We provide a moduli stabilisation mechanism for realising anisotropic string compactifications with one or two large extra dimensions, corresponding to the ADD and Dark Dimension scenarios. This is achieved within the type IIB Large Volume Scenario, where an exponentially large Calabi-Yau volume in string units can naturally generate a parametrically low Kaluza-Klein scale. Anisotropy is realised by considering a Calabi-Yau threefold which is a K3 fibration over a $\mathbb{P}^1$ base. The volume of the 4D K3 fibre is stabilised at relatively small values by perturbative corrections to the effective action, in particular string loops and higher-derivative effects, leaving an exponentially large volume of the 2D $\mathbb{P}^1$ base. We argue that complex structure moduli stabilisation can dynamically deform the $\mathbb{P}^1$ base, corresponding to a Tyurin degeneration limit where the internal geometry effectively develops  a single large 1D cycle. Within a unified description, the ADD case is instead recovered as a symmetric alternative limit. The potential can feature either a dS vacuum or a quintessence runaway, although in both cases some degree of tuning is required to match the observed cosmological constant scale. We also present an explicit Calabi-Yau orientifold example with consistent brane setup, tadpole cancellation and moduli stabilisation. We analyse the resulting moduli spectrum and associated phenomenological constraints, including supersymmetry breaking, cosmological moduli overproduction and fifth force bounds.}

\begin{document}

\maketitle

\section{Introduction}

Large extra dimensions (EDs) in braneworld scenarios have a long history in theoretical physics. In this framework the Standard Model (SM) is confined to a brane, while gravity propagates in the bulk of the EDs. The difficulty of testing Newton's inverse-square law at very short distances results in a relatively weak upper bound on the size of the EDs of order $R\lesssim 40\,\mu\text{m}$ \cite{Lee:2020zjt}, which translates into a lower bound on the associated Kaluza-Klein (KK) scale of order $M_\KK\gtrsim\, 5$ meV.\footnote{See also \cite{Arkani-Hamed:1998sfv,Anchordoqui:2025nmb} for astrophysical and cosmological bounds for models with 1 or 2 EDs.}

This observation opens up the possibility of addressing the Higgs hierarchy problem via TeV-scale fundamental gravity since $M_\KK\sim$ meV correlates with a 6D Planck scale in the TeV range, $M_6\sim$ TeV, in ADD models with 2 large EDs \cite{Arkani-Hamed:1998jmv}. Models with 1 large ED would not solve the hierarchy problem since $M_5\sim 10^8$ GeV, while models with more than 2 large EDs would be ruled out since $M_d\ll$ TeV for $d>2$. 
Interestingly, models with 2 large EDs have also been proposed to address the cosmological constant problem via a very low scale of supersymmetry breaking in the bulk in SLED scenarios \cite{Aghababaie:2003wz}. These constructions stem from the observation that $M_6\sim$ TeV correlates with a gravitino mass of the same order as the dark energy scale, $m_{3/2}\sim$ meV. 

Clearly, braneworlds with large EDs are string inspired models which require a proper string theory embedding to be under quantitative control. A major challenge is moduli stabilisation which determines dynamically the number of large EDs starting from the six internal dimensions of a typical string compactification. A well established mechanism to obtain large EDs in string theory is the type IIB Large Volume Scenario (LVS) where the CY volume is stabilised exponentially large in terms of the fundamental string length \cite{Balasubramanian:2005zx}. Anisotropic LVS compactifications with stabilised moduli have enabled the realisation of ADD and SLED scenarios within string theory \cite{Cicoli:2011yy}, with related global chiral embeddings studied in \cite{Cicoli:2011qg}; the issue of loop corrections remains although open. Indeed, loops of SM open strings, if not cancelled by brane backreaction effects, are expected to yield large contributions to the vacuum energy. Moreover, the bulk theory is not exactly supersymmetric above the 6D KK scale. 

Recently, there has been renewed interest in models with large EDs due to the swampland programme \cite{Vafa:2005ui}. In particular, the Dark Dimension (DD) scenario postulates the existence of a single micron-sized ED \cite{Montero:2022prj}. The main idea behind this scenario is the claim that the observed small positive vacuum energy should correlate with asymptotic infinite-distance limits \cite{Ooguri:2006in} where the scalar potential goes to zero \cite{Lust:2019zwm} and light towers of KK modes emerge \cite{Lee:2019wij}. More precisely, in the asymptotic regime identified by $R\to \infty$, the potential is conjectured to scale as the fourth power of the
cut-off of the theory, identified with the KK scale, $V \sim M_\KK^4 \sim R^{-4}\to 0$, where $R$ is now interpreted as the dynamical radion field. The fact that $V\sim (\text{meV})^4$ then suggests $M_\KK\sim \text{meV}$ which, as we have already pointed out, can be compatible only with 1 or 2 large EDs.  

This interesting proposal currently remains at the level of a qualitative scenario. Promoting it to a fully-fledged model with a concrete realisation in string theory is a hard task, as it must overcome a few major challenges:
\begin{enumerate}
\item There are no known accelerating solutions at the boundaries of moduli space. In particular, neither de Sitter (dS) vacua \cite{Andriot:2026lac} nor phenomenologically viable quintessence models \cite{Brinkmann:2022oxy, Cicoli:2020noz,Cicoli:2021fsd,Shiu:2023nph,Shiu:2023fhb,Shiu:2024sbe} have been successfully constructed in this regime, making it difficult to account for the observed accelerated expansion of the Universe.

\item The assumption that the vacuum energy scales as $V \sim M_\KK^4$ requires further justification. Such a scaling implicitly assumes the absence of a tree-level contribution to the potential. In string compactifications, achieving this may necessitate switching off background fluxes, which are typically essential for moduli stabilisation, unless one restricts attention to no-scale constructions where the tree-level potential cancels due to a combination of supersymmetry, higher dimensional scaling symmetries and axionic shift symmetries \cite{Burgess:2020qsc}. Furthermore, it must be assumed that quantum corrections do not generate contributions with a parametrically smaller dependence on the KK scale, such as $V \sim M_\KK^2 m_{3/2}^2$, which may arise after supersymmetric cancellations. One possible alternative is to consider non-supersymmetric compactifications in which the vacuum energy originates from Casimir-type loop effects. However, in the absence of supersymmetry, maintaining perturbative control over the effective field theory (EFT) becomes substantially more challenging.

\item The contribution of SM loops also requires careful consideration. A generic relation of the form $V \sim m^4$ can be interpreted not only as determining the relevant mass scale $m$ from a given vacuum energy $V$, but also conversely as predicting the size of $V$ once a mass scale is specified. Applying this reasoning to SM particles, and in particular to the top quark mass, would lead to vacuum-energy contributions many orders of magnitude larger than the observed value. Understanding the suppression or cancellation of such contributions therefore remains an essential challenge.
\end{enumerate}

In order to find late-time accelerating solutions which can describe our Universe, one therefore has to move away from the asymptotic regime of moduli space where the potential is expected to be different, and actually much richer, than a simple $V\sim M_\KK^4$ relation. From this perspective, the motivation behind the original DD proposal and its connection to the cosmological constant problem becomes less compelling. Nevertheless, the idea of constructing models with 1 or 2 large EDs and stabilised moduli in a vacuum compatible with the observed cosmic acceleration remains highly attractive, since micron-sized EDs can lead to observable signatures that may be tested in the near future.

With this motivation in mind, in this paper we focus on constructing string vacua with stabilised moduli which realise ADD and DD models. Our strategy is as follows:\footnote{See also \cite{Dudas:2025yqm} for a different approach to derive ADD/SLED models from non-supersymmetric strings, and \cite{Blumenhagen:2026rgu} for a recent attempt to realise the DD scenario in Horava-Witten theory.}
\begin{enumerate}
\item We consider type IIB flux compactifications where moduli stabilisation is best understood. In this framework, the complex structure moduli and the axion-dilaton can be fixed at tree-level by turning on background fluxes. The radion corresponds to one of the K\"ahler moduli whose potential remains exactly flat at tree-level due to the extended no-scale structure: $V_{\text{tree}}=0$.

\item To obtain a stabilised vacuum with large EDs, we work within the LVS where exponentially large compactification volumes arise from the interplay between $\alpha'$ corrections and non-perturbative effects. The minimum lies at $\mathcal{V}\sim e^{2\pi/(N g_s)}\gg 1$ where $\mathcal{V}$ denotes the dimensionless CY volume in string units, $g_s\ll 1$ is the string coupling set by the flux-stabilised dilaton, and $N=1$ for E3-instantons, while $N$ is a positive integer for gaugino condensation on a stack of $N$ D7-branes.

\item The resulting minimum is a non-supersymmetric AdS vacuum which needs to be uplifted through additional positive contributions to the scalar potential, arising for example from anti-branes \cite{Kachru:2003aw}, T-branes \cite{Cicoli:2015ylx}, non-zero F-terms of the complex structure moduli \cite{Gallego:2017dvd}, or non-perturbative effects at singularities \cite{Cicoli:2012fh}.

\item To get an anisotropic solution with 1 or 2 large EDs, $\mathcal{V}$ needs to be controlled by at least 2 K\"ahler moduli. The simplest example (ignoring rigid del Pezzo divisors supporting non-perturbative effects) is a K3-fibred CY with volume $\mathcal{V}\simeq t_\P1\tau_\K3$, where $\tau_\K3$ is the volume of the K3 fibre, while $t_\P1$ is the volume of the $\mathbb{P}^1$ base \cite{Cicoli:2011it}. We parametrise the K\"ahler sector in terms of $\mathcal{V}$ and $\tau_{\K3}$. The leading no-scale breaking contributions to the scalar potential fix $\mathcal{V}$ at exponentially large values while leaving $\tau_{\K3}$ as a flat direction.

\item String loop corrections combined with higher-derivative $\alpha'$ effects can fix $\tau_\K3\sim \mathcal{O}(5)$, generating a highly anisotropic compactification with $\mathcal{V}\sim t_\P1 \sim e^{2\pi/(N g_s)}\gg 1$. If the $\mathbb{P}^1$ base is isotropic, one obtains the ADD scenario with 2 large EDs. On the other hand, if the $\mathbb{P}^1$ base becomes highly anisotropic, like a sphere stretched along one direction, as typically occurs near a Tyurin degeneration limit at infinite distance in complex structure moduli space, one realises a DD scenario with effectively 1 large ED. The shape of the $\mathbb{P}^1$ is controlled by complex structure moduli fixing via $3$-form fluxes. 
\end{enumerate}

Interestingly, in this concrete setup the radion potential scales as $V \sim m_{3/2}^2 \left(M_\KK^{10D}\right)^2$, which corresponds to the typical Coleman-Weinberg scaling of loop corrections in the underlying 10D theory. In the DD scenario this potential reduces to $V\sim W_0^2 \left(M_\KK^{5D}\right)^2 M_p^2 \sim W_0^2 M_p^2/R^2$, whereas in the ADD scenario it becomes $V\sim W_0^2\left(M_\KK^{6D}\right)^3 M_p \sim W_0^2 M_p/R^3$, where $W_0$ is the flux superpotential. For natural $\mathcal{O}(1-10)$ values of $W_0$, one finds in both cases $V\gg R^{-4}$, implying the need of some degree of tuning. Otherwise, matching the observed vacuum energy would require a KK scale below meV, corresponding to EDs larger than $40\,\mu m$. Indeed, for $W_0\sim\mathcal{O}(1-10)$ one can achieve a dS minimum at $\tau_\K3\sim\mathcal{O}(5)$ with an uplifting sector needed to cancel the leading contribution to $V$ in order to generate a vacuum energy of order $\langle V\rangle \sim R^{-4}\sim (\text{meV})^4$. For much smaller and rarer values of $W_0$, one can instead realise a non-interacting quintessence model along a shallow runaway potential of the form $V\simeq V_0\,e^{-c\phi}$ with $c=1/\sqrt{3}$ which would need $W_0\sim\mathcal{O}(10^{-12})$ for the ADD case and $W_0\sim\mathcal{O}(10^{-27})$ for the DD scenario.

The explicit nature of our construction provides a concrete framework in which the theoretical and phenomenological challenges of building DD and ADD models can be analysed in detail. On the theoretical side, the main concern is that unknown quantum corrections could spoil moduli stabilisation. Nevertheless, substantial progress has been made in the computation and in the classification of corrections at different orders in the $\alpha'$ and $g_s$ expansions, exploiting higher dimensional scaling symmetries \cite{Burgess:2020qsc} and F-theory limits \cite{Cicoli:2021rub}. Another issue, which we have not addressed explicitly, concerns the identification of $3$-form flux quanta capable of generating hierarchical vacuum expectation values for the complex structure moduli, thereby driving the $\mathbb{P}^1$ base into a strongly anisotropic regime. In this respect, recent applications of artificial intelligence and machine-learning techniques for exploring the type IIB flux landscape may prove particularly valuable \cite{Ebelt:2023clh,Chauhan:2025rdj,Chauhan:2026gid}.

From a phenomenological perspective, the main challenge is related to the fact that large EDs at the edge of detectability require an exponentially large $\vo$ which, in turn, lowers all the scales in the model. 
This generically leads to several phenomenological problems. Decaying moduli with masses below $\mathcal{O}(30)$ TeV would suffer from the cosmological moduli problem (CMP) \cite{Coughlan:1983ci,Banks:1993en,deCarlos:1993wie}. Gravitationally coupled moduli with masses below $\mathcal{O}(10)$ MeV would be cosmologically stable and could overclose the Universe, due to an excessive dark matter abundance, while moduli lighter than $\mathcal{O}(1)$ meV could mediate unobserved long-range fifth forces. In addition, the exponentially large volume lowers the gravitino mass. In a gravity-mediated scenario this would typically imply soft terms below the TeV scale, and hence superpartners that should already have been observed. The CMP and dark matter overproduction can be avoided in the presence of a mechanism which suppresses the moduli abundance, for example, by enforcing a tiny initial misalignment. Long range fifth forces could instead be absent if the light moduli are effectively decoupled from the SM due to geometrical sequestering \cite{Acharya:2018deu}. Finally, if supersymmetry is non-linearly realised in the SM sector, the superpartners would appear at the string scale, $M_{\text{soft}}\sim M_s$, which for ADD coincides with the TeV-scale, while for the DD scenario it is higher, around $10^8$ GeV. A non-supersymmetric SM brane would, in turn, naively contribute to the vacuum energy as $V\sim M_s^4$. Such a large term would spoil moduli stabilisation, unless it is cancelled by the SM brane backreaction, as advocated in SLED scenarios \cite{Aghababaie:2003wz}.

Our paper is organised as follows. In Sec. \ref{sec:AnisModStab} we discuss the main theoretical and phenomenological features of our moduli stabilisation mechanism, with the aim of illustrating how to realise ADD and DD scenarios as stabilised type IIB vacua without delving into technical details. In Sec. \ref{sec:ExplicitCY} we present a concrete CY model with explicit orientifold involution, brane setup and minimisation of the K\"ahler moduli potential which leads to an anisotropic compactification with either 1 or 2 large EDs. We present our conclusions in Sec. \ref{Concl}, while App. \ref{App} discusses a toy model for the Tyurin degeneration limit.

\section{Anisotropic moduli stabilisation for ADD and the Dark Dimension scenario}
\label{sec:AnisModStab}

In this section we convey the main idea behind our realisation of the ADD and DD scenarios without delving into too many technical details. An explicit CY embedding will be analysed in Sec. \ref{sec:ExplicitCY}.

\subsection{Large extra dimensions from anisotropic string compactifications}

In what follows, we are interested in realising models with EDs at the edge of detectability. Short distance tests of Newton's law require $M_\KK\gtrsim$ meV regardless of the number $d$ of EDs. The $(4+d)$-dimensional fundamental gravity scale $M_{(4+d)}$ is related to $M_\KK$ and the 4D Planck scale $M_p\equiv M_4$ as:
\begin{equation}\label{Eq:M4pd}
M_{(4+d)} \simeq  \left(\frac{M_\KK}{M_p}\right)^{\frac{d}{(d+2)}} M_p\,.
\end{equation}
Setting $M_\KK\sim 10^{-30}\,M_p\sim \text{meV}$, one obtains 
$M_5 \sim  10^{-10} M_p\sim 10^8$ GeV for $d=1$, $M_6 \sim  10^{-15} M_p \sim 1$ TeV for $d=2$, and $M_{(4+d)}\ll 1$ TeV for $d>2$. The results are summarised in Tab. \ref{Tab1}.

\begin{table}[ht]
\centering
\begin{tabular}{c c}
\hline
$d$ & \qquad $M_{(4+d)}$ \\
\hline
1 & \qquad $10^8\,\text{GeV}$ \\
2 & \qquad $10^3\,\text{GeV}$ \\
3 & \qquad $1\,\text{GeV}$ \\
4 & \qquad $10^{-2}\,\text{GeV}$ \\
5 & \qquad $5\cdot 10^{-4}\,\text{GeV}$ \\
6 & \qquad $5\cdot 10^{-5}\,\text{GeV}$ \\
\hline
\end{tabular}
\caption{Fundamental gravity scale $M_{(4+d)}$ for KK scale fixed at $M_\KK\sim \text{meV}$.}
\label{Tab1}
\end{table}

Given that experiments do not show any evidence for a strongly coupled gravitational interaction below the TeV scale, the only cases compatible with $M_\KK\sim$ meV are $d=1$ (DD) and $d=2$ (ADD). 

To realise the DD and ADD scenarios in string theory, we shall work within the framework of fluxed type IIB compactifications on a Calabi-Yau (CY) orientifold $X$ where moduli stabilisation is best understood. In this context, at the semi-classical level, the axio-dilaton $S$ and the complex structure moduli $U_\alpha$, $\alpha=1,...,h^{1,2}_-$, can be fixed by $3$-form fluxes supersymmetrically, while the K\"ahler moduli $T_i$, $i=1,...,h^{1,1}_+$, remain flat due to the no-scale cancellation. These directions can be lifted by a combination of different effects, in particular, D-terms generated by gauge fluxes on D7-branes, and the inclusion of perturbative and non-perturbative quantum corrections to the F-term potential.

Given that string theory requires $d=6$ extra dimensions, to derive effectively a model with $d=1$ or $d=2$, we need to fix the moduli so that the CY volume becomes anisotropic. To this end, we consider a CY threefold $X$ corresponding to a 4D K3 fibration over a 2D $\mathbb{P}^1$ base with total volume:
\begin{equation}
\text{Vol(K3)} = \ell_\K3^4\quad\text{Vol}(\mathbb{P}^1) =\ell_\I\,R
\quad\Rightarrow\quad \text{Vol}(X) = \text{Vol(K3)}\,\text{Vol}(\mathbb{P}^1) = \ell_\K3^4\, \ell_I\,R\,.
\end{equation}
The DD limit corresponds to $\ell_\K3 \sim \ell_\I \ll R$, while the ADD limit to $\ell_\K3 \ll \ell_\I \sim R$. In string theory, the fundamental length is the string length $\ell_s = 2\pi\sqrt{\alpha'}$, and so it is natural to measure any distance in units of $\ell_s$:
\begin{equation}
\text{Vol}(X)  = \mathcal{V}\,\ell_s^6 \,,\qquad \ell_\K3 = \tau_\K3^{1/4}\, \ell_s \,, \qquad \ell_\I = \sqrt{u\, t_\P1}\,\ell_s\,, \qquad R =\sqrt{\frac{t_\P1}{u}}\,\ell_s\,.
\end{equation}
Here, $\tau_\K3 = (\ell_\K3/\ell_s)^4$ is a K\"ahler modulus controlling the volume of the K3 divisor ($4$-cycle) which has to remain small in string units, while $t_\P1 = (\ell_\I\, R)/\ell_s^2$ is related to another K\"ahler modulus parametrising the volume of the $\mathbb{P}^1$ ($2$-cycle) which has instead to become large. The parameter $u = \ell_\I/R$ is the ratio of two complex structure moduli controlling the volumes of two $3$-cycles constructed as the fibration of a complex curve $C$ ($2$-cycle) in K3 over, respectively, a loop $L$ of radius $\ell_\I$ or an interval $I$ of length $R$ in $\mathbb{P}^1$. Note that $u\simeq 1$ leads to the ADD case, whereas $u\ll 1$ realises the DD scenario. The CY volume $\mathcal{V}$ therefore becomes:
\begin{equation}
\mathcal{V}= t_\P1\,\tau_\K3\,.
\label{CYVol}
\end{equation}
Important quantities are the string scale and the KK scales associated to different radii. From dimensional reduction, the string scale $M_s = \ell_s^{-1} \equiv M_{10}$ is related to $M_p$ as:
\begin{equation}
M_s \simeq  \frac{M_p}{\sqrt{\mathcal{V}}}\,.
\label{Ms}
\end{equation}
The 10D, 6D and 5D KK scales are given respectively by:
\begin{eqnarray}
M_\KK^{10D} &=& \frac{1}{\ell_\K3}  = \frac{M_s}{\tau_\K3^{1/4}} \simeq \frac{M_p}{\tau_\K3^{1/4}\, \sqrt{\mathcal{V}}}\,,
\label{MKK10D} \\
M_\KK^{6D} &=& \frac{1}{\ell_\I} = \frac{M_s}{\sqrt{u\,t_\P1}} \simeq \frac{M_p}{\mathcal{V}}\sqrt{\frac{\tau_\K3}{u}} \,, 
\label{MKK6D} \\
M_\KK^{5D} &=& \frac{1}{R}  = \sqrt{\frac{u}{t_\P1}}\,M_s \simeq \frac{M_p}{\vo}\,\sqrt{u\,\tau_\K3} \simeq u\,M_\KK^{6D}\,,
\label{MKK5D}
\end{eqnarray}
where we used $t_\P1=\mathcal{V}/\tau_\K3$. In the ADD limit, $u\to 1$, and $M_\KK^{5D}\to M_\KK^{6D}$, leaving a 6D EFT. On the other hand, in the DD case, $u\ll 1$, showing that $M_\KK^{5D}\ll M_\KK^{6D}$, yielding a genuinely 5D EFT. Moreover, an intermediate 6D EFT exists only if $\ell_\K3 \ll \ell_\I\ll R$, corresponding to the window $\tau_\K3^{3/2}/\vo \ll u\ll 1$. Otherwise, if $\ell_\K3 \sim \ell_\I\ll R$ for $u\to \tau_\K3^{3/2}/\vo$, $M_\KK^{6D}\to M_\KK^{10D}$, leaving just a 10D EFT above the 5D theory.

Note that, for ADD with $u\simeq 1$, the phenomenological constraint $M_\KK^{6D} \simeq 10^{-30}\,M_p$ combined with (\ref{Ms}) implies $M_s \simeq (10^3\,\text{GeV})\,\tau_\K3^{-1/4}$ which requires $\tau_\K3^{1/4}\sim\mathcal{O}(1)$ to avoid strings below the TeV scale. In what follows, we shall show that perturbative K\"ahler moduli stabilisation can naturally lead to $\tau_\K3\sim\mathcal{O}(5)$, which guarantees both that the EFT is under control and that indeed $\tau_\K3^{1/4}\sim\mathcal{O}(1)$. Hence in this case $M_\KK^{10D}\sim M_s\sim M_6$ and, from (\ref{Ms}), the required CY volume to get $M_s\sim 10^3$ GeV is $\vo \sim 10^{30}$. 

The value of the total CY volume in the DD scenario is instead smaller. The smallest allowed value of $\vo$ is obtained for the minimal value of $u$, $u_{\text{min}}\simeq \tau_\K3^{3/2}/\vo$, which, when inserted in (\ref{MKK5D}), gives $M_\KK^{5D} \simeq M_p\,\tau_\K3^{5/4}\,\vo^{-3/2}$. For $\tau_\K3\sim\mathcal{O}(5)$, $M_\KK^{5D}\simeq 10^{-30}\,M_p$ is then obtained for $\vo\simeq 5\cdot 10^{20}$ that, in turn, implies $u\simeq 10^{-20}$. Plugging this value of $\vo$ in (\ref{Ms}) one finds in this case a common scale $M_\KK^{6D}\sim M_\KK^{10D}\sim M_s \sim M_6\sim M_5\sim 10^8$ GeV. Larger values of $u$, which at the microscopic level require less tuning in flux stabilisation of the complex structure moduli, would require larger values of $\vo$, and would lead to the hierarchy $M_\KK^{6D}\ll M_\KK^{10D}\sim M_s \sim M_6\ll M_5$. Some illustrative numerical examples are reported in Tab. \ref{Tab2}. 

\begin{table}[ht]
\centering
\begin{tabular}{c | c c | c c c c c c c}
\hline
& $t_\P1$ & $u$ & $\mathcal{V}$ & $\ell_\K3$ & $\ell_\I$ & $R$ & $M_\KK^{5D}$ & $M_\KK^{6D}$ & $M_s$ \\
\hline
\textbf{ADD} & $10^{30}$ & $1$ & $5\cdot 10^{30}$ & $1.5\,\ell_s$ & $10^{15}\,\ell_s$ & $10^{15}\,\ell_s$ & - & 1\,$\text{meV}$ & $10^3\,\text{GeV}$ \\
\textbf{DD} & $10^{26}$ & $10^{-8}$ & $5\cdot 10^{26}$ & $1.5\,\ell_s$ & $10^9\,\ell_s$ & $10^{17}\,\ell_s$ & 1\,$\text{meV}$ & $0.1\,\text{MeV}$ & $10^5\,\text{GeV}$ \\
\textbf{DD} & $10^{20}$ & $10^{-20}$ & $5\cdot 10^{20}$ & $1.5\,\ell_s$ & $1.0 \,\ell_s$ & $10^{20}\,\ell_s$ & 1\,$\text{meV}$ & - & $10^8\,\text{GeV}$ \\
\hline
\end{tabular}
\caption{Illustrative numerical examples realising the ADD and the DD scenarios. In all cases the K3 fibre volume is stabilised at $\tau_\K3\simeq 5$.}
\label{Tab2}
\end{table}

\subsection{K\"ahler moduli stabilisation}

As explained above, a string theory realisation of the ADD model needs $\vo\sim 10^{30}$, while the DD scenario requires at least $\vo\gtrsim 10^{20}$. The most established way to realise such large values of the CY volume is the type IIB Large Volume Scenario (LVS) which features de Sitter (dS) vacua where $\vo$ is fixed exponentially large in terms of the string coupling: $\vo\sim e^{2\pi/(N\,g_s)}\gg 1$ for $g_s \ll 1$ and $N\in \mathbb{N}$. In principle, an alternative realisation could be based on a run-away potential with a quintessence solution at large field values where $\vo\gg 1$. However, several recent studies have provided growing evidence against dynamical dark energy at the asymptotic boundaries of moduli space \cite{Cicoli:2021fsd}, leaving LVS dS vacua as the most promising string embedding of models with $1$ or $2$ large EDs.

A key feature of LVS models is the presence of del Pezzo divisors which support non-perturbative effects that are needed to stabilise $\vo$ at exponentially large values.\footnote{In the absence of blow-up modes, an exponentially large $\vo$ could also be achieved via the inclusion of logarithmic perturbative corrections to the K\"ahler potential \cite{Antoniadis:2018hqy,Antoniadis:2019rkh}, even if their presence requires a localised region of high curvature in the CY.} These $4$-cycles are shrinkable blow-up modes which arise very frequently and, being rigid, are the best candidates to generate non-perturbative corrections to the superpotential. Following \cite{Cicoli:2016xae}, we shall therefore slightly generalise the volume form (\ref{CYVol}) including a third K\"ahler modulus $\tau_\dP$ controlling the volume of a del Pezzo divisor:
\begin{equation}
\mathcal{V} = t_\P1\, \tau_\K3 - \tau_\dP^{3/2}\simeq  t_\P1\, \tau_\K3\,,
\label{CYVol_new}
\end{equation}
where the last approximation is justified by the fact that $\tau_\dP$ will be fixed at a relatively small value with respect to the CY volume. The total number of K\"ahler moduli, $T_i = \tau_i + i\,\theta_i$, is therefore $h^{1,1}=3$, with $\tau_\P1\simeq t_\P1\sqrt{\tau_\K3}$ and $\theta_i$ the axions obtained from the dimensional reduction of the type IIB RR $4$-form $C_4$ over the different $4$-cycles. 

\subsubsection{Leading order stabilisation}
\label{LOS}

At semi-classical level, all K\"ahler moduli remain flat due to the no-scale structure of the type IIB tree-level K\"ahler potential $K_0=-2\ln\vo$. Moreover, the tree-level superpotential $W_0$ is in general an $\mathcal{O}(1-10)$ constant after axio-dilaton and complex structure fixing via quantised $3$-form fluxes, as shown by scans of type IIB flux vacua \cite{Ebelt:2023clh, Chauhan:2025rdj, Chauhan:2026gid}. The leading quantum corrections that lift the K\"ahler moduli are $\mathcal{O}(\alpha'^3)$ contributions to $K$ \cite{Becker:2002nn} and $T_\dP$-dependent non-perturbative effects in $W$:
\begin{eqnarray}
K &=& K_0 + \delta K_{\alpha'^3} \simeq K_0 -\frac{\xi}{g_s^{3/2}\,\vo}\,, 
\label{Kalphap3} \\
W &=& W_0 + A_\dP\,e^{-a_\dP\,T_\dP}\,,
\label{Wnp}
\end{eqnarray}
where $\xi\sim A_\dP\sim\mathcal{O}(1)$ and $a_\dP = 2\pi/N$ with $N\in\mathbb{N}$. After fixing the axion $\theta_\dP$ at its minimum, these corrections generate the typical LVS potential of the following form (ignoring all $\mathcal{O}(1)$ numerical factors):
\begin{equation}
V_\LVS \simeq  a_\dP^2 \sqrt{\tau_\dP}\, \frac{e^{-2 a_\dP\tau_\dP}}{\mathcal{V}} -   W_0 a_\dP \tau_\dP \,\frac{e^{-a_\dP\tau_\dP}}{\mathcal{V}^2}+ \frac{W_0^2}{g_s^{3/2}\,\mathcal{V}^3} + \frac{c_{\text{up}}}{\vo^\alpha}\,,
\label{LVSPot}
\end{equation}
where we added an uplifting contribution proportional to the positive coefficient $c_{\text{up}}$ which can have different origins (for example, $\alpha=1$ for non-perturbative effects at singularities \cite{Cicoli:2012fh}, $\alpha=4/3$ for $\overline{D3}$-branes at the tip of a warped throat \cite{Kachru:2003aw}, $\alpha=2$ for non-zero F-terms of the complex structure moduli \cite{Gallego:2017dvd}, and $\alpha=8/3$ for T-branes \cite{Cicoli:2015ylx}). This term has two important features: ($i$) the string landscape gives enough freedom to tune $c_{\text{up}}$ to obtain a Minkowski vacuum at this level of approximation; ($ii$) the uplifting depends just on the overall volume $\vo$ without introducing any dependence on additional moduli. 

The LVS potential (\ref{LVSPot}) admits a typical minimum at:
\begin{equation}
\langle \tau_\dP \rangle \simeq 1/g_s \qquad\text{and}\qquad
\langle\mathcal{V}\rangle\simeq W_0\,e^{a_\dP \langle\tau_\dP\rangle} \simeq e^{\frac{2\pi}{N\,g_s}} \,.
\end{equation}
For $g_s\sim\mathcal{O}(0.1)$ in the regime where perturbation theory is under control, $\langle\tau_\dP\rangle\sim\mathcal{O}(10)$ while $\vo$ is exponentially large. This justifies the approximation $\tau_\dP\ll\vo$ and allows to obtain naturally huge values of $\vo$ in the regime $10^{20}\lesssim\vo\lesssim 10^{30}$ that are needed to realise the ADD and the DD scenarios.

Note, in addition, that $V_\LVS$ depends just on $2$ K\"ahler moduli: $\tau_\dP$ and the combination of $t_\P1$ and $\tau_\K3$ corresponding to $\vo\simeq t_\P1\,\tau_\K3$. Hence, at this level of approximation, the potential features a Minkowski minimum with $1$ saxionic and $2$ axionic flat directions given by $\theta_\P1$ and $\theta_\K3$. The saxionic flat direction is orthogonal to $\vo$ and can be identified by the canonical normalisation of the moduli kinetic terms (focusing just on $\tau_\P1$ and $\tau_\K3$):
\begin{equation}
\mathcal{L}_{\text{kin}} = \frac{\partial^2 K_0}{\partial T_i \partial \overline{T}_{\overline{j}}} \partial_\mu T^i \partial^\mu \overline{T}^{\overline{j}}= \frac{2\,\partial_\mu \tau_\P1 \partial^\mu \tau_\P1}{\tau_\P1^2}  + \frac{\partial_\mu \tau_\K3 \partial^\mu \tau_\K3}{\tau_\K3^2}  \,,
\label{KinLagr}
\end{equation}
where we used $K_0\simeq -\ln\tau_\K3-2\ln\tau_\P1$. This kinetic Lagrangian becomes canonical, i.e. $\mathcal{L}_{\text{kin}} = \frac12\,\partial_\mu \chi \partial^\mu \chi + \frac12\,\partial_\mu \phi \partial^\mu \phi$, for \cite{Cicoli:2021skd}:
\begin{equation}
\tau_\K3 = \, e^{\sqrt{\frac23}\,\chi + \frac{2}{\sqrt{3}}\,\phi} \qquad\text{and}\qquad
\tau_\P1 = \, e^{\sqrt{\frac23}\,\chi - \frac{1}{\sqrt{3}}\,\phi} \,.
\label{CanNorm}
\end{equation}
These relations can be easily inverted to find:
\begin{equation}
\mathcal{V} \simeq \, \sqrt{\tau_\K3}\,\tau_\P1 = t_\P1\,\tau_\K3 = e^{\sqrt{\frac32}\,\chi}\qquad\text{and}\qquad
r \equiv \, \frac{\tau_\K3}{\tau_\P1} =\frac{\sqrt{\tau_\K3}}{t_\P1} =  e^{\sqrt{3}\,\phi}\,,
\label{CanNormNew}
\end{equation}
implying that the flat direction in the $(\tau_\P1,\tau_\K3)$ space is $r$ which corresponds to the ratio between the root of the volume of the K3 fibre and the volume of the $\mathbb{P}^1$ base. To get $t_\P1\gg \sqrt{\tau_\K3}$, subdominant corrections need to fix $r$ very small. Without loss of generality, we shall instead parametrise the saxionic flat direction as $\tau_\K3$, and show that subleading perturbative corrections to $K$ can fix $\tau_\K3\sim\mathcal{O}(5)$. Consequently, $t_\P1\simeq\vo /\tau_\K3$ turns out to be exponentially large since $\vo$ is fixed exponentially large at leading order.

\subsubsection{Subleading perturbative stabilisation}
\label{SPS}

Let us introduce subdominant perturbative corrections to fix $\tau_\K3$. Note that the $2$ axions $\theta_\P1$ and $\theta_\K3$ enjoy a shift symmetry that is exact at perturbative level, and so they would not be lifted by these effects. They would instead be stabilised by additional non-perturbative corrections to $W$ of the form $W_{\text{np}} = A_\P1\,e^{-a_\P1 T_\P1} + A_\K3\,e^{-a_\K3 T_\K3}$. However, these effects are tiny, and so would lead to even more subleading contributions to $V$ due to $\tau_\P1 \simeq\vo/\sqrt{\tau_\K3}\gg 1$ (as $\vo\gg 1$) and $a_\K3\gg 1$ (since the K3, being non-rigid, is expected to contribute to the instanton series only at higher order). We shall therefore ignore the $2$ axions given that they are expected to be fixed at next-to-next-to-leading order. 

The leading order perturbative corrections to the LVS potential (\ref{LVSPot}) are: 
\begin{enumerate}
\item \textbf{String loop corrections:} These are open string 1-loop corrections whose leading contribution to $V$ takes the form:
\begin{equation}
V_{\text{1-loop}}\simeq - g_s\, \frac{c_{\text{loop}}\,W_0^2}{\vo^3\sqrt{\tau_\K3}}\,,
\label{V1loop}
\end{equation}
where $c_{\text{loop}}$ is expected to be a 1-loop factor of order $c_{\text{loop}}\simeq 1/(16\pi^2)$. This term is subdominant with respect to the leading LVS potential (\ref{LVSPot}) as can be seen from considering the ratio between the 1-loop contribution (\ref{V1loop}) and the $\mathcal{O}(\alpha'^3)$ term in (\ref{LVSPot}):
\begin{equation}
\frac{|V_{\text{1-loop}}|}{V_{\alpha'^3}} \simeq \frac{c_{\text{loop}}\,g_s^{5/2}}{\sqrt{\tau_\K3}}\ll 1 \quad\text{for}\quad g_s\ll 1\quad\text{and}\quad c_{\text{loop}}\ll 1\,.
\end{equation}
In particular, for $g_s\sim\mathcal{O}(0.1)$, $c_{\text{loop}}\sim\mathcal{O} (10^{-2})$ and $\tau_\K3\sim\mathcal{O}(5)$, one would get $|V_{\text{1-loop}}|/V_{\alpha'^3}\sim \mathcal{O}(10^{-5})$.

\item \textbf{Higher-derivative $\alpha'$ corrections:} These are $\mathcal{O}(\alpha'^3)$ corrections arising at higher $F^4$ order in the superspace derivative expansion and their leading expression reads:
\begin{equation}
V_{F^4} \simeq \sqrt{g_s}\, \frac{c_{F^4}\,W_0^4}{\vo^3\,\tau_\K3}\,,
\label{VF4}
\end{equation}
where $c_{F^4}$ is a positive topological factor which for a K3 divisor is expected to be of order $c_{F^4}\simeq 0.001$. Also this term is subleading with respect to the LVS potential (\ref{LVSPot}) since:
\begin{equation}
\frac{V_{F^4}}{V_{\alpha'^3}} \simeq \frac{c_{F^4}\,g_s^2\,W_0^2}{\tau_\K3}\ll 1\quad\text{for}\quad g_s\ll 1\quad\text{and}\quad c_{F^4}\ll 1\,.
\end{equation}
In particular, for $g_s\sim\mathcal{O}(0.1)$, $c_{F^4}\sim\mathcal{O}(10^{-3})$, $W_0\sim\mathcal{O}(1)$ and $\tau_\K3\sim\mathcal{O}(5)$, one would get $V_{F^4}/V_{\alpha'^3}\sim \mathcal{O}(10^{-5})$.
\end{enumerate}
Including both (\ref{V1loop}) and (\ref{VF4}), the subleading perturbative potential takes the form:
\begin{equation}
V_{\text{sub}} \simeq \frac{W_0^2}{\mathcal{V}^3}\left(-\frac{A}{\sqrt{\tau_\K3}}+\frac{B}{\tau_\K3}\right) \,,
\label{Vsub}
\end{equation}
where $A \equiv g_s\,c_{\text{loop}}$ and $B\equiv \sqrt{g_s}\,c_{F^4}\,W_0^2$. Being subdominant with respect to $V_\LVS$, the potential $V_{\text{sub}}$ can lift the flat direction $\tau_\K3$ without destroying the leading order stabilisation of $\vo$ at exponentially large values. The potential $V_{\text{sub}}$ admits a minimum at:
\begin{equation}
\langle\tau_\K3\rangle \simeq \left(\frac{2B}{A}\right)^2  = \frac{1}{g_s}\left(\frac{2\,c_{F^4}\,W_0^2}{c_{\text{loop}}}\right)^2\,,
\label{tauK3Min}
\end{equation}
which can naturally lie at $\langle\tau_\K3\rangle\sim\mathcal{O}(5)$ for $g_s\sim\mathcal{O}(0.1)$, $c_{\text{loop}}\sim\mathcal{O}(10^{-2})$, $c_{F^4}\sim\mathcal{O}(10^{-3})$ and $W_0\sim\mathcal{O}(1)$. Note that this minimum is AdS, with depth:
\begin{equation}
\langle V_{\text{sub}}\rangle \simeq - \frac{B\,W_0^2}{\vo^3\langle\tau_\K3\rangle}\,.
\end{equation}
For the ADD scenario, using (\ref{MKK6D}) with $u\simeq 1$, the vacuum energy scales as (for $W_0\sim\mathcal{O}(1)$):
\begin{equation}
\langle V_{\text{sub}}\rangle \sim -\left(M_\KK^{6D}\right)^3 \,M_p\sim -\frac{M_p}{R^3}\,,
\end{equation}
implying that, in order to obtain the right positive value of the cosmological constant, the uplifting coefficient $c_{\text{up}}$ has to be chosen to cancel this contribution to get at the end $\langle V_{\text{tot}}\rangle \sim \left(M_\KK^{6D}\right)^4 \sim M_p^4/\vo^4 \sim (\text{meV})^4$ for $\vo\sim 10^{30}$.

For the DD scenario the relevant 5D KK scale is given by (\ref{MKK5D}) with $\tau_\K3^{3/2}/\vo \ll u\ll 1$. In the extreme limit $u\to \tau_\K3^{3/2}/\vo$ where the 6D KK scale approaches the 10D KK scale, one finds $M_\KK^{5D}\simeq M_p\,\tau_\K3^{5/4}/\vo^{3/2}$, giving (again for $W_0\sim\mathcal{O}(1)$):
\begin{equation}
\langle V_{\text{sub}}\rangle \sim -\left(M_\KK^{5D}\right)^2 \,M_p^2\sim -\frac{M_p^2}{R^2}\,.
\label{Vsubvac}
\end{equation}
Hence in this case $c_{\text{up}}$ would have to be tuned to cancel this contribution, remaining with $\langle V_{\text{tot}}\rangle \sim \left(M_\KK^{5D}\right)^4 \sim M_p^4/\vo^6 \sim (\text{meV})^4$ for $\vo\sim 10^{20}$. Note that, for larger values of $u$, the power of $M_\KK^{5D}$ in (\ref{Vsubvac}) would increase from $2$ up to $3$ when $u\to 1$ and the DD scenario reduces to the ADD one.

Let us finally comment on the possibility to construct ADD and DD scenarios as dynamical dark energy instead of a dS vacuum. In this approach, quintessence could be realised along a shallow runaway with all the other directions stabilised. This can be done consistently only if the quintessence field is either the saxion $r$ as in \cite{Cicoli:2012tz}, or one of the two axions $\theta_\P1$ and $\theta_\K3$ as in \cite{Cicoli:2021skd,Cicoli:2024yqh}. However, the scale of the potential (\ref{Vsub}) would generically be too large for $r$ (as explained above in the absence of uplifting), and too small for the axions (due to the huge exponential suppression). For the case of saxion quintessence, the leading potential for $\tau_\K3$ could be a runaway generated by either string loops or higher-derivative $\alpha'$ corrections.\footnote{See however \cite{Cicoli:2012tz} for quintessence driven by a K\"ahler modulus in a K3 fibred LVS model where the potential is generated by poly-instantons under the assumption that loops are suppressed.} Using the canonical normalisation (\ref{CanNorm}), it would scale as $V \sim V_0\,e^{-c\,\phi}$, with $c=1/\sqrt{3}$ for (\ref{V1loop}) and $c=2/\sqrt{3}$ for (\ref{VF4}). In principle, the slope $c=1/\sqrt{3}$ would match observations extremely well \cite{Akrami:2025zlb}, but obtaining $V\sim \left(\text{meV}\right)^4$ for $g_s\sim 0.1$, $c_{\text{loop}}\sim 0.01$ and $\tau_\K3\sim 5$, would require extremely tuned values of $W_0$ of order $W_0\sim 10^{-12}$ for the ADD case and $W_0\sim 10^{-27}$ for the DD scenario. Such small values of the flux superpotential could be achieved in perturbatively flat vacua \cite{Carta:2022oex} but would be highly non-generic. Moreover, without a minimum that can fix $\tau_\K3$, there is no dynamical mechanism which justifies why quintessence occurs in the region in moduli space around $\tau_\K3\sim 5$. Note that if dark matter consists of 6D (in ADD) or 5D (in DD) KK gravitons \cite{Gonzalo:2022jac}, one could try to build an interacting dark energy-dark matter model where $M_\KK^{5D}\sim M_{\KK,0}^{5D} \, e^{-c'\,\phi}$ with $c'=-1/c$. However, $|c'|$ would be at least one order of magnitude too large to fit observations \cite{Bedroya:2025fwh}. This implies that dark matter can be 6D/5D KK gravitons only in a dS vacuum, while in a quintessence model driven by $r$ dark energy would not interact with dark matter which, as we shall see in Sec. \ref{Challenges}, could be given by the gravitino or the overall volume modulus (whose masses, in fact, do not depend on $r$).

\subsection{Complex structure stabilisation}
\label{CxStrStab}

To realise the DD scenario one further step is needed: the $\mathbb{P}^1$ base has to become anisotropic dynamically via moduli stabilisation, so that effectively only $1$ large extra dimension remains for $u = \ell_\I/R\ll 1$. Intuitively, the $\mathbb{P}^1$ base, which is the Riemann sphere, must be stretched so that one direction becomes much longer than the transverse one. This deformation is controlled by the complex structure moduli of the CY threefold and corresponds to a particular infinite-distance limit in complex structure moduli space, $u\to 0$, corresponding to a so-called \emph{Tyurin degeneration} that we will describe in detail in Sec. \ref{sec:base-deformation}.

Concretely, in a Tyurin degeneration the CY splits into two pieces $Z_a$ and $Z_b$ which meet along a common K3 surface $S$. In this limit the $\mathbb{P}^1$ base pinches, effectively splitting into two spheres joined at a point, and the points of the base where the K3 fibre becomes singular (due to the collapse of certain $2$-cycles of the K3 fibre) separate into two groups, with one group belonging to each of the two parts.

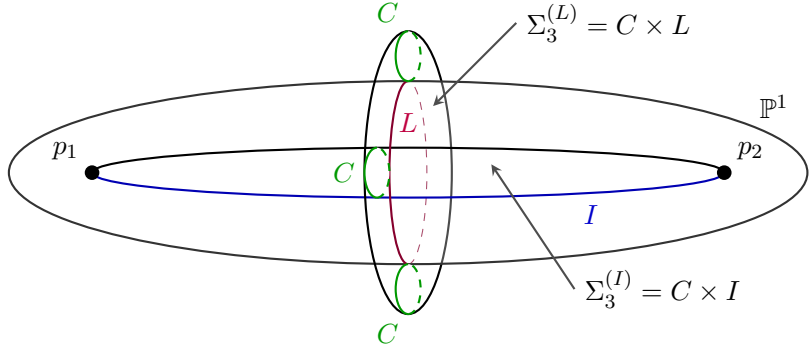
\begin{figure}[htbp]
\centering
\begin{tikzpicture}[scale=1.1, every node/.style={scale=0.95}]

        \draw[thick ,black] (0,-1.7) arc (-90:90:0.525cm and 1.7cm);
    
    \draw[dashed, purple!70!black] (0,-1.1) arc (-90:90:0.225cm and 1.1cm);
    
     \draw[thick, black!80, fill=white, fill opacity=0.30] (0,0) ellipse (4.8cm and 1.1cm);
    \node[black] at (4.4, 0.8) {$\mathbb{P}^1$};

     \draw[thick, black] (-3.8,0) arc (180:0:3.8cm and 0.3cm);
    \draw[thick, blue!70!black] (-3.8,0) arc (180:360:3.8cm and 0.3cm);

    \fill[black] (-3.8,0) circle (2.5pt) node[above left=2pt] {$p_1$};
    \fill[black] (3.8,0) circle (2.5pt) node[above right=2pt] {$p_2$};
    
    \node[blue!80!black] at (2.2, -0.5) {$I$};

    \draw[thick, black] (0,1.7) arc (90:270:0.525cm and 1.7cm);
    \draw[thick, purple!70!black] (0,1.1) arc (90:270:0.225cm and 1.1cm);

    \node[purple] at (0, 0.6) {$L$};

    \fill[white] (-0.375, 0) ellipse (0.15cm and 0.3cm);
    \draw[thick, dashed, green!60!black] (-0.375, 0) + (-90:0.15cm and 0.3cm) arc (-90:90:0.15cm and 0.3cm);
    \draw[thick, green!60!black] (-0.375, 0) + (90:0.15cm and 0.3cm) arc (90:270:0.15cm and 0.3cm);
    \node[green!60!black, left] at (-0.525, 0) {$C$};

    \fill[white] (0, 1.4) ellipse (0.15cm and 0.3cm);
    \draw[thick, dashed, green!60!black] (0, 1.4) + (-90:0.15cm and 0.3cm) arc (-90:90:0.15cm and 0.3cm);
    \draw[thick, green!60!black] (0, 1.4) + (90:0.15cm and 0.3cm) arc (90:270:0.15cm and 0.3cm);
    \node[green!60!black, above left] at (0, 1.7) {$C$};
 
    \fill[white] (0, -1.4) ellipse (0.15cm and 0.3cm);
    \draw[thick, dashed, green!60!black] (0, -1.4) + (-90:0.15cm and 0.3cm) arc (-90:90:0.15cm and 0.3cm);
    \draw[thick, green!60!black] (0, -1.4) + (90:0.15cm and 0.3cm) arc (90:270:0.15cm and 0.3cm);
    \node[green!60!black, below left] at (0, -1.7) {$C$};

    \draw[stealth-, thick, black!70] (0.3, 0.8) -- (1.3, 1.8) node[right, black] {$\Sigma_3^{(L)}=C\times L$};
    \draw[stealth-, thick, black!70] (1.0, 0.1) -- (2.0, -1.4) node[right, black] {$\Sigma_3^{(I)}=C\times I$};

\end{tikzpicture}
\caption{Pictorial view of the anisotropic limit of the $\mathbb{P}^1$ base that realises the DD scenario.}
\label{Fig1}
\end{figure}

The important result is that in this degeneration limit the $\mathbb{P}^1$ base does not just split but it becomes a long thin tube connecting two regions. This is detected using two special $3$-cycles of the CY, $\Sigma_3^{(L)}$ and $\Sigma_3^{(I)}$, whose size depends on the geometry of the $\mathbb{P}^1$ base. These $3$-cycles are built from a $2$-cycle in the K3 fibre $C$ times paths in the base. For $\Sigma_3^{(L)}=C\times L$ the path in the base is the loop $L$ ($1$-cycle) which shrinks to zero size when the $\mathbb{P}^1$ base pinches. Note that $\Sigma_3^{(L)}$ defines a global $3$-cycle since $C$ comes back to itself when transported along $L$ given that the total monodromy of the K3 fibration around the pinching loop $L$ is trivial. On the other hand, for $\Sigma_3^{(I)}=C\times I$ the path in the base is an interval $I$ connecting two points where the K3 fibre becomes singular due to the collapse of the curve $C$. The two endpoints of $I$, $p_1$ and $p_2$, lie in the two different components, i.e. $p_1\in Z_a$ and $p_2\in Z_b$. The fact that $C$ shrinks at the endpoints of $I$ guarantees that $\Sigma_3^{(I)}$ is a well-defined $3$-cycle. The $3$-cycle volumes are computed by integrating the CY holomorphic $(3,0)$-form $\Omega_3$ over them. The Tyurin degeneration is then a limit in complex structure moduli space where
\begin{equation}
\text{Vol}(\Sigma_3^{(I)})\simeq\int_{C\times I} \Omega_3 \gg \text{Vol}(\Sigma_3^{(L)}) \simeq \int_{C\times L} \Omega_3 \,,
\label{3cycleDegen}
\end{equation}
corresponding to the case where in the base the circular direction of radius $\ell_\I$ becomes small, while the `length direction' of size $R$ becomes large, $u = \ell_\I/R\ll 1$, so that the $\mathbb{P}^1$ base effectively turns into a long tube, as illustrated pictorially in Fig. \ref{Fig1}. Hence the model features just a single large extra dimension.

Realising the DD scenario, therefore, amounts to finding appropriate background flux quanta which can fix the complex structure moduli creating a hierarchy between $3$-cycle volumes as in (\ref{3cycleDegen}). We leave the detailed study of flux stabilisation to future work. Here we simply stress that such an anisotropic limit exists in complex structure moduli space. The two scenarios, ADD and DD, are therefore two endpoints of the same construction, distinguished only by the values of the volume $\vo$ and by whether the complex structure modulus $u$ is chosen to keep the base isotropic or to make one direction of the base much longer than the transverse one.

\subsection{Theoretical and phenomenological challenges}
\label{Challenges}

Let us now discuss some of the theoretical and phenomenological challenges involved in realising models with 1 or 2 large EDs:
\begin{itemize}
\item \textbf{Further quantum corrections:} The main concern associated with any moduli stabilisation mechanism which exploits corrections beyond the semi-classical level is the robustness of the resulting vacuum against any kind of quantum corrections. In particular, one should ensure that no quantum corrections are larger than the string loop contribution (\ref{V1loop}) and the $F^4$ term (\ref{VF4}). Perturbative corrections are organised in two expansions, one in $g_s$ and another in $\alpha'$, as classified systematically in \cite{Burgess:2020qsc, Cicoli:2021rub, Cicoli:2024bwq}, with no correction known to arise at $\mathcal{O}(\alpha')$. Given that (\ref{V1loop}) is a term of $\mathcal{O}(g_s^2\alpha'^4)$, while (\ref{VF4}) is of $\mathcal{O}(\alpha'^3)$ but at $F^4$ level, the most dangerous corrections would be those scaling as $\alpha'^2$ at some order in the $g_s$ expansion. The first of these corrections seem to arise at $\mathcal{O}(g_s \alpha'^2)$. It is, however, still under debate whether these effects generate a genuine correction to $V$, as argued in \cite{Weissenbacher:2019mef,Weissenbacher:2020cyf}, or whether they can be entirely absorbed into a moduli redefinition, as found in \cite{Grimm:2013gma}. Moreover, $\mathcal{O}(\alpha'^2)$ corrections at any order in $g_s$ enjoy the extended no-scale cancellation \cite{Cicoli:2007xp}, and so would not contribute to $V$, unless they involve logarithmic contributions as those arising due to gauge threshold corrections to the gauge kinetic function of the field theory living on D7-branes wrapped around $4$-cycles. These have been computed just for blow-up modes in \cite{Conlon:2009kt,Conlon:2009xf}, finding a logarithmic correction only for blow-ups of orientifolded singularities. No worldsheet computation has been performed for bulk cycles like $\tau_\K3$ and $\tau_\P1$. However, ref. \cite{Klaewer:2020lfg} used F-theory/heterotic duality to infer the presence of $\mathcal{O}(\alpha'^2)$ logarithmic corrections of the form: 
\begin{equation}
\tau_i^{\text{new}} = \tau_i + \beta_i\,\ln\vo \qquad i=\text{K3},\,\P1\,,
\label{LogCorr}
\end{equation}
where $\beta_i =\beta_{0,i}/(8\pi)$ depends on the 1-loop $\beta$-function coefficient $\beta_{0_i}$ of the gauge theory on D7s wrapped around the $4$-cycle with volume $\tau_i$, $i=\text{K3},\,\P1$. Note that these effects would yield an actual correction to $V$ only under the assumptions that: ($i$) both the K3 fibre and the divisor including the $\mathbb{P}^1$ base are wrapped by D7-branes; ($ii$) the logarithmic corrections (\ref{LogCorr}) are indeed present; ($iii$) the effects (\ref{LogCorr}) induce corrections to $K$ obtained just by replacing $\tau_i$ with $\tau_i^{\text{new}}$. If these conditions are met then one would generate new contributions to $V$ of the form:
\begin{equation}
\delta V_\P1 \sim \beta_\P1\,\frac{W_0^2\,\sqrt{\tau_\K3}}{\vo^3}\qquad\text{and}\qquad
\delta V_\K3 \sim \beta_\K3\,\frac{W_0^2}{\vo^2\,\tau_\K3}\,.
\label{dVnew}
\end{equation}
Interestingly, $\delta V_\P1$, when combined with (\ref{V1loop}) and (\ref{VF4}), could yield a dS minimum for $\beta_\P1>0$, while $\delta V_\K3$ would be problematic since it would induce a minimum at exponentially large $\tau_\K3$.

Other interesting logarithmic corrections might arise at $\mathcal{O}(g_s^2\alpha'^3)$ if the bulk features a localised region of high curvature \cite{Antoniadis:2018hqy,Antoniadis:2019rkh}. In this case the $\alpha'^3$-corrected K\"ahler potential (\ref{Kalphap3}) would become:
\begin{equation}
\delta K_{\alpha'^3}=-\frac{\xi}{g_s^{3/2}\vo}\left(1- c\,g_s^2\,\ln\vo\right),
\end{equation}
and the new logarithmic correction would allow to get an LVS-like minimum at exponentially large volume without the need to introduce a dP divisor supporting non-perturbative effects \cite{Leontaris:2022rzj}.

\item \textbf{Hierarchical flux stabilisation:} As explained in Sec. \ref{CxStrStab}, a realisation of the DD scenario requires a hierarchical structure among the volumes of the $3$-cycles $\Sigma_3^{(L)}$ and $\Sigma_3^{(I)}$ which control the level of anisotropy of the $\mathbb{P}^1$ base: $u=\ell_\I/R\ll 1$. This limit should be achieved dynamically via complex structure moduli stabilisation. The primary source for the potential of these modes is given by quantised $3$-form fluxes. Tadpole cancellation constrains these flux quanta to be in general $\mathcal{O}(1-10)$ numbers, with the possibility to have some individual flux quanta of $\mathcal{O}(10-100)$. Hence, it is reasonable to expect that flux quanta can generate only modest hierarchies, at most of order $\mathcal{O}(10^{-3})\lesssim u < 1$ which would reproduce a DD scenario with 6D KK scale relatively close to the 5D KK scale. To get a much larger 6D KK scale, as can be seen from Tab. \ref{Tab2}, one needs much smaller values of $u$ with the limiting value $u\gtrsim\mathcal{O}(10^{-20})$ corresponding to $M_\KK^{6D}\sim M_\KK^{10D}\sim M_s$. This situation might be reproduced by focusing on flux vacua where the fixing of $u$ requires balancing the perturbative superpotential against worldsheet instantons, mimicking an LVS-type stabilisation in complex structure moduli space where $u$ turns out to be exponentially small in terms of flux quanta. More generally, stabilising complex structure moduli close to a prescribed locus in moduli space suggests an inverse approach to flux stabilisation. A related strategy has been developed in F-theory, where $G_4$ fluxes constructed from algebraic cycles can stabilise moduli at prescribed complex structures \cite{Braun:2020jrx}. It would be interesting to understand whether analogous ideas can be adapted to the type IIB setting relevant for our construction.

\item \textbf{Explicit uplifting sector:} The explicit CY orientifold example that we will present in Sec. \ref{sec:ExplicitCY} lacks a concrete sector responsible to uplift the vacuum energy to dS. The simplest deformation of our construction that could induce a positive uplifting contribution would involve a non-supersymmetric stabilisation of the complex structure moduli \cite{Gallego:2017dvd}. The main challenge in this case would be to find flux configurations which lead to tiny non-zero F-terms of the complex structure moduli, in order to avoid a runaway. Recent applications of machine-learning techniques to mapping the type IIB flux landscape could be very helpful in this search \cite{Ebelt:2023clh,Chauhan:2025rdj,Chauhan:2026gid}.

Another option is to include $\overline{D3}$-brane uplift in a setup where the orientifold involution yields O3-planes that can be moved on top of each other at the tip of the throat associated with a deformed conifold singularity \cite{Kallosh:2015nia,Garcia-Etxebarria:2015lif}. Explicit K3-fibred CY realisations of LVS models with this type of uplifting were found in \cite{AbdusSalam:2022krp,Cicoli:2024bxw}, following \cite{Crino:2020qwk}. The main problem in these constructions is the tension between tadpole cancellation and the requirement to find a minimum at very large $\vo$ without inducing a runaway.

Finally, another common source of uplifting in concrete CY orientifold models with D7-branes, gauge and background fluxes is the one based on T-branes \cite{Cicoli:2015ylx}. The corresponding uplifting contribution would be induced by considering D-term stabilisation where charged matter fields develop non-zero VEVs. Explicit LVS CY models with T-brane uplifting can be found in \cite{Cicoli:2012vw,Cicoli:2013mpa,Cicoli:2013cha,Cicoli:2017shd,Cicoli:2021dhg}. Again the main challenge is to find flux configurations which can induce a minimum at large enough CY volume.

\item \textbf{Supersymmetry breaking:} Given that $\vo$ sets the size of all mass scales, lowering the 5D/6D KK scale to the meV range tends to give a very low scale of supersymmetry breaking. Indeed, the gravitino mass is related to the KK scale as:
\begin{equation}
m_{3/2}\sim \frac{W_0\,M_p}{\vo}\sim W_0\left(\frac{M_\KK}{M_p}\right)^{\frac{2d}{(2+d)}} M_p\,.
\end{equation}
Using the values of Tab. \ref{Tab2} with $W_0\sim\mathcal{O}(1\,$-$\,5)$, we have:
\begin{eqnarray}
&&d=1:\qquad m_{3/2} \sim \left(\frac{M_p}{M_\KK^{5D}}\right)^{1/3} M_\KK^{5D} \sim 10\,\text{MeV} \\
&&d=2:\qquad m_{3/2}\sim M_\KK^{6D} \sim 1\,\text{meV} 
\end{eqnarray}
In gravity mediation, soft terms are of order $M_{\text{soft}}\lesssim m_{3/2}$, implying that superpartners should have already been observed for such a small value of the gravitino mass. An alternative way to communicate supersymmetry breaking to the visible sector is gauge mediation which, if dominant over gravity mediation, might lead to soft terms above the TeV scale. This is, however, incompatible with our construction, as a viable gauge mediation scenario relies on 3 conditions that cannot all be met simultaneously. In fact, if the spurion superfield responsible for breaking supersymmetry  is $X=x+ \theta^2 F^X$, we need:
\begin{eqnarray}
&&(i)\qquad M_{\text{soft}}^{\text{gauge}}\simeq\left(\frac{\alpha}{4\pi}\right)\frac{F^X}{x}\gtrsim 1\,\text{TeV}\,, \\
&&(ii)\qquad F^X<x^2\,, \\
&& (iii)\qquad |F^X|^2\lesssim m_{3/2}^3\,M_p\,.
\end{eqnarray}
The first condition is to have superpartners above the TeV scale, the second to avoid tachyonic messengers, and the third to prevent a runaway in the potential caused by the large F-term of the spurion field. Combining $(i)$ with $(ii)$ yields $x>10^5$ GeV, while combining $(i)$ with $(iii)$ gives $x<10$ GeV for DD and $x<0.01$ meV for ADD.

The only possible way out seems to consider a non-supersymmetric SM brane where supersymmetry is realised non-linearly. In this case, the mass of the superpartners would be of the same order as the string scale, $M_{\text{soft}}\sim M_s$, i.e. $M_{\text{soft}}\sim 10^8$ GeV for DD and $M_{\text{soft}}\sim 10^3$ GeV for ADD. The main challenge is that the brane configuration is now non-supersymmetric. As a result, one expects a sizeable contribution to the vacuum energy from loops of SM open strings, of order $V\sim M_s^4$ which would generically spoil moduli stabilisation. Avoiding this outcome would require this contribution to be cancelled by the backreaction of the SM brane, as has been argued in SLED scenarios \cite{Aghababaie:2003wz}.

\item \textbf{Cosmological moduli overproduction and fifth forces:} Other important scales which are affected by the exponentially large CY volume are the moduli masses. The K\"ahler moduli mass spectrum looks like:
\begin{eqnarray}
m_{\tau_\dP}&\sim& m_{\theta_\dP}\sim m_{3/2}\,\ln\left(\frac{M_p}{m_{3/2}}\right)\qquad m_\vo\sim m_{3/2}\,\sqrt{\frac{m_{3/2}}{M_p}} \\
m_r&\sim&\frac{m_\vo}{\sqrt{\langle\tau_\K3\rangle}}\qquad m_{\theta_\P1}\sim m_{\theta_\K3} \sim 0\,.
\end{eqnarray}
As typical in the LVS framework, the complex structure moduli, including $u$, are expected to develop masses of order the gravitino mass (for example when $\mathcal{O}(10^{-3})\lesssim u < 1$) or even smaller, if their stabilisation is due to worldsheet instantons (as might be required to be the case for $u\gtrsim\mathcal{O}(10^{-20})$).  In particular, if the F-term of $u$ is non-zero in the vacuum, one can obtain a lower bound on its mass by demanding the absence of a runaway direction in the LVS potential: $m_\vo \lesssim m_u\lesssim m_{3/2}$.

Using again the values of Tab. \ref{Tab2}, we find the moduli mass spectra shown in Tab. \ref{TabNew}.

\begin{table}[ht]
\centering
\begin{tabular}{c | c c | c c c c c }
\hline
& $u$ & $\mathcal{V}$ & $m_\dP$ & $m_{3/2}$ & $m_\vo$ & $m_r$ \\
\hline
\textbf{ADD} & $1$ & $5\cdot 10^{30}$ & $100\,\text{meV}$ &  $1\,\text{meV}$ & $10^{-18}\,\text{eV}$ & $10^{-18}\,\text{eV}$  \\
\textbf{DD} & $10^{-8}$ & $5\cdot 10^{26}$ & $10^3\,\text{eV}$ & $1\,\text{eV}$ & $10^{-12}\,\text{eV}$ & $10^{-12}\,\text{eV}$ \\
\textbf{DD} & $10^{-20}$ & $5\cdot 10^{20}$ & $1\,\text{GeV}$ & $10\,\text{MeV}$ & $1\,\text{meV}$ & $1\,\text{meV}$ \\
\hline
\end{tabular}
\caption{Moduli mass spectra for the ADD and the DD scenarios for typical $\mathcal{O}(5)$ values of $W_0$.}
\label{TabNew}
\end{table}

Three cosmologically important mass scales for the moduli are the following ones: 
\begin{enumerate}
\item \textbf{Cosmological moduli problem:} A necessary condition for gravitationally coupled moduli to decay before BBN is $m_{\text{mod}}\gtrsim \mathcal{O}(50)$ TeV.

\item \textbf{Dark matter overproduction:} Moduli are cosmologically stable for $m_{\text{mod}}\lesssim \mathcal{O}(10)$ MeV. In this case, they contribute to dark matter via their oscillations around the minimum of the potential in the standard misalignment mechanism.

\item \textbf{Fifth forces:} Moduli which couple to ordinary matter with Planckian strength do not lead to unobserved fifth forces only if $m_{\text{mod}}\gtrsim \mathcal{O}(1)$ meV.
\end{enumerate}

In the ADD scenario, $\vo$ and $r$ generically violate fifth force bounds unless their coupling to visible sector fields is highly suppressed, for example by realising the SM on a sequestered brane \cite{Acharya:2018deu}, or by some screening mechanism. Moreover, all moduli are stable, and so each of them would contribute to dark matter as:
\begin{equation}
\Omega_\phi\,h^2\sim 5\cdot 10^{13}\left(\frac{\phi_{\text{in}}}{M_p}\right)^2\sqrt{\frac{m_\phi}{1\,\text{MeV}}}\,,   
\end{equation}
where $\phi_{\text{in}}$ is the initial misalignment. Requiring that dark matter is not overproduced, i.e. $\Omega_\phi\,h^2\lesssim 0.12$, turns into the following upper bound on $\phi_{\text{in}}$:
\begin{equation}
\phi_{\text{in}}\lesssim \phi_{\text{in}}^{\text{max}}\simeq 5\cdot 10^{-8}\,M_p\left(\frac{1\,\text{MeV}}{m_\phi}\right)^{1/4}\,.
\end{equation}
Tab. \ref{TabThree} shows an estimate of the limit on the initial moduli misalignment for the moduli masses of our interest shown in Tab. \ref{TabNew}. Note that achieving initial misalignments as small as $\phi_{\text{in}}\lesssim\mathcal{O}(10^{-5}-10^{-6})\,M_p$ might be a challenge as explicit computations for concrete inflationary models tend to give $\phi_{\text{in}}\sim\mathcal{O}(0.1-0.01)\,M_p$ \cite{Cicoli:2016olq}.

\begin{table}[ht]
\centering
\begin{tabular}{ | c  | c |}
\hline
$m_\phi$ & $\phi_{\text{in}}^{\text{max}}$ \\
\hline
$100\,\text{meV}$ & $5\cdot 10^{-6}\,M_p$    \\
$1\,\text{meV}$ & $10^{-5}\,M_p$    \\
$10^{-18}\,\text{eV}$ & $0.05\,M_p$    \\
\hline
\end{tabular}
\caption{Maximal value of moduli initial misalignment to avoid dark matter overproduction.}
\label{TabThree}
\end{table}

In the DD scenario, $\tau_\dP$ and $\theta_\dP$ suffer from the CMP only in the extreme case with $u\simeq 10^{-20}$.\footnote{Note that in perturbatively stabilised LVS vacua these moduli are simply absent \cite{Leontaris:2022rzj}.} For larger values of $u$ they tend to become stable, as all the other moduli, which therefore all contribute to dark matter. Note that, for $u\simeq 10^{-20}$, all moduli are heavier than the meV scale, resulting in no problem with fifth forces. However, complex structure moduli with $m_{\text{mod}}\sim 10$ MeV would overproduce dark matter unless $\phi_{\text{in}}\lesssim 3\cdot 10^{-8}\,M_p$, which seems very hard to achieve, unless some dilution mechanism is at play. Increasing $u$ makes moduli lighter, reducing the tuning on the initial misalignment parameter to avoid dark matter overproduction, but introduces the problem with fifth forces mediated by $\vo$ and $r$.

These considerations show that making ADD and DD scenarios compatible with both cosmological moduli overproduction and fifth force bounds is rather challenging. Note that similar phenomenological problems, as pointed out in \cite{Cicoli:2021skd}, are shared by models where quintessence is driven by a saxion.

\end{itemize}

\section{An explicit Calabi-Yau example}
\label{sec:ExplicitCY}

We now illustrate the results of Sec.~\ref{sec:AnisModStab} in more detail by focusing on a specific Calabi-Yau example.

\subsection{Toric data and divisor volumes}
\label{sec:toric-data}

We introduce an explicit CY example taken from \cite{Cicoli:2017axo}. This model has $h^{1,1}=4$ but D-term stabilisation makes one K\"ahler modulus very heavy, reducing the volume form effectively to (\ref{CYVol_new}). The GLSM data are given by:
\begin{table}[H]
  \centering
 \begin{tabular}{|c|cccccccc|}
\hline
     & $x_1$  & $x_2$  & $x_3$  & $x_4$  & $x_5$ & $x_6$  & $x_7$   & $x_8$    \\
    \hline
4 & 0 & 0 & 0 & 1 & 1 & 0 & 0 & 2 \\
4 & 0 & 0 & 1 & 0 & 0 & 1 & 0 & 2 \\
4 & 0 & 1 & 0 & 0 & 0 & 0 & 1 & 2 \\
8 & 1 & 0 & 0 & 1 & 0 & 1 & 1 & 4 \\ \hline
  & dP$_7$  & NdP$_{11}$ & NdP$_{11}$ &  K3 & NdP$_{11}$ & K3  &  K3 & SD  \\
    \hline
  \end{tabular}
 \end{table}
 \noindent
The Hodge numbers are $(h^{1,2}, h^{1,1}) = (98, 4)$, and therefore the Euler number is $\chi(X)= 2(h^{1,1} - h^{1,2}) = -188$. The Stanley-Reisner ideal reads
\begin{equation}
    \mathrm{SR} = \{x_1 x_4, \, x_1 x_6, \, x_1 x_7, \, x_2 x_7, \, x_3 x_6, \, x_4 x_5 x_8, \, x_2 x_3 x_5 x_8 \} \, .
    \label{SRidealofCY}
\end{equation}
This CY threefold corresponds to polytope ID $\#1206$ in the database of \cite{Altman:2014bfa}. We choose the following integral basis of toric divisors:
\begin{equation}\label{Eq:IntegrBasis}
    \{D_1, D_4, D_6, D_7\}\:.
\end{equation}
The remaining toric divisors can be expressed in terms of this basis as:
\begin{equation}
    \begin{aligned} 
        D_2 =& ~ D_7 - D_1 \, , \quad &&D_3 = D_6 - D_1\, , \\
        D_5 =& ~ D_4 - D_1 \, ,  \quad &&D_8 = 2 \left(D_4 + D_6 + D_7 - D_1 \right) \,.
    \end{aligned}
\end{equation}
In the chosen basis, the intersection polynomial is
\begin{equation}
    I_3 =2\, D_4 D_6 D_7 + 2\, D_1^3 \,.
    \label{intersecpolyofCYformodel}
\end{equation}
Oguiso's theorem states that, if the intersection polynomial is linear in $D_i$, then $D_i$ is a K3 (or a $T^4$) fibration over a $\mathbb{P}^1$ base \cite{oguiso:theorempaper}. This result implies that our CY threefold features $3$ K3 fibrations where the $3$ K3 divisors are $D_4$, $D_6$ and $D_7$. 

Using a basis of Poincar\'e dual $(1,1)$-forms $\hat D_i=\mathrm{PD}[D_i]$, the second Chern class is
\begin{equation}
c_2(X) = 4\,\hat{D}_1^2 + 12 \left( \hat{D}_4 \hat{D}_6 + \hat{D}_4 \hat{D}_7 + \hat{D}_6 \hat{D}_7\right),
\label{secondChernclassofCYforourmodel}
\end{equation}
while the K\"ahler form can be expanded as
\begin{equation}
J = t_1 \hat{D}_1 + t_4 \hat{D}_4 + t_6 \hat{D}_6 + t_7 \hat{D}_7 \,,
\label{expansofkahlerformondivisorbasisCYmodel}
\end{equation}
where the coefficients $t^i$ denote $2$-cycle volumes. The CY volume is given by
\begin{equation}
\mathcal{V} = \frac{1}{6} \int_X J \wedge J \wedge J = \frac{1}{6} \left(\int_X \hat{D}_i \wedge \hat{D}_j \wedge \hat{D}_k\right) t^i t^j t^k \equiv \frac{1}{6}\, \kappa_{ijk}\,t^i t^j t^k \,,
 \label{CYvolumewithintersecnumber}
\end{equation}
where $\kappa_{ijk}$ are the triple intersection numbers that can be read as the coefficients of each term of (\ref{intersecpolyofCYformodel}). Hence, the CY volume reduces to
\begin{equation}
\mathcal{V} = 2\, t_4 t_6 t_7 + \frac{1}{3}\, t_1^3 \,.
\label{volumeCYmodel2cyclesvolumes}
\end{equation}
The $4$-cycle volumes $\tau_i$ are expressed as:
\begin{equation}
\tau_i \equiv \frac{\partial \mathcal{V}}{\partial t^i} = \frac12 \int_X J \wedge J \wedge \hat{D}_i = \frac12\, \kappa_{ijk}\, t^j t^k \,.
\end{equation}
Using again (\ref{intersecpolyofCYformodel}), we obtain:
\begin{equation}
\tau_1 = t_1^2\,, \qquad \tau_4 = 2\, t_6 t_7 \,, \qquad\tau_6 = 2\, t_4 t_7 \,, \qquad \tau_7 = 2\, t_4 t_6 \,.
\label{4cyclesvolumesCYmodel}
\end{equation}
Note that the volume \eqref{volumeCYmodel2cyclesvolumes} can equivalently be written as:
\begin{equation}
\mathcal{V} = \, t_4 \tau_4 - \frac{1}{3} \tau_1^{3/2} = t_6 \tau_6 - \frac{1}{3} \tau_1^{3/2}
= \, t_7 \tau_7 - \frac{1}{3} \tau_1^{3/2}           = \, \frac{1}{\sqrt{2}} \sqrt{\tau_4 \tau_6 \tau_7} - \frac{1}{3} \tau_1^{3/2} \,,
\label{volumeCYmodel_h11=4}
\end{equation}
showing that this CY threefold $X$ admits indeed $3$ K3 fibrations over distinct $\mathbb{P}^1$ bases, as expected from Oguiso's theorem.

Let us finally report the conditions that define the K\"ahler cone:
\begin{equation}
t_1<0\,,\qquad t_1+t_7>0\,,\qquad t_1+t_4>0\,,\qquad t_1+t_6>0\,. 
\label{KahlerCone}
\end{equation}

\subsection{Orientifold involution and brane setup} \label{subsec:orientif_and_branesetup}

We now discuss the orientifold involution and the associated brane setup. We focus on the involution, already considered in \cite{Cicoli:2017axo}, which reflects the coordinate $x_8$. Its fixed-point locus is summarised in Tab.~\ref{table:involution} and it consists of a single O7-plane wrapping the divisor $D_8$ and no O3-planes, a feature that will be important in what follows.

\begin{table}[H]
  \centering
 \begin{tabular}{|c|c|c|c|}
\hline
&  &  &    \\
$\sigma$ & O7  & O3    & $\chi(\text{O7})$  \\
&  &  &    \\
\hline
\hline
$x_8 \to -x_8$ &  $D_8$ & $\emptyset$  & 208  \\
\hline
\end{tabular}
\caption{Fixed-point locus of the involution reflecting the coordinate $x_8$.}
\label{table:involution} 
\end{table}
\noindent

There is one rigid dP$_7$ divisor invariant under the orientifold involution, namely $D_1$. We will later require this divisor to support an invariant, i.e. $O(1)$, E3-instanton needed to fix the CY volume exponentially large. For this to be possible, the gauge-invariant flux on the E3 worldvolume must vanish, $\mathcal{F}_1=0$, where $\mathcal{F}_1=F_1-\iota^*_{D_1} B$, with $B$ the type IIB $B$-field, $\iota^*_{D_1}$ the pullback from $X$ to $D_1$, and $F_1$ the gauge flux on the E3 wrapping $D_1$. For Freed-Witten anomaly cancellation \cite{Freed:1999vc}, $F_1$ must satisfy the shifted quantisation condition
$F_1+\frac{c_1(D_1)}{2}=F_1-\frac{\iota^*_{D_1}\hat{D}_1}{2}\in H^2(D_1,\mathbb{Z})$. To be able to set $\mathcal{F}_1=0$, we then need $B$ to be properly half-quantised, choosing:
\begin{equation}
B = \frac{\hat{D}_1}{2}\,.
\end{equation}
This choice allows us to choose a properly quantised flux $F_1$ that makes $\mathcal{F}_1=0$.

We next turn to D7-tadpole cancellation. The D7-charge of the O7-plane must be cancelled by an appropriate set of D7-branes whose total D7-charge is $-8[\text{O7}]=-8D_8$. We consider two such setups. The first is the one used in \cite{Cicoli:2017axo}, while the second is a variation designed to generate a larger negative D3-charge which is useful for accommodating a larger number of background fluxes needed for moduli stabilisation.

\paragraph{First D7-brane setup:} The D7-charge of the O7-plane can be cancelled by taking:
\begin{equation}
\label{Eq:FirstD7setup}
\begin{aligned}
    &\mbox{$8+8$ D7s wrapping }D_2\,,\\
    &\mbox{$8+8$ D7s wrapping }D_4\,,\\
    &\mbox{$8+8$ D7s wrapping }D_6\,,
\end{aligned}
\end{equation}
where `$8+8$' denotes $8$ branes together with their $8$ orientifold images. Since $2(D_2+D_4+D_6)=D_8$, the total D7-charge is $8D_8$, as required.

The contribution of this D7/O7 system to the D3-charge is computed in the standard way (see \cite{Crino:2022zjk} for a review of D3-charge computations in our conventions where $Q^{\text{D3}}$ denotes the D3-charge in the double-cover CY threefold). The O7-plane charge is
\begin{equation}
\label{Eq:O7D3charge}
Q^{\text{D3}}_{\text{O7}} = -\frac{\chi(\text{O7})}{6} = -\frac{\chi(D_8)}{6} = - \frac{104}{3}\,,
\end{equation}
where, using the adjunction formula, $\chi(D_8)=\int_{D_8}c_2(D_8) = \int_{X}(\hat{D}_8^3+c_2(X)\wedge \hat{D}_8)=208$. On the other hand, a D7-brane wrapping a divisor $D$ and carrying worldvolume flux $\mathcal{F}$ contributes to the D3-charge as:
\begin{equation}
Q^{\text{D3}}_{\text{D7}} = -\frac{\chi(D)}{24} - \frac12 \int_D \mathcal{F}^2\:. 
\end{equation}
In our model $\chi(D_2)=14$ and $\chi(D_4)=\chi(D_6)=24$. Moreover, since $h^{1,1}_-(X)=0$, the D3-charge of a D7-brane and its image are the same (as $D'=D$ and $\mathcal{F}'=-\mathcal{F}$).

The total D3-charge of our brane setup is then
\begin{equation}
Q^{\text{D3}}_{\text{tot}} = Q^{\text{D3}}_{\text{O7}} + 16 \left(Q^{\text{D3}}_{\text{D7},D_2} + Q^{\text{D3}}_{\text{D7},D_4} + Q^{\text{D3}}_{\text{D7},D_6}\right) = -76\,\, + Q^{\text{D3}}_{\text{flux}} \,,
\label{D3First}
\end{equation}
where $Q^{\text{D3}}_{\text{flux}}$ denotes the contribution of the flux on the D7-branes in \eqref{Eq:FirstD7setup}. When we switch on the same Abelian flux on all the branes in the stack, the flux contribution becomes $ Q^{\text{D3}}_{\text{flux}}= - 8 \sum_{i=2,4,6} \int_{D_i} \mathcal{F}_i^2$.

\paragraph{Second D7-brane setup:}  Relative to the first D7-brane setup, we now recombine most of the branes, leaving the following configuration:
\begin{equation}
\label{Eq:SecondD7setup}
\begin{aligned}
    &\mbox{$1+1$ D7s wrapping }D_2\,,\\
    &\mbox{$1+1$ D7s wrapping }D_4\,,\\
    &\mbox{$1+1$ D7s wrapping }D_6\,,\\
    &\mbox{$1$ Whitney D7 wrapping }7D_8\:.
\end{aligned}
\end{equation}
The stacks wrapping the invariant divisors $D_2$, $D_4$ and $D_6$ support $Sp(1)$ gauge groups. The last object is a Whitney brane \cite{Collinucci:2008pf,Collinucci:2008sq}, namely an invariant D7-brane wrapping $D_W=2 D_P$. The worldvolume of the Whitney brane has the typical form of a \textit{Whitney umbrella}:
\begin{equation}
\eta^2(x) - x_8^2 \,\left( \psi^2(x) - \rho(x)\tau(x)  \right) = 0 \,,
\end{equation}
where $\eta$, $\psi$, $\rho$, $\tau$ are sections, respectively, of $\mathcal{O}(D_P)$, $\mathcal{O}(D_P-[\text{O7}])$, $\mathcal{O}(2(D_P-S_F-B)-[\text{O7}])$ and $\mathcal{O}(2(S_F+B)-[\text{O7}])$. Here $S_F$ determines the flux class on the Whitney brane. The contribution of a Whitney brane to the D3-charge can be computed straightforwardly (see \cite{Collinucci:2008sq,Braun:2011zm,Crino:2022zjk}):
\begin{equation}
Q^{\text{D3}}_W = -\frac{\chi(D_P)}{12} - \int_X \hat{D}_P\wedge \left( \hat{S}_F-\frac{\hat{D}_P}{2} - B \right)^2\,.
\end{equation}
The integral $2$-form $\hat{S}$ must satisfy the following condition, assuming that the sections $\eta$, $\psi$, $\rho$ and $\tau$ are generic and that the Whitney brane is stable against splitting into disjoint branes \cite{Collinucci:2008sq}:
\begin{equation}
\label{Eq:WD7condStab}
\frac{\mathrm{PD}[\text{O7}]}{2}+B \,\leq\, \hat{S}\, \leq\, \hat{D}_P-\frac{\mathrm{PD}[\text{O7}]}{2}+B\,.
\end{equation}
In the setup \eqref{Eq:SecondD7setup}, we have $D_W=2D_P=7D_8$ and $[\text{O7}]=D_8$, which gives $\chi(D_P)=3878$.\footnote{Notice from the toric data that $D_8$ is an even class.} Moreover, the condition \eqref{Eq:WD7condStab} becomes
\begin{equation}
\label{Eq:WD7condStab2}
- \frac{5\hat{D}_8}{4} \,\leq\, \hat{S} - \frac{\hat{D}_P}{2}-B\, \leq\, \frac{5\hat{D}_8}{4}\,.
\end{equation}
Taking into account the integrality properties of $\hat{S}$ and $\hat{D}_P/2$, together with $B=\hat{D}_1/2$, the maximal absolute value of $Q^{\text{D3}}_W$ is obtained for $\hat{S}-\frac{\hat{D}_P}{2} - B=\frac{5\hat{D}_8}{4}-\frac{\hat{D}_1}{2}$. Putting everything together, with this value of $\hat{S}$ we obtain:
\begin{equation}
Q^{\text{D3}}_W =  - \, \frac{4333}{6}\,.
\end{equation}
Adding now all contributions, we end up with:
\begin{equation}
\label{Eq:totQ3secondoD7setup}
Q^{\text{D3}}_{\text{tot}} = Q^{\text{D3}}_{\text{O7}} + 2 \left(Q^{\text{D3}}_{\text{D7},D_2} + Q^{\text{D3}}_{D7,D_4} +  Q^{\text{D3}}_{\text{D7},D_6}\right) + Q_W^{\text{D3}} = -762\,\, -  \sum_{i=2,4,6} \int_{D_i} \mathcal{F}_i^2 \,,
\end{equation}
where $\mathcal{F}_i$ are the Abelian worldvolume fluxes on the $Sp(1)$ stacks on $D_2$, $D_4$ and $D_6$.

\subsection{D-term stabilisation}

As explained in \cite{Cicoli:2017axo}, D-term stabilisation effectively reduces our explicit CY example to a model with $h^{1,1}=3$. To show that this is indeed the case, let us start recalling that worldvolume gauge fluxes induce moduli-dependent Fayet-Iliopoulos (FI) terms on the D7-branes \cite{Jockers:2005zy}:
\begin{equation}
\xi_\FI = \frac{1}{4\pi\mathcal{V}} \int_{D} J\wedge \mathcal{F}\,.
\label{FI-term}
\end{equation}
In our model, we choose the following worldvolume fluxes:
\begin{equation}
\label{Eq:cF2flux}
\mathcal{F}_2=f_4\, \hat{D}_4 + f_6\, \hat{D}_6 \qquad \text{and}\qquad \mathcal{F}_4=\mathcal{F}_6=0\,.
\end{equation}
This choice satisfies the flux quantisation condition which requires
$F-\frac{\iota_D^*\hat{D}}{2}$ to be an integral $2$-form. Indeed, $D_4$ and $D_6$ are K3 surfaces, and so $F$ is integrally quantised and can be set to zero. Moreover, $\iota^*_{D_4}B=\iota^*_{D_6}B=0$ since $B=\hat{D}_1/2$. Regarding $D_2$, we have $\iota^*_{D_2}\hat{D}_1=-\iota^*_{D_2}\hat{D}_2$, and so we obtain just an integral contribution after adding the $B$-field to the half-integral part of $F_2$. Hence, $f_4,f_6\in\mathbb{Z}$ and the flux choice (\ref{Eq:cF2flux}), when used in (\ref{FI-term}), yields a single non-trivial FI-term:
\begin{equation}
\xi_\FI = \frac{1}{2\pi\mathcal{V}}\left(f_4\,t_6+f_6\,t_4\right),
\end{equation}
where we used \eqref{expansofkahlerformondivisorbasisCYmodel}. For vanishing VEVs of charged matter fields, as it is generically the case for non-tachyonic soft terms generated by $3$-form fluxes, the D-term potential associated to the diagonal $U(1)$ on the D7-stack wrapping $D_2$ scales as:
\begin{equation}
V_D\simeq g_2^2\,\xi_\FI^2 \simeq \frac{1}{\left(\tau_7-\tau_1\right)}   \frac{1}{4\pi^2\mathcal{V}^2}\left(f_4\,t_6+f_6\,t_4\right)^2\sim \frac{1}{\vo^2}\left(f_4\sqrt{\frac{\tau_4}{\tau_6}} + f_6\sqrt{\frac{\tau_6}{\tau_4}}\right)^2\,,
\end{equation}
where $g_2^{-2}\simeq \tau_2=\tau_7-\tau_1$ is the corresponding gauge coupling and we used the relations (\ref{4cyclesvolumesCYmodel}). This D-term potential behaves as $V_D\gtrsim \mathcal{O}(\vo^{-2})$, and so it is dominant with respect to the LVS F-term potential (\ref{LVSPot}) which scales as $V_\LVS\sim\mathcal{\vo}^{-3}$. Thus, at leading order, D-term fixing leads to a supersymmetric vacuum with vanishing FI-terms.

The condition $\xi_\FI=0$ then fixes the following relation among the K\"ahler moduli:
\begin{equation}
t_4 = \alpha\, t_6\,\qquad\mbox{with}\qquad \alpha \equiv -\frac{f_4}{f_6}\,.
\label{eq:2-cycle-linear-rel-gauge-flux}
\end{equation}
Given that the K\"ahler cone conditions (\ref{KahlerCone}) require both $t_4>0$ and $t_6>0$, $f_4$ and $f_6$ need to have opposite signs to find a solution inside the K\"ahler cone.

Plugging \eqref{eq:2-cycle-linear-rel-gauge-flux} into (\ref{volumeCYmodel_h11=4}) makes the volume depend only on $3$ K\"ahler moduli:
\begin{equation}
\mathcal{V} = \, 2 \alpha\, t_6^2\, t_7 + \frac{1}{3}\, t_1^3 = \frac{1}{\sqrt{2 \alpha}}\, \sqrt{\tau_7}\,\tau_6 - \frac{1}{3}\, \tau_1^{3/2} 
= t_7\,\tau_7 - \frac{1}{3}\, \tau_1^{3/2}   \,.
\label{volumeCYmodel}
\end{equation}
After D-term stabilisation, the CY volume takes therefore the effective form (\ref{CYVol_new}) representing a single K3 fibration, with volume $\tau_7\equiv \tau_\K3$, over a $\mathbb{P}^1$ base, with volume $t_7\equiv t_\P1$, together with a diagonal del Pezzo divisor, with volume $\tau_1\equiv \tau_\dP$.

Finally, for the flux choice \eqref{Eq:cF2flux}, the contribution to the D3-charge is proportional to
\begin{equation}
\int_{D_2}\mathcal{F}_2^2 = 4\,f_4\,f_6\,.
\end{equation}
For opposite signs of $f_4$ and $f_6$, this leads to a positive contribution to the total D3-charge which is minimised for $f_4=-f_6=\pm1$, giving $t_4=t_6$ or, equivalently, $\tau_4=\tau_6$. Using (\ref{Eq:totQ3secondoD7setup}), in the second D7-brane setup one obtains $Q^{\text{D3}}_{\text{tot}}=-758$. Such a large negative D3-charge provides ample freedom to turn on $3$-form fluxes for axio-dilaton and complex structure stabilisation. Analogous considerations apply to the first D7-brane setup, leading to $Q^{\text{D3}}_{\text{tot}}=-44$ from (\ref{D3First}).

\subsection{LVS stabilisation of overall volume}
\label{sec:LVS-review}

At semi-classical level, $2$-form gauge fluxes (inducing moduli-dependent FI-terms) and $3$-form background fluxes (generating soft terms for matter fields and a potential for the axio-dilaton $S$ and the $h^{1,2}$ complex structure moduli $U$) are expected to fix all moduli except for the $3$ K\"ahler moduli which appear in the volume (\ref{volumeCYmodel}). These $3$ moduli remain flat due to the no-scale structure of the tree-level K\"ahler potential $K_0 = -2\ln\vo$. 

The leading no-scale breaking quantum corrections which lift these modes are shown in (\ref{Kalphap3}) and (\ref{Wnp}). In our explicit example, we have:
\begin{equation}
\xi = \, -\frac{\chi(X) \zeta(3)}{2(2 \pi)^3}  = \frac{188 \,\zeta(3)}{2(2 \pi)^3}\simeq 0.46\qquad\text{and}\qquad
a_1 \equiv a_\dP=2\pi\,,
\end{equation}
while both the tree-level flux superpotential $W_0$ and the Pfaffian $A_1\equiv A_\dP$ are expected to be $\mathcal{O}(1)$ constants. 

After fixing the axion $\theta_1\equiv\theta_\dP$ at its minimum, the resulting F-term scalar potential takes the typical LVS form:
\begin{equation}
V_\LVS = \frac{16 \pi^2 A_\dP^2 \, g_s \sqrt{\tau_\dP} \,  e^{- 4\pi \tau_\dP}}{\mathcal{V}} - \frac{4 \pi A_\dP g_s W_0  \tau_1 e^{- 2\pi \tau_\dP}}{\mathcal{V}^2}  + \frac{3 W_0^2 \xi}{8 \sqrt{g_s} \, \mathcal{V}^3} \,,
\label{LVSscalarpotmodel}
\end{equation}
which is minimised at:
\begin{equation}
\langle \tau_\dP \rangle  \simeq \left(\frac{3\xi}{2}\right)^{2/3}\frac{1}{g_s} \qquad\text{and}\qquad
\langle \mathcal{V} \rangle \simeq \frac{W_0 \sqrt{\langle \tau_\dP \rangle} \, e^{2\pi \langle \tau_\dP \rangle }}{8\pi A_\dP}\,,
\label{LVSStab}
\end{equation}
where we have neglected corrections of order $(2\pi\langle\tau_\dP\rangle)^{-1}\ll 1$. The values of Tab. \ref{Tab2} can be reproduced for natural choices of the underlying parameters. In particular, $g_s=0.1$, $W_0=3$ and $A_\dP=1$ realise a typical DD scenario with $\langle\vo\rangle\simeq 4.9\cdot 10^{20}$ and $\langle\tau_\dP\rangle\simeq 7.8$ (with $(2\pi\langle\tau_\dP\rangle)^{-1}\simeq 0.02$), while $g_s=0.07$, $W_0=22$ and $A_\dP=1$ realise a typical ADD scenario with $\langle\vo\rangle\simeq 5.1\cdot 10^{30}$ and $\langle\tau_\dP\rangle\simeq 11.1$ (with $(2\pi\langle\tau_\dP\rangle)^{-1}\simeq 0.01$).

Note that the minimum (\ref{LVSPot}) is AdS, and so, as explained in Sec. \ref{LOS}, it needs to be raised to Minkowski by the inclusion of a suitable uplifting sector. This new contribution would slightly shift the exact location of the minimum (\ref{LVSStab}) but without changing the qualitative LVS picture (the UV parameters can be easily readjusted to obtain the same values of $\langle\vo\rangle$ and $\langle\tau_\dP\rangle$ also in the Minkowski minimum).

At this level of approximation, besides the two ultra-light axions $\theta_7\equiv\theta_\K3$ and $\theta_6\equiv\theta_\P1$, the LVS potential (\ref{LVSscalarpotmodel}) features a saxionic flat direction. As can be seen from (\ref{CanNormNew}), this flat direction is $r = \sqrt{\tau_7}/t_7\equiv\sqrt{\tau_\K3}/t_\P1$ which gives the degree of anisotropy of the CY volume since it controls the relative size of the volumes of the K3 fibre and the $\mathbb{P}^1$ base.

\subsection{Anisotropic stabilisation}
\label{sec:stabilisation}

As mentioned in Sec. \ref{SPS}, the leading effects responsible for lifting the flat direction $r$ are string loop corrections \cite{Berg:2005ja, Berg:2007wt} and higher order contributions in the superspace derivative expansion \cite{Ciupke:2015msa}. Without loss of generality, from now on, we shall parametrise the flat direction as $\tau_\K3$, and show that these corrections can fix $\tau_\K3\sim\mathcal{O}(5)$ so that the $\mathbb{P}^1$ base becomes exponentially large due to $t_\P1\simeq \vo/\tau_\K3\gg 1$. Before minimising the potential, let us first introduce both kinds of corrections and then interpret them from an EFT viewpoint.

\subsubsection{String loop corrections}
\label{subsec:string-loop-correc}

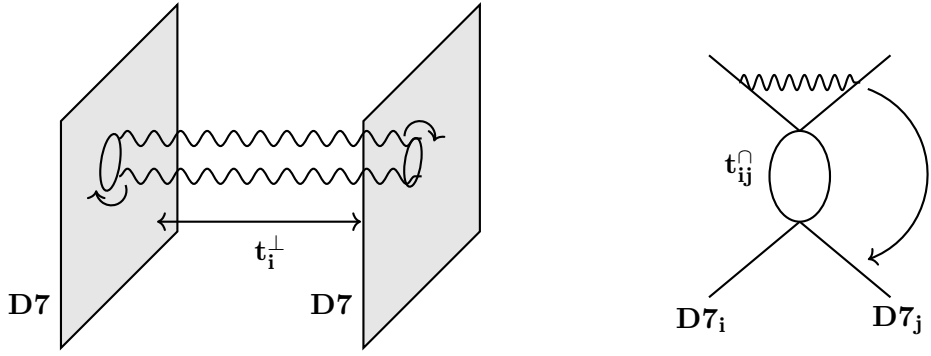
\begin{figure}[ht]
\centering
\begin{tikzpicture}
\begin{scope}[shift={(-6,0)}]
\begin{scope}[canvas is yz plane at x=0]
\filldraw[fill=gray!20, draw=black, thick] (-1.5,2) -- (-1.5,-2) -- (1.5,-2) -- (1.5,2) -- cycle;
\node at (-0.5,3.1) {\textbf{D7}};
\draw[thick] (0.3,0.3) circle (0.35);
\end{scope}
            
\begin{scope}[canvas is yz plane at x=4]
\filldraw[fill=gray!20, draw=black, thick] (-1.5,2) -- (-1.5,-2) -- (1.5,-2) -- (1.5,2) -- cycle;
\node at (-0.5,3.1) {\textbf{D7}};
\draw[thick] (0.3,0.3) circle (0.3);
\end{scope}
            
\draw[thick, decorate, decoration={snake, amplitude=0.1cm}] (0,0.5,0) -- (2,0.5,0) -- (4,0.5,0);
\draw[thick, decorate, decoration={snake, amplitude=0.1cm}] (0,0,0) -- (2,0,0) -- (4,0,0);
            
\draw[->, thick] (0.2,0,0.3) arc[start angle=20, end angle=-180, radius=0.2];
\draw[<-, thick] (4.2,0.5,0.1) arc[start angle=-20, end angle=180, radius=0.2];
            
\draw[<->, thick] (0.5,-0.6,0) -- (3.2,-0.6,0);
\node at (2,-1,0) {$\mathbf{t^\perp_i}$};
\end{scope}
                
\begin{scope}[shift={(3,0)}, scale=2]

\draw[thick] (-0.6, 0.8) -- (0, 0.3);
\draw[thick] (0, 0.3) -- (0.6, 0.8);

    \draw[thick] (-0.6, -0.8) -- (0, -0.3);
    \draw[thick] (0, -0.3) -- (0.6, -0.8);

    \node at (-0.65, -0.95) {$\mathbf{D7_i}$};
    \node at (0.65, - 0.95) {$\mathbf{D7_j}$};

    \draw[
        thick,
        decoration={snake, segment length=2mm, amplitude=1mm},
        decorate
    ] (-0.4,0.62) -- (0.4,0.62);

    \draw[thick] (0,0) ellipse (0.2 and 0.3);

    \draw[<-, thick] (0.45, -0.55)
        arc[start angle=-70, end angle=70, radius=0.6];
\node at (-0.4, 0.05, 0) {$\mathbf{t_{ij}^\cap}$};
\end{scope}
\end{tikzpicture}  
\caption{Pictorial view of 1-loop open string corrections. LHS: 1-loop of open strings stretching between two parallel D7s. RHS: 1-loop of open strings at the intersection between two D7s.}
\label{Fig2}
\end{figure}

Starting from the explicit computation in toroidal orientifolds performed in \cite{Berg:2005ja}, ref. \cite{Berg:2007wt} proposed an educated guess for the form of open string 1-loop corrections to $K$ for a generic CY compactification:
\begin{eqnarray}
\delta K^\KK_{g_s} &=& \, g_s \sum_i \frac{C_i^\KK t_i^{\perp}}{\mathcal{V}}\,,
\label{KKcorrectoKahlpot} \\
\delta K^\W_{g_s} &=& \, \sum_{ij} \frac{C_{ij}^\W}{\mathcal{V}\, t_{ij}^{\cap}}\,,
\label{WcorrectoKahlpot}
\end{eqnarray}
where $C_i^\KK$ and $C_{ij}^\W$ are functions of the complex structure moduli, which are expected to become $\mathcal{O}(1)$ tunable constants after such moduli have been stabilised by $3$-form fluxes. $\delta K^\KK_{g_s}$ is an $\mathcal{O}(g_s^2 \alpha'^2)$ correction arising from the tree-level exchange of KK closed strings between stacks of parallel O7/D7-branes, or between an O7/D7-stack and O3/D3-branes, with $t_i^{\perp}$ denoting the volume of the $2$-cycle perpendicular to the O7/D7-stack. On the other hand, $\delta K^\W_{g_s}$ is an $\mathcal{O}(g_s^2 \alpha'^4)$ correction due to the tree-level exchange of closed strings winding around a non-contractible $1$-cycle at the intersection between two stacks of O7/D7-branes, with $t_{ij}^{\cap}$ denoting the volume of the $2$-cycle where the $i$-th and the $j$-th stack intersect. A pictorial view of these loop effects is presented in Fig. \ref{Fig2}.

Inserting (\ref{KKcorrectoKahlpot}) and (\ref{WcorrectoKahlpot}) in the general expression of the supergravity F-term scalar potential, one obtains the following corrections to $V_\LVS$:
\begin{eqnarray}
\delta V^\KK_{g_s} &=& \, \frac{g_s^3}{2} \frac{W_0^2}{\mathcal{V}^2} \sum_{i,j} C_i^\KK C_j^\KK K_{ij}^0
\label{KKcorrectoscalpot} \, , \\
\delta V^\W_{g_s} &=& \, - g_s \frac{W_0^2}{\mathcal{V}^3} \sum_{ij} \frac{C_{ij}^\W}{t_{ij}^{\cap}}
\label{Wcorrectoscalpot} \,,
\end{eqnarray}
where $K_{ij}^0$ is the tree-level K\"ahler metric and $t_{ij}^\cap$ is computed as:
\begin{equation}
t_{ij}^\cap = \int_X J \wedge \hat{D}_i \wedge \hat{D}_j \,.
\label{intersection2cycledefinition}
\end{equation}
Note that, although $\delta K^\KK_{g_s}$ is an $\mathcal{O}(g_s^2 \alpha'^2)$ effect, $\delta V^\KK_{g_s}$ is just an $\mathcal{O}(g_s^4 \alpha'^4)$ correction due to the \emph{extended no-scale structure} \cite{Cicoli:2007xp}.\footnote{To estimate the order in $\alpha'$ and $g_s$ of a correction, one has to factor out the tree-level contribution proportional to $g_s/\vo^2$, and then has to go from Einstein to string frame using the fact that $\vo_{\text{Ein}}=\vo_{\text{str}}/g_s^{3/2}=\text{Vol(CY)}/\left((2\pi)^6 g_s^{3/2}\alpha'^3\right)$.} Hence $\delta V^\KK_{g_s}$ scales effectively as a 2-loop contribution, which we shall neglect for the following reasons:
\begin{enumerate}
\item The KK loop correction $\delta K^\KK_{g_s}$ is expected to arise only in the presence of parallel stacks of O7/D7-branes. However, in our explicit CY example, all O7/D7-branes intersect each other. Moreover, the fixed locus of the orientifold involution does not include any O3-plane, and D3-tadpole cancellation can be achieved without introducing any D3-brane. Hence this type of correction can be argued to be absent by construction. This is the main reason for choosing this particular CY example.

\item The extended no-scale structure can be understood by noticing that $\delta K^\KK_{g_s}$, if present by construction, corresponds to the first term of the Taylor expansion of a corrected K\"ahler potential which, however, is still of no-scale type. This might indicate a generic cancellation of all loop corrections to $K$ of KK type if 2-loop corrections, which are presently unknown, keep respecting the form of the Taylor expansion.

For simplicity, let us illustrate this argument for the simple case with just a single K\"ahler modulus where $\vo\simeq t^3\simeq \tau^{3/2}$, even if our considerations hold for a generic number of K\"ahler moduli. In the single modulus case: 
\begin{equation}
\delta K^\KK_{g_s}\simeq g_s \frac{C^\KK}{\tau}\,,
\end{equation}
can be seen to arise from the expansion of the K\"ahler potential (for $k\equiv g_s\,C^\KK/3$):
\begin{equation}
K_{\text{no-scale}} = -3 \ln\left(\tau-k\right) = -3\ln\tau - 3\ln\left(1-\frac{k}{\tau}\right) \stackrel{k/\tau \ll 1}{\simeq} -3\ln\tau + \frac{3\,k}{\tau} + \mathcal{O}\left(\frac{1}{\tau^2}\right).
\label{Ktot1modulnoscalestruc}
\end{equation}
However, $K_{\text{no-scale}} = -3 \ln\left(\tau-k\right)$ is a known no-scale type K\"ahler potential which generates no contribution to the scalar potential.  
\end{enumerate}

\subsubsection{Higher-derivative corrections}
\label{subsec:higher-derivative-corrections}

The $\mathcal{O}(\alpha'^3)$ correction to $K$ (\ref{Kalphap3}) arises from the dimensional reduction of the 10D term $\alpha'^3\int d^{10}x \left(\mathcal{R}^4+\mathcal{R}^3 G_3^2\right)$, where the $\mathcal{R}^4$ term gives rise just to a correction to the moduli kinetic terms, while the correction to $V$ comes from the term $\mathcal{R}^3 G_3^2$. However, the 10D action features additional contributions at the same order in $\alpha'$ which lead to further corrections. The most important one for us originates from the 10D term $\alpha'^3\int d^{10}x\, \mathcal{R}^2 G_3^4$ whose dimensional reduction generates higher-derivative corrections to the scalar potential. These are non-K\"ahler corrections which behave effectively as $F^4$-terms in the superspace derivative expansion of the form:  
\begin{equation}
\delta V_{F^4} = - \frac{\lambda \sqrt{g_s} }{4} \frac{W_0^4}{\mathcal{V}^4} \sum_{i=1}^{h^{1,1}}\,\Pi_i\, t_i\,,
\label{F4correcscalarpot}
\end{equation}
where $\lambda$ is a combinatorial factor which, in the single modulus case, has been computed to be $\lambda = - 3.5 \cdot 10^{-4}$ \cite{Grimm:2017okk}. As argued in \cite{Cicoli:2023njy}, $\lambda$ is expected to remain small also in the case with several moduli since the general structure of the corrections implies that $\lambda$ is always proportional to $\zeta(3)/(2\pi)^4\simeq 7.7\cdot 10^{-4}$. In fact, $\delta V_{F^4}$, being an $F^4$ correction, can be seen to come from a tensor coupling to four derivatives:
\begin{equation}
\delta V_{F^4} = e^{-2K} \mathcal{T}^{i j \overline{k} \overline{l}} D_i W D_j W D_{\overline{k}} \overline{W} D_{\overline{l}} \overline{W}\,,
\end{equation}
where the tensor scales schematically as:
\begin{equation}
\mathcal{T}_{i j \overline{k} \overline{l}} \simeq \frac{c}{\mathcal{V}^{8/3} g_s^{3/2}}\left(\frac{\zeta(3)}{(2\pi)^4}\right) \int_X c_2(X) \wedge J  \,,
\end{equation}
with $c$ a combinatorial factor that in the single modulus cases is $11/384$ \cite{Grimm:2017okk}. 

The $\Pi_i$'s are topological quantities which depend on the second Chern class of the CY threefold $X$ as:
\begin{equation}
\Pi_i \equiv \int_X c_2(X) \wedge \Hat{D}_i \,.
\label{defsecondchernnumber}
\end{equation}
For our explicit CY example, we have $\Pi_1 = 8$ and $\Pi_4 = \Pi_6 = \Pi_7=24$ \cite{Cicoli:2017axo}.

\subsubsection{Field theory interpretation of perturbative corrections}

Following \cite{vonGersdorff:2005bf,Cicoli:2007xp,Gao:2022uop, Bansal:2024uzr, Cicoli:2026bqo}, in this section we provide an EFT interpretation of the perturbative corrections discussed above. In a supersymmetric theory, loop corrections to $K$ should give rise to loop corrections to $V$, as well as to a renormalisation of the moduli kinetic terms. From an EFT point of view, 1-loop corrections to $V$ are expected to take the typical Coleman-Weinberg form:
\begin{equation}
\delta V_{\text{1-loop}}^\CW \simeq \frac{1}{16\pi^2}\,\Lambda^2\,\text{Str}\left(\mathcal{M}^2\right)\,, 
\label{CWV}
\end{equation}
where in a supergravity framework the supertrace of the mass-squared matrix scales as the gravitino mass $m_{3/2}$:
\begin{equation}
\text{Str}\left(\mathcal{M}^2\right) \simeq m_{3/2}^2\simeq \frac{W_0^2}{\mathcal{V}^2}\,M_p^2\,,
\end{equation}
and it is natural to take the cut-off $\Lambda$ of order the 10D KK scale (\ref{MKK10D}), since the theory does not become exactly supersymmetric above the 5D or 6D KK scale (implying that loops of heavy bosons are not cancelled by loops of heavy fermions):
\begin{equation}
\Lambda \simeq M_\KK^{10D} \simeq \frac{M_p}{\tau_\K3^{1/4}\,\sqrt{\mathcal{V}}}\,.  
\end{equation}
Combining these expressions, we expect a 1-loop potential that scales as:
\begin{equation}
\delta V_{\text{1-loop}} \simeq \frac{W_0^2}{\mathcal{V}^3\,\sqrt{\tau_\K3}}\,,
\label{VloopScaling}
\end{equation}
which indeed reproduces the scaling with the K\"ahler moduli of $\delta V^\W_{g_s}$ in (\ref{Wcorrectoscalpot}) for $t^{\cap}\simeq t_6 \simeq\sqrt{\tau_\K3}$. This correction, in the closed string channel, originates from the tree-level exchange of winding strings. This EFT analysis shows that, in the open string channel, it can be instead interpreted as arising from loops of KK open strings at the intersection between stacks of O7/D7-branes.

Further contributions would arise when taking $\Lambda\simeq M_\KK^{6D}$ or $\Lambda\simeq M_\KK^{5D}$ but these would be subdominant. As can be seen from (\ref{MKK6D}) and (\ref{MKK5D}), $M_\KK^{6D}$ and $M_\KK^{5D}$ have the same K\"ahler moduli dependence, while they differ in their dependence on the complex structure modulus $u$ which controls the anisotropy of the volume of the $\mathbb{P}^1$ base. Given that we are interested just in the dependence of loop corrections to $V$ on the K\"ahler moduli, we can substitute in (\ref{CWV}) either $M_\KK^{6D}$ or $M_\KK^{5D}$, finding:
\begin{equation}
\delta V_{\text{1-loop}} \simeq \frac{W_0^2\,\tau_\K3}{\mathcal{V}^4}\,,
\label{VscalingSub}
\end{equation}
which matches again the scaling of $\delta V^\W_{g_s}$ in (\ref{Wcorrectoscalpot}),  this time however for $t^{\cap}\simeq t_\P1 \simeq\vo/\tau_\K3$. This observation provides additional support in favour of the EFT interpretation of these corrections as due to loops of KK modes. Note, moreover, that this EFT analysis would apply also to loops of KK replicas of closed strings, implying the presence of these loop corrections to $V$ even in the absence of open strings on branes.

\begin{figure}[htbp]
    \centering
        \begin{tikzpicture}[very thick]
         \def\radius{1.0}
      \draw[dashed] (0,0) circle (\radius);
      \node[above] (1) at (0,\radius) {$H$};
      \node[below] (2) at (0,-\radius) {$H$};
      \filldraw
      (-2*\radius,0) -- (-\radius,0) circle (2pt)
      (\radius,0) circle (2pt) -- (2*\radius,0);
      \draw (-3*\radius,0) -- (-\radius,0) node[midway,above] {$L$};
      \draw (\radius,0) -- (3*\radius,0) node[midway,above] {$L$};
      \node at (-\radius-0.2,-0.4) {$g$};
      \node at (\radius+0.2,-0.4) {$g$};
    \end{tikzpicture}
\caption{Loop corrections to the 2-point function of $L$ due to heavy modes $H$ running in the loop.}
\label{FigLoop}
\end{figure}
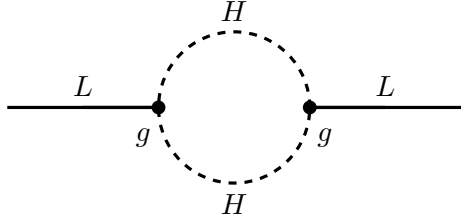

Let us now show that the scaling of the potential (\ref{VloopScaling}) can be inferred also from 1-loop corrections to the moduli 2-point function. In fact, light moduli $L$ couple to heavy modes $H$ with mass $M$ as:
\begin{equation}
\mathcal{L} = \frac12\partial_\mu L \partial^\mu L + \frac12\partial_\mu H \partial^\mu H  + \frac12 M^2 H^2 + g L H^2 \,.
\label{lagrangian-light-heavy_new}
\end{equation}
Any generic mass $M$ (string, KK or winding scale) has a power law dependence on the K\"ahler moduli, and so becomes an exponential function of $L$ once $L$ is identified with either of the $2$ canonically normalised fields $\phi$ and $\chi$ in (\ref{CanNorm}). Expanding $L$ around the minimum of its potential as $L=\langle L\rangle+\hat L$, one can infer the value of the coupling $g$ in (\ref{lagrangian-light-heavy_new}):
\begin{equation}
M^2(L) H^2 = M^2(\langle L \rangle) H^2 + \frac{\partial}{\partial L} M^2 \Bigg |_{\langle L \rangle} \,\frac{\hat{L}}{M_p}\,  H^2 \qquad \Rightarrow \qquad g\simeq \frac{M^2}{M_p} \,.
\label{lagrangian-mass-term-moduli_new}
\end{equation}
This coupling would induce a correction to the kinetic terms of the light moduli due to heavy modes with mass $M$ running in the loop (see Fig. \ref{FigLoop}):      
\begin{equation}
\mathcal{L}_\text{kin} = \frac12\partial_\mu L \partial^\mu L \left[ 1 + \frac{1}{16 \pi^2} \left( \frac{g}{M}\right)^2 \right] \simeq \frac12\partial_\mu L \partial^\mu L \left[ 1 + \left( \frac{M}{M_p}\right)^2 \right].
\label{lagrangian-1-loop-interac_new}
\end{equation}

This correction has to arise from a loop contribution $\delta K$ to the K\"ahler potential, $K=K_0+\delta K$, which can be inferred as follows:
\begin{equation}
\mathcal{L}_\text{kin} = \left(K_{0,ij} + \delta K_{ij}\right) \partial_\mu \tau^i\partial^\mu\tau^j \simeq \frac12 \partial_\mu L \partial^\mu L \left(1+ K_{0,ij}^{-1} \delta K_{ij}\right)\simeq  \frac12 \partial_\mu L \partial^\mu L \left(1+ \delta K\right).
\end{equation}
Comparing this expression with (\ref{lagrangian-1-loop-interac_new}), we infer that $\delta K$ has to depend on $M$:
\begin{equation}
\delta K \simeq \left(\frac{M}{M_p} \right)^2 \,.
\end{equation}
Following \cite{Cicoli:2026bqo}, let us consider different heavy modes running in the loop:
\begin{enumerate}
\item If $M = M_s$:
\begin{equation}
M = M_s \simeq \frac{M_p}{\sqrt{\mathcal{V}}} \qquad \Rightarrow \qquad \delta K \simeq \frac{1}{\mathcal{V}}\,,
\end{equation}
which reproduces the typical volume scaling of an $\mathcal{O}(\alpha'^3)$ correction to $K$.

\item If $M = M_\KK^{10D}$:
\begin{equation}
M = M_\KK^{10D} \simeq\frac{M_p}{\tau_\K3^{1/4}\,\sqrt{\mathcal{V}}} \qquad \Rightarrow \qquad \delta K \simeq \frac{1}{ \mathcal{V}\, \sqrt{\tau_\K3}} \,,
\end{equation}
which replicates the scaling of $\delta K^\W_{g_s}$ in (\ref{WcorrectoKahlpot}) for $t^{\cap}\simeq t_6 \simeq\sqrt{\tau_\K3}$. 

\item If $M = M_\KK^{6D}$ or $M = M_\KK^{5D}$:
\begin{equation}
M = M_\KK^{6D} \simeq\frac{M_p\,\sqrt{\tau_\K3}}{\mathcal{V}}  \qquad \Rightarrow \qquad \delta K \simeq \frac{\tau_\K3}{ \mathcal{V}^2} \,,
\end{equation}
which captures the scaling of $\delta K^\W_{g_s}$ in (\ref{WcorrectoKahlpot}) for $t^{\cap}\simeq t_\P1 \simeq\vo/\tau_\K3$. 
        
\item If $M=M_\W$:
\begin{equation}
M = M_\W \simeq M_s\, \tau_\K3^{1/4} \simeq \frac{M_p\, \tau_\K3^{1/4}}{\sqrt{\mathcal{V}}} \qquad \Rightarrow \qquad \delta K \simeq \frac{\sqrt{\tau_\K3}}{\mathcal{V}} \,,
\end{equation}
which matches the scaling of $\delta K^\KK_{g_s}$ in (\ref{KKcorrectoKahlpot}) for $t^{\perp}\simeq t_6 \simeq\sqrt{\tau_\K3}$ that would, however, give rise to no contribution to $V$ due to the extended no-scale cancellation. A similar cancellation applies to the contribution to $V$ from loops of heavy modes winding around the base with mass $M_\W \simeq M_s\, \sqrt{t_\P1}$.
\end{enumerate}

Another source of perturbative corrections is provided by higher order terms in the superspace derivative expansion at tree-level. If $|F^T|$ is the F-term of the light fields, after integrating out heavy modes with mass $M$, the EFT naively contains an expansion in powers of $g\,|F^T|/M^3$ \cite{Cicoli:2013swa} where:
\begin{equation}
|F^T| = \sqrt{K_{i\overline{j}} F^{T_i} F^{\overline{T}_{\overline{j}}}}\simeq m_{3/2}M_p\qquad\Rightarrow\qquad\frac{g\,|F^T|}{M^3}\simeq \frac{m_{3/2}}{M}\ll 1\,.    
\end{equation}
This analysis implies that the $\mathcal{O}(\alpha'^3 F^2)$ contribution to $V$, that scales as $W_0^2/\mathcal{V}^3$, should receive higher-derivative corrections at $\mathcal{O}(\alpha'^3 F^4)$ which are expected to scale as:
\begin{equation}
\delta V_{F^4}\simeq \delta V_{F^2}\left(\frac{m_{3/2}}{M}\right)^2 \simeq \frac{W_0^4}{\mathcal{V}^5} \left(\frac{M_p}{M}\right)^2\,.
\end{equation}
Let us consider two cases for $M$, corresponding respectively to the 10D and 6D/5D KK scales:
\begin{enumerate}
\item If $M = M_\KK^{10D}$:
\begin{equation}
M = M_\KK^{10D} \simeq\frac{M_p}{\tau_\K3^{1/4}\,\sqrt{\mathcal{V}}} \qquad \Rightarrow \qquad \delta V_{F^4}\simeq\frac{W_0^4\,\sqrt{\tau_\K3}}{\mathcal{V}^4}\,,
\end{equation}
which reproduces indeed the scaling with the K\"ahler moduli of (\ref{F4correcscalarpot}) for $t\simeq t_6\simeq\sqrt{\tau_\K3}$.

\item If $M = M_\KK^{6D}$ or $M = M_\KK^{5D}$:
\begin{equation}
M = M_\KK^{6D} \simeq\frac{M_p\sqrt{\tau_\K3}}{\mathcal{V}}  \qquad \Rightarrow \qquad \delta V_{F^4}\simeq\frac{W_0^4}{\mathcal{V}^3\,\tau_\K3} \,,
\label{VF4Scaling}
\end{equation}
which matches the scaling with the K\"ahler moduli of (\ref{F4correcscalarpot}) for $t\simeq t_\P1 \simeq\vo/\tau_\K3$.
\end{enumerate}
Given that we are interested in finding a minimum in the region $\vo\gg 1$ and $\tau_\K3\sim\mathcal{O}(5)$, we realise that the leading terms are the loop correction (\ref{VloopScaling}) and the $F^4$ term (\ref{VF4Scaling}). In the next section we will show that these two terms can indeed compete to induce a minimum for $\tau_\K3$ at values of $\mathcal{O}(5)$ where the CY volume becomes very anisotropic.

\subsubsection{Minimisation}

Let us now use (\ref{Wcorrectoscalpot}) and (\ref{F4correcscalarpot}) to compute the explicit form, respectively, of $\delta V^\W_{g_s}$ and $\delta V_{F^4}$ for our CY example. As motivated in Sec. \ref{subsec:string-loop-correc}, we can ignore $\delta V^\KK_{g_s}$. We will focus on the first brane setup with $\lambda=-|\lambda|$ and $f_4=-f_6=\pm1$ which gives $\alpha=1$. A similar analysis reaching the same conclusions can be performed also for the second brane setup. 

All relevant intersection $2$-cycle volumes for the first brane setup have been computed in \cite{Cicoli:2017axo} and are given by (for $t_4=t_6$ from D-term stabilisation):
\begin{eqnarray}
t_{22}^\cap  &=& 2\, t_1 \, , \qquad
t_{24}^\cap = t_{26}^\cap = 2\, t_6 \,, \qquad
t_{28}^\cap = \, 4 \left(t_1 + 2 t_6\right) \\
t_{46}^\cap &=& 2\, t_7 \,, \qquad
t_{48}^\cap = t_{68}^\cap = 4 \left(t_6 + t_7\right) \, , \qquad
t_{88}^\cap = 8\left(t_1 + 2 (2 t_6 + t_7)\right).
\end{eqnarray}
Identifying $\tau_1\equiv\tau_\dP$ and $\tau_7\equiv\tau_\K3$, and defining the useful parameters:
\begin{equation}
c_\K3 \equiv \sqrt{\frac{\langle \tau_\dP \rangle}{2\tau_\K3}}\sim\mathcal{O}(1)  \qquad\text{and}\qquad\epsilon_\K3\equiv \frac{\tau_\K3^{3/2}}{\sqrt{2}\,\vo} \ll 1\,, 
\end{equation}
the general expressions (\ref{Wcorrectoscalpot}) and (\ref{F4correcscalarpot}) reduce to:
\begin{eqnarray}
\delta V_{g_s}^\W &=& - g_s\,\frac{c_{\text{loop}}\, W_0^2}{ \vo^3\,\sqrt{\tau_\K3}} - g_s\,\frac{C_{22}^\W\,W_0^2}{2\, \vo^3\,\sqrt{\langle \tau_\dP \rangle}} \,,
\label{Wcorrec_ourmodel} \\
\delta V_{F^4} &=& \sqrt{g_s}\, \frac{c_{F^4}\,W_0^4}{\vo^3\,\tau_\K3} +\sqrt{g_s}\, \frac{2\,|\lambda|\, W_0^4}{\vo^4} \sqrt{\langle \tau_\dP \rangle}  \, ,
\label{F4correc_ourmodel}
\end{eqnarray}
where:
\begin{eqnarray}
c_{\text{loop}}&\equiv& \sqrt{2}\left( C_{24}^\W + C_{26}^\W 
+ \frac{C_{28}^\W}{4\left(1+ c_\K3\right)} \right) +\epsilon_\K3\sqrt{2}\left(C_{46}^\W +\frac{\left(C_{48}^\W + C_{68}^\W\right)}{2\left(1+\epsilon_\K3 \right)}
+ \frac{C_{88}^\W}{16\left(1+ \epsilon_\K3( 2 + c_\K3) \right) }\right) \nonumber \\
&\simeq& \sqrt{2}\left( C_{24}^\W + C_{26}^\W 
+ \frac{C_{28}^\W}{4\left(1+ c_\K3\right)} \right),
\label{cloop}
\end{eqnarray}
and:
\begin{equation}
c_{F^4}\equiv  6\,|\lambda| \left(1+2\epsilon_\K3\right)\simeq 6\,|\lambda|\,.   
\label{cF4}
\end{equation}
Note that the last terms on the RHS of both (\ref{Wcorrec_ourmodel}) and (\ref{F4correc_ourmodel}) do not depend on $\tau_\K3$, and so we can safely ignore them given that we are interested in stabilising $\tau_\K3$. Hence $\delta V_{g_s}^\W$ in (\ref{Wcorrec_ourmodel}) reduces exactly to (\ref{V1loop}) with $c_{\text{loop}}$ given in terms of the underlying parameters as in (\ref{cloop}). Similarly, $\delta V_{F^4}$ in (\ref{F4correc_ourmodel}) takes the same form as (\ref{VF4}) with $c_{F^4}$ given in (\ref{cF4}). The leading order contribution to the $\tau_\K3$-dependent potential is therefore given by (\ref{Vsub}) which is minimised at (\ref{tauK3Min}). For convenience, we report here both $V_{\text{sub}} = \delta V_{g_s}^\W + \delta V_{F^4}$ and $\langle\tau_\K3\rangle$ for ease of reference:
\begin{equation}
V_{\text{sub}} \simeq \frac{W_0^2}{\mathcal{V}^3}\left(-\frac{A}{\sqrt{\tau_\K3}}+\frac{B}{\tau_\K3}\right)
\qquad\Rightarrow\qquad
\langle\tau_\K3\rangle \simeq \left(\frac{2B}{A}\right)^2  = \frac{1}{g_s}\left(\frac{2\,c_{F^4}\,W_0^2}{c_{\text{loop}}}\right)^2\,.
\end{equation}
Natural choices of the UV parameters can lead to $\langle\tau_\K3\rangle\sim\mathcal{O}(5)$, allowing for an explicit realisation of both the DD and the ADD scenarios, depending on the value of $\vo$. Some illustrative choices of parameters are shown in Tab. \ref{Tab4} where we report also the values used in Sec. \ref{sec:LVS-review} to stabilise $\vo$.

\begin{table}[ht]
\centering
\begin{tabular}{c | c c c c c c c | c c c}
\hline
& $g_s$ & $W_0$ & $A_\dP$ & $|\lambda|$ & $C_{24}^\W$ & $C_{26}^\W$ & $C_{28}^\W$ & $\langle\tau_\dP\rangle$ & $\langle\vo\rangle$ & $\langle\tau_\K3\rangle$  \\
\hline
\textbf{DD} & $0.1$ & $3$ & $1$ & $2.0\cdot 10^{-3}$ & $0.1$ & $0.1$ & $0.1$ & $7.8$ & $4.9\cdot 10^{20}$ & $5.1$ \\
\textbf{DD} & $0.08$ & $14$ & $1$ & $3.3\cdot 10^{-4}$ & $0.4$ & $0.4$ & $0.4$ & $9.7$ & $5.0\,\cdot 10^{26}$ & $5.2$ \\
\textbf{ADD} & $0.07$ & $22$ & $1$ & $3.4\cdot 10^{-4}$ & $1.1$ & $1.1$ & $1.1$ & $11.1$ & $5.1\,\cdot 10^{30}$ & $5.1$ \\
\hline
\end{tabular}
\caption{Illustrative values of the microscopic parameters which realise both the DD and the ADD scenarios. For our explicit CY example we have $\xi\simeq 0.456$ and $\alpha=1$.}
\label{Tab4}
\end{table}

\subsection{\texorpdfstring{Deformation of the $\mathbb{P}^1$ base}
                           {Deformation of the P1 base}}
\label{sec:base-deformation}

The moduli stabilisation procedure described in the previous section leads to a situation where the $\mathbb{P}^1$ base is much larger than the K3 fibre, so that the CY manifold $X$ has effectively $2$ large dimensions. To realise the DD scenario, however, the $\mathbb{P}^1$ base should become highly anisotropic and elongated along one direction, so that only $1$ extra dimension remains large. While the volume of the $\mathbb{P}^1$ is controlled by a K\"ahler parameter, its shape is controlled by the complex structure of $X$. In the following, we argue that many K3 fibred CYs admit such a limit in their complex structure moduli space, and explicitly write down such a limit for the CY example of Sec.~\ref{sec:ExplicitCY}.

\subsubsection{Tyurin degenerations of K3 fibred Calabi-Yau threefolds}

CY threefolds generally permit many infinite distance limits in their moduli space in which the CY undergoes a degeneration. This is of great relevance to the swampland program and has been studied in this context in \cite{Grimm:2018ohb,Monnee:2025ynn}, which provide a physicist's introduction to the topic. 

For a CY threefold $X$ depending on a parameter $z$, it is useful to consider the family: 
\begin{equation}
    \mathcal{X} := \{ X(z) \,|\, z \in \mathbb{C} \}\,,
\end{equation}
which is a non-compact algebraic variety of complex dimension four. A general feature of the degenerations we are interested in is that $X(z)$ not only becomes singular as $z \to 0$, but in fact becomes reducible, with several components meeting transversely. At a technical level, this typically requires performing a blow-up of $\mathcal{X}$ over $z=0$ in order to make the degeneration sufficiently `stable'. Note that this does not affect $X(z)$ for any $z \neq 0$; rather, it only changes the algebraic description of the degeneration, leaving the original geometry under discussion unchanged.

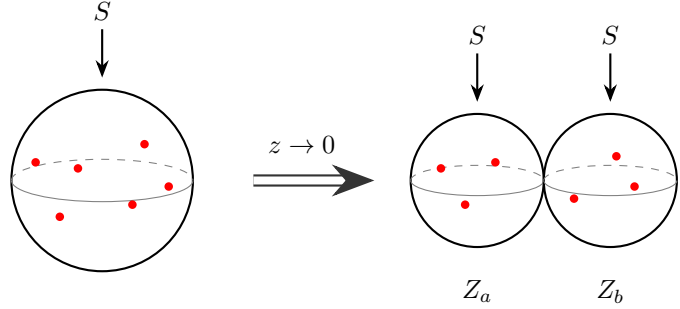
\begin{figure}[htbp]
\centering
\begin{tikzpicture}[scale=0.8, every node/.style={scale=0.9}]

    \draw[thick] (-4,0) circle (1.5cm);
    \draw[dashed, gray] (-5.5,0) arc (180:0:1.5cm and 0.35cm);
    \draw[gray] (-5.5,0) arc (180:360:1.5cm and 0.35cm);
    
    \draw[{Stealth}-, thick, black] (-4, 1.7) -- (-4, 2.5) node[above] {$S$};
    
    \foreach \point in {(-5.1,0.3), (-4.7,-0.6), (-4.4,0.2), (-3.5,-0.4), (-3.3,0.6), (-2.9,-0.1)} {
        \fill[red] \point circle (2pt);
    }

    \draw[double, double distance=3pt, -{Stealth[scale=1.6]}, thick, black!80] (-1.5,0) -- (0.5,0);
    \node[above] at (-0.7, 0.3) {$z \to 0$};

    \draw[thick] (2.2,0) circle (1.1cm);
    \draw[dashed, gray] (1.1,0) arc (180:0:1.1cm and 0.25cm);
    \draw[gray] (1.1,0) arc (180:360:1.1cm and 0.25cm);
    \node[below=1.2cm] at (2.2,0) {$Z_a$};
    
     \draw[thick] (4.4,0) circle (1.1cm);
    \draw[dashed, gray] (3.3,0) arc (180:0:1.1cm and 0.25cm);
    \draw[gray] (3.3,0) arc (180:360:1.1cm and 0.25cm);
    \node[below=1.2cm] at (4.4,0) {$Z_b$};
    
     \draw[{Stealth}-, thick, black] (2.2, 1.3) -- (2.2, 2.1) node[above] {$S$};
    \draw[{Stealth}-, thick, black] (4.4, 1.3) -- (4.4, 2.1) node[above] {$S$};

    \foreach \point in {(1.6,0.2), (2.0,-0.4), (2.5,0.3)} {
        \fill[red] \point circle (2pt);
    }
    \foreach \point in {(3.8,-0.3), (4.5,0.4), (4.8,-0.1)} {
        \fill[red] \point circle (2pt);
    }
\end{tikzpicture}
\caption{A K3 fibred CY threefold which degenerates by a pinching of its $\mathbb{P}^{1}$ base. The red dots indicate singular K3 fibres.}
\label{fig:pinchie}
\end{figure}

The mildest type of such a degeneration is called a Tyurin degeneration \cite{Tyurin2004} (see \cite{Doran_Harder_Thompson_2017} for a discussion in the context of mirror symmetry and \cite{Hassfeld:2025uoy} for the relevance to the emergent string proposal). Here, $X$ degenerates into only two components:
\begin{equation}
\label{eq:Xdeglimit}
    X(z) \rightarrow X(0) =  Z_a \cup_S Z_b\,,
\end{equation}
which furthermore meet along a K3 surface $S$. A specific class of such degenerations are given by CY threefolds fibred by K3 surfaces. Here, the base is necessarily a $\mathbb{P}^1$ and a Tyurin degeneration can be thought of as a pinching of the base, such that it splits into two $\mathbb{P}^1$s meeting in a point, as illustrated in Fig. \ref{fig:pinchie}. The two threefolds $Z_a$ and $Z_b$ are no longer CY but satisfy $c_1(Z_a) = c_1(Z_b) = [S]$. This implies that the singular K3 fibres of $X$ separate into two groups, one for each of the components $Z_a$ and $Z_b$. There cannot be any non-trivial monodromy acting on the fibres over the circle in the base $\mathbb{P}^1$ that pinches in the degeneration, so that a necessary requirement for such a degeneration to exist is that the singular fibres can be separated into two groups such that the product of monodromies for each group is the identity map.  

For CY threefolds realised as hypersurfaces in toric varieties, all of this can be made fairly explicit using reflexive polytopes. For a CY threefold $X_{(\Delta^\circ, \Delta)}$ obtained from a pair $(\Delta^\circ, \Delta)$ of reflexive polytopes, the existence of a K3 fibration can be inferred from a reflexive sub-polytope $\Delta_F^\circ \subset \Delta^\circ$ of dimension $3$. This separates $\Delta^\circ$ into two halves, $\lozenge_a$ and $\lozenge_b$, called tops. Under the technical condition of these tops being `projecting', the degeneration limit \eqref{eq:Xdeglimit} together with the algebraic threefolds $Z_a$ and $Z_b$ can be written down in full generality.\footnote{Here, `projecting' means that a projection of the top $\lozenge^\circ$ onto the hyperplane defined by $\Delta^\circ_F$ does not map $\lozenge^\circ$ outside of $\Delta^\circ_F$. This is a sufficient but not necessary condition.} See \cite{Braun:2016igl} for the construction of the threefolds $Z$ appearing in the degenerations from tops $\lozenge$, and \cite{Davis:2014:STS,Cvetic:2015uwu, Braun:2017ryx} for a discussion of degeneration limits of K3 fibred toric hypersurfaces. 

So far we have sketched Tyurin degenerations of K3 fibred CYs from the perspective of algebraic and toric geometry. As $X$ admits a Ricci-flat metric, it is natural to ask how such a degeneration develops metrically, and we wish to argue that the base $\mathbb{P}^1$ gets elongated in this limit. Let us assume that $C$ is a curve in the K3 fibre which shrinks to a point over two points $p_1$ and $p_2$ of the $\mathbb{P}^1$ base. 

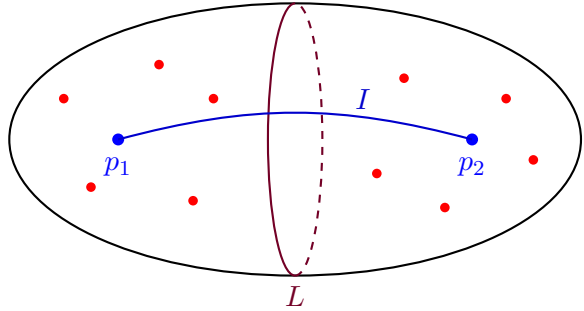
\begin{figure}[htbp]
\centering
\begin{tikzpicture}[scale=0.9]

    \draw[thick] (0,0) ellipse (4.2cm and 2.0cm);

    \draw[thick, purple!60!black] (0, 2.0) arc (90:270:0.4cm and 2.0cm);
    \draw[thick, dashed, purple!60!black] (0, -2.0) arc (270:450:0.4cm and 2.0cm);
    \node[purple!60!black, below] at (0, -2.0) {$L$};

    \coordinate (P1) at (-2.6, 0);
    \coordinate (P2) at (2.6, 0);
    
    \fill[blue] (P1) circle (2.5pt) node[below=3pt] {$p_1$};
    \fill[blue] (P2) circle (2.5pt) node[below=3pt] {$p_2$};
 
    \draw[thick, blue!80!black] (P1) to[out=15, in=165] (P2);
    \node[blue!80!black, above] at (1, 0.3) {$I$};

    \foreach \p in {(-3.4,0.6), (-3.0,-0.7), (-2.0,1.1), (-1.5,-0.9), (-1.2,0.6)} {
        \fill[red] \p circle (2pt);
    }

    \foreach \p in {(1.2,-0.5), (1.6,0.9), (2.2,-1.0), (3.1,0.6), (3.5,-0.3)} {
        \fill[red] \p circle (2pt);
    }

\end{tikzpicture}
\caption{Shrinkable $1$-cycle $L$ and interval $I$ in the $\mathbb{P}^{1}$ base used to construct $3$-cycles of $X$ that allow to measure the degeneration of the base.}
\label{fig:cycinbase}
\end{figure}

This allows us to construct a $3$-cycle $\Sigma_3^{(I)}$ by fibering $C$ over an interval $I$ connecting $p_1$ and $p_2$, as can be seen in Fig. \ref{fig:cycinbase}. As there is no monodromy acting on the K3 fibre over the circle $L$ which pinches in the degeneration limit, we can form another $3$-cycle by taking $C$ over $L$. As $L$ collapses to a point when $z \rightarrow 0$, it must be that in this limit:
\begin{equation}
   \left. \int_{C \times L} \Omega_3  \middle/ \int_{C \times I} \Omega_3 \right. \rightarrow 0\,.
\end{equation}
This implies that the ratio of the circle $L$ and any interval connecting the two ends of the base $\mathbb{P}^1$ goes to zero in this limit.

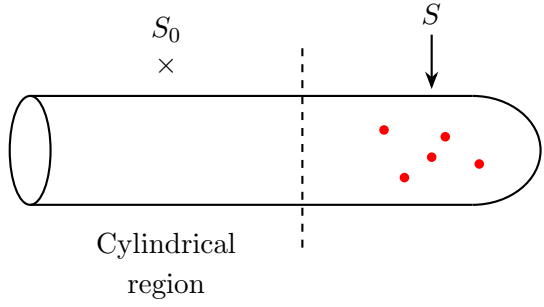
\begin{figure}[htbp]
\centering
\begin{tikzpicture}[scale=0.9]

    \draw[thick] (-4,0) ellipse (0.3cm and 0.8cm);
    
    \draw[thick] (-4, 0.8) -- (2.5, 0.8);
    \draw[thick] (-4, -0.8) -- (2.5, -0.8);
    
    \draw[thick] (2.5, 0.8) arc (90:-90:1.0cm and 0.8cm);

    \draw[dashed, black, thick] (0, 1.5) -- (0, -1.5);

    \node[align=center, black] at (-2, -1.7) {Cylindrical\\region};
    \node[above=0.4cm] at (-2, 1) {$S_0$};
    \node[above=0.4cm] at (-2, 0.5) {$\times$};
    
    \node[right] at (1.6, 2) {$S$};
    \draw[-{Stealth}, thick, black] (1.9, 1.7) -- (1.9, 0.9);

    \foreach \point in {(1.2,0.3), (1.5,-0.4), (2.1,0.2), (2.6,-0.2), (1.9,-0.1)} {
        \fill[red] \point circle (2pt);
    }

\end{tikzpicture}
\caption{A cartoon of an asymptotically cylindrical CY threefold $X_a$. $X_a$ contains a region which is not cylindrical to the right, where the K3 fibration is non-trivial. On the left, the fibration becomes trivial and $X_a$ metrically approaches a cylinder times $S_0$.}
\label{fig:acylCY}
\end{figure}

An alternative perspective, informed by differential geometry, leads to the same conclusion: in the limit under consideration, $X$ can be viewed as being glued from two pieces. Using the construction of \cite{Braun:2016igl}, let us directly obtain the associated K3 fibred threefold $Z_a$ of a projecting top $\lozenge_a$. Upon excising a single K3 fibre we find a non-compact threefold:
\begin{equation}
X_a := Z_a \setminus S_{0,a}\,,
\end{equation}
which turns out to be an asymptotically cylindrical CY threefold \cite{Corti:2013ers}. This means that $X_a$ is a CY threefold that metrically asymptotes to a cylinder times a K3 surface, $\mathbb{R}_+ \times  S^1 \times S_{0,a}$, as shown in Fig. \ref{fig:acylCY}. Crucially, the cylindrical region can be made (metrically) arbitrarily long. 

Two asymptotically cylindrical CY threefolds $X_a$ and $X_b$ can in turn be glued along their cylindrical regions to a compact K3 fibred CY if the asymptotic K3 surfaces $S_{0,a}$ and $S_{0,b}$ are isomorphic as complex manifolds \cite{doi_yotsutani_2014,talbot2017gluing}. Due to the cylindrical regions being arbitrarily long, this implies that the resulting CY threefold is K3 fibred over an elongated $\mathbb{P}^1$ base. Hence, a K3 fibred CY threefold with a Tyurin degeneration as above not only gives rise to a pair of asymptotically cylindrical CY threefolds, $X_a = Z_a \setminus S$ and $X_b = Z_b \setminus S$, but also allows the asymptotic K3 fibres of their cylindrical regions to be chosen as the common K3 surface $S = Z_a \cap Z_b$. These fibres are therefore isomorphic by construction. This means we can think of any $X$ with such a degeneration as being glued from two asymptotically cylindrical pieces. This gluing construction also has an analogue in algebraic geometry discussed in \cite{KawamataNamikawa1994}, where it is shown how a suitable variety $Z_a \cup_S Z_b$ with normal crossings can be smoothed to a CY. 

We can therefore conclude that stretching limits of the $\mathbb{P}^1$ base of K3 fibred CY threefolds are a rather generic feature which exists under mild assumptions. As the vast majority of CY threefolds admit K3 fibrations \cite{Candelas:2012uu,Anderson:2017aux}, this makes 
such limits a generic feature of CY threefolds.

\subsubsection{Anisotropic limit for explicit CY model}

Let us now consider the CY threefold introduced in Sec.~\ref{sec:ExplicitCY} and describe one of its degeneration limits explicitly. It is useful to start by writing its weight system, along with the ray generators $\nu_i$ associated to the following homogeneous coordinates.

\begin{equation}
\label{eq:weightsandrays}
\begin{array}{c|cccccccc|c}
 & x_1 & x_2 & x_3 & x_4 & x_5 & x_6 & x_7 & x_8  &   \tabularnewline \hline 
   &  0  &  0  &  0  &  1  &  1  &  0  &  0  & 2 & 4 \tabularnewline 
   &  0  &  0  &  1  &  0  &  0  &  1  &  0 & 2 & 4 \tabularnewline
   &  0  &  1  &  0  &  0  &  0  &  0  &  1  & 2 & 4 \tabularnewline
   &  1  &  0  &  0  &  1  &  0  &  1  &  1  & 4 & 8 \tabularnewline\hline 
   &  \nu_1  & \nu_2   & \nu_3  & \nu_4  &\nu_5   &\nu_6   &  \nu_7  &\nu_8  &  \tabularnewline\hline 
   & \begin{pmatrix}
       0 \\ -1 \\ -2 \\ 1
   \end{pmatrix}
   & \begin{pmatrix}
       1 \\ 0 \\ 0 \\ 1
   \end{pmatrix}
   & \begin{pmatrix}
      -1 \\ 0 \\ -2 \\ 0
   \end{pmatrix}
   & \begin{pmatrix}
       0 \\ 1 \\ 0 \\ 0
   \end{pmatrix}
   & \begin{pmatrix}
       0 \\ -1 \\ -2 \\ 0
   \end{pmatrix}
   & \begin{pmatrix}
       1 \\ 0 \\ 0 \\ 0
   \end{pmatrix}
   & \begin{pmatrix}
       -1 \\0  \\ -2 \\ -1
   \end{pmatrix}
   & \begin{pmatrix}
       0 \\ 0 \\ 1 \\ 0
   \end{pmatrix}
   &
\end{array}\:
\end{equation}
The CY threefold $X$ is then defined as the vanishing locus of a sufficiently generic polynomial of degree $(4,4,4,8)$ in these homogeneous coordinates:
\begin{equation}
X: \qquad Q_{4,4,4,8}(x_i) = 0 \,.
\label{eq:defX}
\end{equation}
It is a hypersurface in a toric variety $\mathbb{P}_\Sigma$, whose fan $\Sigma$ is obtained from a suitable triangulation of the polytope
\begin{equation}
\Delta^\circ = \left\{\nu_1, \cdots, \nu_8  \right\}_{\text{conv}}  \,,
\end{equation}
constructed as the convex hull of the $\nu_i$. Note that $\Delta^\circ$ is reflexive and furthermore contains a reflexive sub-polytope
\begin{equation}
\Delta^\circ = \left\{\nu_3,\nu_4,\nu_5,\nu_6,\nu_8 \right\}_{\text{conv}}  \:,
\end{equation}
which gives rise to a K3 surface $S$ which is a double cover of $\mathbb{P}^1_{[x_4:x_5]} \times \mathbb{P}^1_{[x_3:x_6]}$. The existence of this sub-polytope signals a fibration of $S$ over a $\mathbb{P}^1$ with homogeneous coordinates $[x_1 x_2 : x_7]$. We can think of this as a K3 fibration with a single reducible fibre with the two components $D_1$ and $D_2$. 

As $\Delta^\circ$ has a sub-polytope $\Delta^\circ_F$ of codimension $1$, we can decompose $\Delta^\circ$ into the two tops: 
\begin{equation}
\begin{aligned}
      \lozenge^\circ_a  &= \left\{\nu_1, \nu_2,\nu_3, \nu_4, \nu_5, \nu_6, \nu_8  \right\}_{\text{conv}}   \\
      \lozenge^\circ_b  &= \left\{\nu_3, \nu_4, \nu_5, \nu_6, \nu_7, \nu_8   \right\}_{\text{conv}} \,.
    \end{aligned}
\end{equation}
We now want to engineer a degeneration in which the $\mathbb{P}^1$ base becomes very long. In particular, we want to see whether we can stretch it such that the distance between $x_1 x_2=0$ and $x_7=0$ becomes large. Furthermore, we want to see that there is no monodromy acting on the K3 fibre when going around the circle $S^1$ given by $|x_1 x_2/x_7|^2 = 1$.

To describe a limit in which the $\mathbb{P}^1_{[x_1 x_2 : x_7]}$ base of the K3 fibration of $X$ degenerates into two $\mathbb{P}^1$s, we follow \cite{Davis:2014:STS,Cvetic:2015uwu, Braun:2017ryx} and appropriately parametrise a slice through the complex structure moduli space of $X$. This is done analogously to the toy example discussed in App. \ref{App}. We first introduce a dependence of the coefficient of \eqref{eq:defX} on a parameter $\zeta_a$ such that the degeneration we are interested in happens at $\zeta_a = 0$. Such a degeneration is not of stable type: thinking of $\zeta_a$ not as a parameter in the defining equation of $X$ but as a coordinate, the resulting 4D space $\mathcal{X}$, i.e. the family $\mathcal{X} = \{ X(\zeta_a)\,|\, \zeta_a \in \mathbb{C} \}$, is singular. This implies that we cannot properly track what happens to $X$ when letting $\zeta_a = 0$, so that we need to blow up the family $\mathcal{X}$ to turn this into a stable degeneration, which introduces a second parameter $\zeta_b$. 

We can describe the above procedure at once by using toric techniques as follows. We first lift the ray generators in \eqref{eq:weightsandrays} into $\mathbb{R}^5$ where we introduce two new rays:
\begin{equation}
\nu_a = (0,0,0,1,1)\,, \qquad \nu_b = (0,0,0,0,1)\,,
\end{equation}
corresponding to the homogeneous coordinates $\zeta_a$ and $\zeta_b$, along with the higher-dimensional cones: 
\begin{equation}
\langle \sigma_a , \nu_a \rangle\,, \qquad \langle \sigma_b , \nu_b \rangle\,,     \qquad\langle \sigma_a ,\sigma_b , \sigma_F \rangle\,,
\end{equation}
where $\sigma_a$ and $\sigma_b$ are cones of $\Sigma$ with generators contained in $\lozenge^\circ_a$ and $\lozenge^\circ_b$, and $\sigma_F$ are cones of $\Sigma$ with generators purely contained in $\Delta^\circ_F$. The above implies that there is also a further $\mathbb{C}^*$ action with weights:
\begin{equation}
    \begin{array}{cccccccccc}
x_1 & x_2 & x_3 & x_4 & x_5 & x_6 & x_7 & x_8 & \zeta_a & \zeta_b \tabularnewline \hline 
0   & 0   & -1  & 0   & 0   & 0   & 1   & 0   & 1       & -1 
    \end{array}
\end{equation}
such that the defining equation of $\mathcal{X}$ has weight $0$ under this scaling. In practice, this amounts to dressing all monomials in \eqref{eq:defX} by powers of $\zeta_a$ and $\zeta_b$ such that each term has weight $0$. This results in the family $\mathcal{X}$ of threefolds:
\begin{equation}
    X(\zeta_a,\zeta_b): \quad Q_{4,4,4,8,0}(x_i,\zeta_a,\zeta_b) = 0 \,.
\end{equation}

We can now describe the degeneration limit of the $\mathbb{P}^1$ base as letting $z = \zeta_a \zeta_b \rightarrow 0$. For all values $z \neq 0$, this still gives a smooth CY threefold, and the degeneration occurs only at the origin. Note that $z = 0$ is reducible and has two components $Z_a$ and $Z_b$. As we will now argue, each of these is a threefold with a K3 fibration over $\mathbb{P}^1$ such that $c_1(Z_a) = c_1(Z_b) = [S]$ where $S$ is the K3 surface $\zeta_a = \zeta_b = 0$.

\subsection*{$\mathbf{Z_b:\,\zeta_b = 0}$}

First note that $\zeta_b$ cannot vanish simultaneously with $x_1$ or $x_2$ as these do not share any cones, so we can set these two homogeneous coordinates to $1$ by gauge fixing the last two $\mathbb{C}^*$ actions in 
\eqref{eq:weightsandrays}. The resulting threefold is then described by:
\begin{equation}
    Q_{4,4,1}(x_3,x_4,x_5,x_6,x_7,x_8,\zeta_a) = 0\,,
\end{equation}
in an ambient toric variety with weights:
\begin{equation}
\label{eq:weightsZb}
\begin{array}{ccccccc}
 x_3 & x_4 & x_5 & x_6 & x_7 & x_8  & \zeta_a  \tabularnewline \hline 
  0  &  1  &  1  &  0  &  0  & 2 & 0\tabularnewline 
  1  &  0  &  0  &  1  &  0 & 2 & 0 \tabularnewline
-1 & 0 & 0 & 0 & 1 & 0 & 1 
\end{array}   
\end{equation}
This is a fibration of the same algebraic K3 surface, a double cover of $\mathbb{P}^1 \times \mathbb{P}^1$ over a $\mathbb{P}^1$ with homogeneous coordinates $[x_7:\zeta_a]$, such that $c_1(Z_b) = [S]$. 

\subsection*{$\mathbf{Z_a:\, \zeta_a = 0}$}

As $\zeta_a$ cannot vanish simultaneously with $x_7$ we can now gauge fix $x_7=1$. We obtain:
\begin{equation}
Z_a: \quad    Q_{4,4,4,0} = 0\,,
\end{equation}
as a hypersurface in an ambient toric variety with weights:
\begin{equation}
\label{eq:weightsZbNew}
\begin{array}{cccccccc}
 x_1 & x_2 & x_3 & x_4 & x_5 & x_6 & x_8  & \zeta_b  \tabularnewline \hline 
  0  &  0  &  0  &  1  &  1  & 0 & 2 & 0 \tabularnewline 
  0  &  0  &  1  &  0  &  0 & 1 & 2 & 0 \tabularnewline
  1 & -1 & 0 & 1 & 0 & 1 & 2 & 0   \tabularnewline
  0 &  1 & 0 & 0 & 0 & -1 & 0 & 1
\end{array}   \, ,
\end{equation}
where we have formed some linear combinations of the $\mathbb{C}^*$ action to facilitate the gauge fixing. 
Now this describes a K3 fibration over a $\mathbb{P}^1$ with homogeneous coordinates $[x_1 x_2 : \zeta_b ]$, with the K3 fibre from the same algebraic family as above. Furthermore, $c_1(Z_a)$ is easily seen to be $c_1(Z_a) = [S]$. 

\subsection*{$\mathbf{S = Z_a \cap Z_b}$}

When both $\zeta_a = \zeta_b = 0$ we can gauge fix $x_1 = x_2 = x_7 = 1 $ and find a hypersurface:
\begin{equation}
S: \quad    Q_{4,4} = 0 \,,
\end{equation}
in a toric ambient space with weights:
\begin{equation}
\label{eq:weightsZbNewNew}
\begin{array}{ccccc}
x_3 & x_4 & x_5 & x_6 & x_8  \tabularnewline \hline 
0  &  1  &  1  &  0 & 2 \tabularnewline 
1  &  0  &  0  &  1 & 2 \tabularnewline
\end{array}   \, ,
\end{equation}
i.e. a K3 surface realised as a double cover of  $\mathbb{P}^1\times  \mathbb{P}^1$.

In summary, we have therefore written down a limit of $X$ in which $X$ splits into $Z_a \cup Z_b$. Specifically, this happens as the $\mathbb{P}^1_{x_1 x_2 : x_7}$ base of the K3 fibration on $X$ splits into two $\mathbb{P}^1$s with homogeneous coordinates $[x_1 x_2: \zeta_b]$ and $[x_7:\zeta_a]$. These two $\mathbb{P}^1$s touch in a point. The existence of $Z_a$ and $Z_b$ together with the smoothness of the fibre where they touch implies that there cannot be any monodromy when encircling the locus $\zeta_b = 0$ in $\mathbb{P}^1_a$ or $\zeta_a=0$ in $\mathbb{P}^1_b$ on a sufficiently small loop. These loops become homologous when going away from $\zeta_a \zeta_b=0$. As no singular fibres cross such a loop when deforming to a finite and sufficiently small $\zeta_a \zeta_b$, there is still a trivial monodromy around such a loop.

Finally, we may think of the present setup in terms of gluing: excising $\zeta_b = 0$ from $Z_a$ and $\zeta_a = 0$ from $Z_b$ we find two asymptotically cylindrical CY threefolds $X_a$ and $X_b$ with arbitrarily long necks. As the K3 fibres over these neck regions are both from the same algebraic family, we can choose them to be isomorphic, and then glue $X_a$ and $X_b$ together along their neck regions to reconstruct $X$. Crucially, the two neck regions are glued together to form the middle part of the $\mathbb{P}^1$ base of the K3 fibration on $X$, which again shows the existence of a stretching limit.

\section{Conclusions}
\label{Concl}

In this work we have presented a concrete proposal for realising anisotropic string compactifications with 1 or 2 large EDs, corresponding respectively to the ADD and Dark Dimension scenarios. Our approach is based on fluxed type IIB CY orientifold compactifications within the LVS framework, combined with perturbative and non-perturbative effects that stabilise all moduli in a controlled regime.

The key ingredient of our construction is the realisation of anisotropy in both K\"ahler and complex structure moduli spaces. On the K\"ahler side, we have shown how an exponentially large overall volume can be generated through the standard LVS mechanism, while leaving a flat direction associated with the size of a K3 fibre. This direction is subsequently lifted by subleading perturbative effects, i.e. string loop correction and $F^4$ terms, leading to a stabilisation of the fibre volume at moderately small values for natural values of the underlying parameters, such as $g_s\sim \mathcal{O}(0.1)$ and $W_0\sim\mathcal{O}(1-10)$. The resulting vacuum realises a hierarchical structure in which different effective gravitational scales emerge, depending on the anisotropy parameter $u$, allowing for both ADD- and DD-type limits within a unified setup.

On the complex structure side, we have identified a degeneration limit in which the anisotropy of the $\mathbb{P}^1$ base can be generated, provided the fluxes stabilise the complex structure moduli near this regime. This drives the geometry towards a Tyurin degeneration limit in which the base becomes elongated, with one direction much longer than the transverse one. This mechanism provides a geometric realisation of the DD scenario in terms of a controlled infinite-distance limit in complex structure moduli space, thereby connecting the EFT description of large EDs with the global structure of the CY moduli space.

We have also presented an explicit CY example with a consistent orientifold involution, brane configuration and D-term stabilisation. In this model, fluxed D7-brane sectors reduce the effective K\"ahler moduli space to three light directions, while simultaneously generating a negative D3-tadpole of sufficient magnitude to allow for full flux stabilisation of the axio-dilaton and complex structure moduli. This explicit construction demonstrates that anisotropic LVS vacua realising large EDs can be embedded consistently within global string compactifications with controlled tadpoles and well-defined EFTs.

From a phenomenological perspective, the resulting setup exhibits a rich spectrum of light moduli whose masses are tightly correlated with the size of the EDs. While this opens the possibility of observable signatures such as deviations from Newtonian gravity at micron scales, it also introduces significant challenges, including the cosmological moduli problem, potential fifth force constraints, and dark matter overproduction. We have discussed how these issues may be addressed through suppression of initial misalignment, geometric sequestering, and dilution mechanisms. Moreover, the gravitino mass turns out to be too low to be compatible with a spontaneous breaking of supersymmetry. Indeed, supersymmetric particles above the TeV scale require a non-supersymmetric SM brane whose tension, if not cancelled by additional effects, would naively generate a dangerously large contribution to the scalar potential. 

Several open questions remain. In particular, a more systematic understanding of flux choices capable of engineering the required complex structure hierarchies is needed, as well as a more complete control over higher-order quantum corrections that could affect moduli stabilisation in the anisotropic regime. Furthermore, the interplay between uplifting mechanisms and the structure of the scalar potential requires further investigation to ensure full consistency of de Sitter vacua in this class of models. Without an appropriate tuning from the uplifting sector, the scalar potential at the minimum would scale as $V\sim\left(M_\KK^{5D}\right)^2\,M_p^2$ in the DD scenario, and as $V\sim\left(M_\KK^{6D}\right)^3\,M_p$ in the ADD case, thereby weakening the motivation for these models as solutions to the cosmological constant problem.

Let us stress that, ADD and DD scenarios can be embedded in string theory not only as dS vacua but also as dynamical dark energy models. Indeed, we have shown that quintessence can be realised along a shallow runaway with all the other directions stabilised. Reproducing the correct vacuum energy requires, however, tuning $W_0$ to exponentially small values. Moreover, in the absence of a minimum at $\tau_\K3\sim 5$, there is no dynamical mechanism which singles out this region of moduli space. 

A completely viable cosmological history should include also inflation. Constructing inflationary models that are compatible with a very low KK scale is notoriously difficult. The best inflaton candidate is the overall volume mode $\vo$ which could evolve from a relatively small value during inflation, yielding large energy scales, to an exponentially large value after the end of inflation, hence lowering all mass scales. However, given that the $\vo$ potential does not feature any plateau-like region, inflation would need to occur around a tuned inflection point \cite{Conlon:2008cj}. Moreover, achieving a very large difference between the value of $\vo$ during inflation and today can be challenging in explicit models \cite{Cicoli:2015wja}. See also \cite{Anchordoqui:2022svl,Anchordoqui:2023etp} for similar ideas for realising inflation within the DD scenario.

Despite these challenges, our results provide a concrete framework in which both ADD and DD scenarios can be embedded within string theory. This offers a unified perspective on large EDs, moduli stabilisation, and late-time cosmology, and opens the door to further explorations of their phenomenological and observational consequences.

\acknowledgments

We would like to thank Alfio Bonanno, Federico Carta, Francisco Gil Pedro, Dieter L\"ust, Miguel Montero, Georges Obied, Tony Padilla, Cumrum Vafa, Alexander Westphal for useful discussions. MC, RM and RV acknowledge support by INFN Iniziativa Specifica ST\&FI.

\appendix

\section{A Tyurin degeneration limit of an elliptic K3 surface}
\label{App}

As a pedagogical toy example of a fibred CY in which the $\mathbb{P}^1$ base is stretched, let us consider a K3 surface fibred by elliptic curves. This degeneration limit is well-known in the physics literature as it is a crucial ingredient in the duality between F-theory and the $E_8 \times E_8$ heterotic string.

Consider an elliptic K3 surface described by a Weierstrass model with two $II^*$ fibres given by:
\begin{equation}
  y^2 = x^3 + z_1^4 z_2^4 a x w^4 + z_1^5 z_2^5 (d z_1^2 + b z_1 z_2 + d' z_2^2) w^6 \, .
\end{equation}
Here $[z_1:z_2]$ are coordinates on the $\mathbb{P}^1$ base,
the $II^*$ fibres are located at $z_1=0$ and $z_2=0$, and the complex structure is parametrized by $a,b,d,d' \in \mathbb{C}$.
There are $4$ additional singular elliptic fibres located at the $4$ points $p_i$ for which:
\begin{equation}
 P_4 = 27d'^2 z_2^4 + 54d'b z_2^3 z_1 +(4a^3+27b^2+54d'd) z_1^2 z_2^2 +54 db z_2 z_1^3 + 27d^2 z_1^4  = 0\, .
\end{equation}

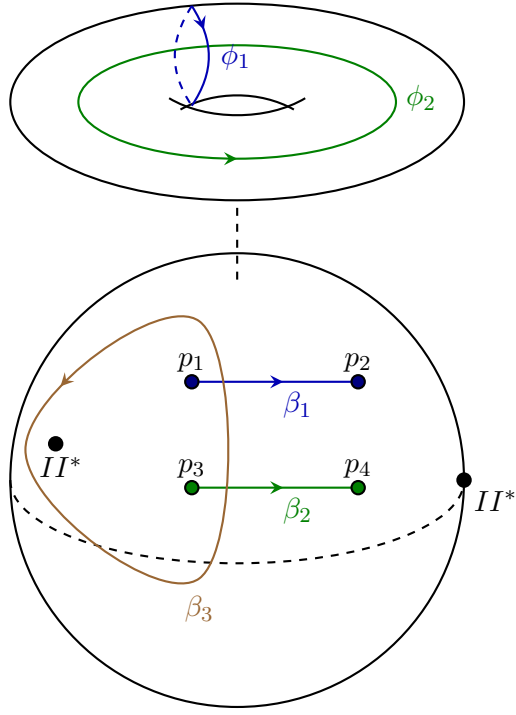
\begin{figure}[!h]
\begin{center}

\begin{tikzpicture}[
    every node/.style={font=\normalsize},
    arrowmark/.style={decoration={markings, mark=at position #1 with {\arrow[scale=1.2]{stealth}}}, postaction={decorate}}
]

\draw[thick] (0, 5) ellipse (3cm and 1.3cm);

\draw[thick] (-0.9, 5.05) .. controls (-0.4, 4.75) and (0.4, 4.75) .. (0.9, 5.05);
\draw[thick] (-0.75, 4.9) .. controls (-0.3, 5.15) and (0.3, 5.15) .. (0.75, 4.9);

\draw[thick, green!50!black, arrowmark=0.75] (0, 5.0) ellipse (2.1cm and 0.75cm);
\node[right, green!50!black] at (2.1, 5.05) {$\phi_2$};

\draw[thick, blue!70!black, arrowmark=0.25] (-0.6, 6.27) .. controls (-0.3, 5.8) and (-0.3, 5.4) .. (-0.6, 4.95);

\draw[thick, blue!70!black, dashed] (-0.6, 6.27) .. controls (-0.9, 5.8) and (-0.9, 5.4) .. (-0.6, 4.95);
\node[right, blue!70!black] at (-0.35, 5.6) {$\phi_1$};

\draw[thick, dashed] (0, 3.6) -- (0, 2.55);

\draw[thick] (0,0) circle (3cm);

\draw[thick, dashed] (-3,0) arc (180:360:3cm and 1.1cm);

\draw[thick, blue!70!black, arrowmark=0.55] (-0.6, 1.3) -- (1.6, 1.3);
\node[below, blue!70!black] at (0.8, 1.3) {$\beta_1$};

\draw[thick, green!50!black, arrowmark=0.55] (-0.6, -0.1) -- (1.6, -0.1);
\node[below, green!50!black] at (0.8, -0.1) {$\beta_2$};

\draw[thick, brown!80!black, arrowmark=0.90] plot[smooth cycle, tension=0.8] 
    coordinates {(-0.5, 2.1) (-2.8, 0.4) (-0.5, -1.3) };
\node[below, brown!80!black] at (-0.5, -1.4) {$\beta_3$};

\filldraw[fill=black, draw=black, thick] (-2.4, 0.48) circle (2.5pt);
\node[below, xshift=2pt, yshift=-1pt] at (-2.4, 0.48) {$II^*$};
 
\filldraw[fill=black, draw=black, thick] (3, 0) circle (2.5pt);
\node[below right] at (3, 0) {$II^*$};

\filldraw[fill=blue!50!black, draw=black, thick] (-0.6, 1.3) circle (2.5pt);
\node[above] at (-0.6, 1.3) {$p_1$};

\filldraw[fill=blue!50!black, draw=black, thick] (1.6, 1.3) circle (2.5pt);
\node[above] at (1.6, 1.3) {$p_2$};

\filldraw[fill=green!50!black, draw=black, thick] (-0.6, -0.1) circle (2.5pt);
\node[above] at (-0.6, -0.1) {$p_3$};

\filldraw[fill=green!50!black, draw=black, thick] (1.6, -0.1) circle (2.5pt);
\node[above] at (1.6, -0.1) {$p_4$};

\end{tikzpicture}

\caption{A cartoon depicting an elliptic K3 surface with two singularities of type $II^*$.
\label{e8e8figure}}
\end{center}
\end{figure}

The limit we are interested in has been discussed in
\cite{Greene:1989ya,Morrison:1996pp}, and it is given by
letting $a,b \rightarrow \infty$ with $a^3/b^2,d,d'$ constant. In this limit, $P_4$ becomes:
\begin{equation}
 P_4 \quad\rightarrow\quad (4a^3+27b^2)\, z_1^2 z_2^2 = 0\,.
\end{equation}
so that two of the four roots of $P_4$ have moved to $z_1=0$, and the other $2$ have moved to $z_2=0$. In this limit the $\tau$ function of the elliptic fibre approaches:
\begin{equation}
 j(\tau) \sim \frac{a^3}{z_1^{12} z_2^{12}(4a^3+27b^2)}\,,
\end{equation}
i.e. there is no non-trivial behaviour away from $z_1=0$ and $z_2=0$.

The same limit can be considered after slightly deforming the $II^*$ fibres. Again, all singular fibres will localise close to the $2$ points $z_1=0$ and $z_2=0$ at the tips of the $\mathbb{P}^1$ base.

Let us try to see a little more clearly how the base becomes stretched in this limit. To do so, we parametrise the holomorphic $2$-form as:
\begin{equation}
\label{omegaexpe8e8}
\Omega_2= \alpha_1' - \tau\sigma \alpha_1 + \tau \alpha_2'
 + \sigma \alpha_2\,,
\end{equation}
with $\alpha_i,\alpha_i'$ generators of the transcendental lattice of $S$.

We can construct the dual cycles explicitly as follows. Let us denote the $4$ roots of $P_4$ by $p_i$, $i=1..4$. The $1$-cycles in the $T^2$ fibre that collapse over these $4$ points are pairwise the same, for $p_1$ and $p_2$ a cycle $\phi_1$ shrinks and for $p_3$ and $p_4$ a $1$-cycle $\phi_2$ shrinks. We may then choose a basis such that $\phi_1$ and $\phi_2$ are as depicted in Fig. \ref{e8e8figure}. Hence we may construct a $2$-cycle $\gamma_1$ by fibring $\phi_2$ over the interval $\beta_1$ connecting $p_1$ and $p_3$. This cycle has the topology of a $2$-sphere and therefore its self-intersection number is $-2$. Similarly, a second sphere is made of the fibration of
$\phi_1$ over the path $\beta_2$ connecting $p_2$ and $p_4$. Hence we may suggestively write
\begin{align}
 \gamma_1 = \beta_1\ltimes \phi_2 \quad\quad \gamma_2 = \beta_2\ltimes\phi_1 \, .
\end{align}
Furthermore, the $SL(2,\mathbb{Z})$ monodromies of this configuration are such that the monodromy map induced along
the loop $\beta_3$ is trivial. This can be seen from the $\mathbb{Z}_2$ symmetry which exchanges the internal and external regions of the sphere surrounded by $\beta_3$. It implies that the monodromy around $\beta_3$ should coincide with its inverse. Hence we can fibre any $1$-cycle in the elliptic fibre over $\beta_3$ to obtain a $2$-cycle. Using the same basis of cycles in the fibre as before, we can hence form the two cycles:
\begin{align}
 \alpha_1 = \beta_3\ltimes\phi_1 \quad\quad\alpha_2=\beta_3\ltimes \phi_2\, ,
\end{align}
so that
\begin{equation}
 \alpha_i\cdot\gamma_j=\delta_{ij} \,.
\end{equation}
As the fibration of the elliptic fibre along $\beta_3$ is trivial, these cycles are just $S^1\times S^1$, i.e. they are $2$-tori, so that:
\begin{equation}
 \alpha_1^2=\alpha_2^2=0\,.
\end{equation}
Finally, we may define
$\alpha_1'=\gamma_1+\alpha_1$, $ \alpha_2'=\gamma_2+\alpha_2$
so that the four $2$-cycles $\alpha_i$ have the intersection form:
\begin{equation}
 T_X= \left(\begin{array}{cc}
  0 & 1\\
  1 & 0
 \end{array}\right) \oplus
 \left(\begin{array}{cc}
  0 & 1\\
  1 & 0
 \end{array}\right) = U^{\oplus 2}\, .
 \label{T_X_e8e8}
\end{equation}

The relationship of $\tau$ and $\sigma$ with $a,b,d,d'$ is given by \cite{ElkiesKumar2014,LopesCardoso:1996hq}:
\begin{equation}
\label{eqj1j2k3e8e8}
\begin{aligned}
   - \frac{a^3}{27\, d\, d'} &=    j_1\, j_2 \\
     \frac{b^2}{4\, d\, d'} &=   ( j_1-1)( j_2-1)
\end{aligned}
\end{equation}
with $j_i={j (\tau_i)/ 1728}$  and $\tau_i=(\sigma,\tau)$.

For special Lagrangian $2$-cycles of a K3 surface, we can measure the volume by an integral of the holomorphic $2$-form. For all other cycles in the same class, this then gives a lower bound on the volume. Let us first examine the complex structure of the elliptic fibre over $\beta_3$. We can find this by comparing the two cycles $\phi_1$ and $\phi_2$ over $\beta_3$, i.e. by considering:
\begin{equation}
 \frac{\int_{\alpha_2}\Omega_2}{\int_{\alpha_1}\Omega_2}
 = \tau\,.
\end{equation}
In the limit we are considering this ratio is fixed.

We can find a lower bound for the distance between $p_1$ and $p_2$ relative to $\beta_3$ as:
\begin{equation}
d(p_1,p_2) \geq \left|\frac{\int_{\gamma_1}\Omega_2}{\int_{\alpha_2} \Omega_2}\right|
= \left|\sigma\right|\,.
\end{equation}
Similarly, we can give a lower bound for distance between $p_3$ and $p_4$ as:
\begin{equation}
d(p_3,p_4) \geq  \left|\frac{\int_{\gamma_2}\Omega_2}{\int_{\alpha_1} \Omega_2}\right|
 = \left|\sigma\right|\,.
\end{equation}
In the limit $a\rightarrow \infty$, $b \rightarrow \infty$ we find that $j_2 \rightarrow \infty$ by \eqref{eqj1j2k3e8e8} using that $j_1$ stays finite. This implies that $\sigma \rightarrow \infty$, which in turn implies that the distances $d(p_1,p_2)$ and $d(p_3,p_4)$ become arbitrarily large. As $\tau$ stays finite and the volume of the $\mathbb{P}^1$, which is given by the integral of the K\"ahler form, also stays finite, this implies that the $\mathbb{P}^1$ becomes elongated.

In algebraic geometry we do not have direct access to a metric, and the stretching limit can be seen as a degeneration limit as follows. From \eqref{eqj1j2k3e8e8} it is clear that we may, for example, let $d \rightarrow 0$ and set $d'=1$ instead of sending
$a,b\to\infty$. Let us make this explicit by
promoting $d$ to a coordinate $\zeta_1$ taking values in $\mathbb{C}$. Let us therefore consider:
\begin{equation}
\label{eq:k3lambda}
  y^2 = x^3 +  z_1^4 z_2^4 a x w^4 + z_1^5 z_2^5 (\zeta_1 z_1^2 + b z_1 z_2 +  z_2^2) w^6\, .
\end{equation}
There is a severe degeneration for which the whole family $S\hookrightarrow X \rightarrow \mathbb{C}_d$ is singular at $y=x=z_2=\zeta_1=0$ even after resolving the $II^*$ fibres.\footnote{In F-theory, this gives rise to a `$(4,6,12)$-point' which realises the singular geometry of the E-string SCFT.}

This can be turned into a stable degeneration in which we can keep track of what happens to the geometry after performing a weighted crepant blow-up (at $y=x=z_1=\zeta_1=0)$ with weights
$(3,2,1,1;-1)$ of this family. The weight system becomes:
\begin{equation}
\begin{tabular}{ccccccc}
$y$ & $x$ & $w$ & $z_1$ & $z_2$ & $\zeta_1$ & $\zeta_2$ \\
\hline
$3$ & $2$ & $1$ & $0$ & $0$ & $0$ & $0$\\
$6$ & $4$ & $0$ & $1$ & $1$ & $0$ & $0$ \\
$3$ & $2$ & $0$ & $0$ & $1$ & $1$ & $-1$
\end{tabular}
\end{equation}
and the proper transform of \eqref{eq:k3lambda} is:
\begin{equation}
\label{eq:k3e8e8bu}
 y^2 = x^3 + a z^4 z_1^4 z_2^4 x w^4 + z_1^5 z_2^5( \zeta_1 z_1^2 + b z_1 z_2 + \zeta_2 z_2^2) w^6\,.
\end{equation}
Note that after the blow-up, the locus $\zeta_1=0$ is replaced by $\zeta_1 \zeta_2 =0$, i.e. the K3 surface $S$ now degenerates into two surfaces. Each of these is a dP$_9$, i.e. an elliptic fibration over $\mathbb{P}^1$ with one $II^*$ and two $I_1$ fibres. Let us make this explicit for $\zeta_2=0$: setting $\zeta_2=0$ allows us to set $z_1 =1$, as the two coordinates cannot vanish simultaneously. We are therefore left with a homogeneous equation of degrees $(6,6)$, namely
\begin{equation}
\label{eq:dp9}
 y^2 = x^3 + a z_2^4 x w^4  +  z_2^5 (\zeta_1 +  b z_2) \, 
 \end{equation}
in an ambient toric space with weights
\begin{equation}
\label{eq:dp9weights}
\begin{tabular}{ccccc}
$y$ & $x$ & $z$ & $z_2$ & $\zeta_1$ \\
\hline
$3$ & $2$ & $1$ & $0$ & $0$   \\
$3$ & $2$ & $0$ & $1$  & $1$
\end{tabular}
\end{equation}
which is precisely a rational elliptic surface dP$_9$. Note that the $E_8$ singularity is preserved in this process. 

The other component of the K3 surface in this degeneration, which is given by $\zeta_1 = 0$, can be described in exactly the same way upon swapping $z_1 \leftrightarrow z_2$. These two surfaces meet in the elliptic curve given by $\zeta_1 = \zeta_2 =0$. This also shows again that there is no monodromy along the path $\beta_3$, as this path can be contracted into the point $\zeta_1 = \zeta_2=0$.

We have therefore seen that the stretching limit can be realised in an algebro-geometric description as a process where the $\mathbb{P}^1$ base of the elliptic K3 pinches and degenerates into two $\mathbb{P}^1$s.

\bibliographystyle{JHEP}
\bibliography{refs}

@article{Cicoli:2016xae,
  author = {Cicoli, Michele and Muia, Francesco and Shukla, Pramod},
  title = {Global Embedding of Fibre Inflation Models},
  journal = {JHEP},
  volume = {11},
  year = {2016},
  pages = {182},
  doi = {10.1007/JHEP11(2016)182},
  eprint = {1611.04612},
  archivePrefix = {arXiv},
  primaryClass = {hep-th}
}

@article{Kallosh:2015nia,
    author = "Kallosh, Renata and Quevedo, Fernando and Uranga, Angel M.",
    title = "{String Theory Realizations of the Nilpotent Goldstino}",
    eprint = "1507.07556",
    archivePrefix = "arXiv",
    primaryClass = "hep-th",
    reportNumber = "DAMTP-2015-36, IFT-UAM-CSIC-15-079",
    doi = "10.1007/JHEP12(2015)039",
    journal = "JHEP",
    volume = "12",
    pages = "039",
    year = "2015"
}

@article{Cicoli:2024bwq,
    author = "Cicoli, Michele and Hughes, Christopher and Kamal, Ahmed Rakin and Marino, Francesco and Quevedo, Fernando and Ramos-Hamud, Mario and Villa, Gonzalo",
    title = "{Back to the origins of brane{\textendash}antibrane inflation}",
    eprint = "2410.00097",
    archivePrefix = "arXiv",
    primaryClass = "hep-th",
    reportNumber = "CERN-TH-2024-128",
    doi = "10.1140/epjc/s10052-025-13982-9",
    journal = "Eur. Phys. J. C",
    volume = "85",
    number = "3",
    pages = "315",
    year = "2025"
}

@article{Conlon:2008cj,
    author = "Conlon, Joseph P. and Kallosh, Renata and Linde, Andrei D. and Quevedo, Fernando",
    title = "{Volume Modulus Inflation and the Gravitino Mass Problem}",
    eprint = "0806.0809",
    archivePrefix = "arXiv",
    primaryClass = "hep-th",
    reportNumber = "DAMTP-2007-36, CAV-HEP-08-07, YITP-08-42",
    doi = "10.1088/1475-7516/2008/09/011",
    journal = "JCAP",
    volume = "09",
    pages = "011",
    year = "2008"
}

@article{Cicoli:2015wja,
    author = "Cicoli, Michele and Muia, Francesco and Pedro, Francisco Gil",
    title = "{Microscopic Origin of Volume Modulus Inflation}",
    eprint = "1509.07748",
    archivePrefix = "arXiv",
    primaryClass = "hep-th",
    doi = "10.1088/1475-7516/2015/12/040",
    journal = "JCAP",
    volume = "12",
    pages = "040",
    year = "2015"
}

@article{Anchordoqui:2023etp,
    author = "Anchordoqui, Luis A. and Antoniadis, Ignatios",
    title = "{Large extra dimensions from higher-dimensional inflation}",
    eprint = "2310.20282",
    archivePrefix = "arXiv",
    primaryClass = "hep-ph",
    doi = "10.1103/PhysRevD.109.103508",
    journal = "Phys. Rev. D",
    volume = "109",
    number = "10",
    pages = "103508",
    year = "2024"
}

@article{Anchordoqui:2022svl,
    author = "Anchordoqui, Luis A. and Antoniadis, Ignatios and Lust, Dieter",
    title = "{Aspects of the dark dimension in cosmology}",
    eprint = "2212.08527",
    archivePrefix = "arXiv",
    primaryClass = "hep-ph",
    reportNumber = "MPP-2022-285, LMU-ASC 55/22",
    doi = "10.1103/PhysRevD.107.083530",
    journal = "Phys. Rev. D",
    volume = "107",
    number = "8",
    pages = "083530",
    year = "2023"
}

@article{Bedroya:2025fwh,
    author = "Bedroya, Alek and Obied, Georges and Vafa, Cumrun and Wu, David H.",
    title = "{Evolving Dark Sector and the Dark Dimension Scenario}",
    eprint = "2507.03090",
    archivePrefix = "arXiv",
    primaryClass = "astro-ph.CO",
    month = "7",
    year = "2025"
}

@article{Carta:2022oex,
    author = "Carta, Federico and Mininno, Alessandro and Shukla, Pramod",
    title = "{Systematics of perturbatively flat flux vacua for CICYs}",
    eprint = "2201.10581",
    archivePrefix = "arXiv",
    primaryClass = "hep-th",
    reportNumber = "ZMP-HH/22-4",
    doi = "10.1007/JHEP08(2022)297",
    journal = "JHEP",
    volume = "08",
    pages = "297",
    year = "2022"
}

@article{Akrami:2025zlb,
    author = "Akrami, Yashar and Alestas, George and Nesseris, Savvas",
    title = "{Has DESI detected exponential quintessence?}",
    eprint = "2504.04226",
    archivePrefix = "arXiv",
    primaryClass = "astro-ph.CO",
    reportNumber = "IFT-UAM/CSIC-25-36",
    month = "4",
    year = "2025"
}

@article{Blumenhagen:2026rgu,
    author = "Blumenhagen, Ralph and Paraskevopoulou, Antonia",
    title = "{Towards the Realization of the Dark Dimension Scenario in Ho{\v{r}}ava-Witten Theory}",
    eprint = "2605.11068",
    archivePrefix = "arXiv",
    primaryClass = "hep-th",
    reportNumber = "MPP-2026-80",
    month = "5",
    year = "2026"
}

@article{Cicoli:2021dhg,
    author = "Cicoli, Michele and Etxebarria, I{\~n}aki Garc{\'\i}a and Quevedo, Fernando and Schachner, Andreas and Shukla, Pramod and Valandro, Roberto",
    title = "{The Standard Model quiver in de Sitter string compactifications}",
    eprint = "2106.11964",
    archivePrefix = "arXiv",
    primaryClass = "hep-th",
    doi = "10.1007/JHEP08(2021)109",
    journal = "JHEP",
    volume = "08",
    pages = "109",
    year = "2021"
}

@article{Cicoli:2017shd,
    author = "Cicoli, Michele and Garc{\`\i}a-Etxebarria, I{\~n}aki and Mayrhofer, Christoph and Quevedo, Fernando and Shukla, Pramod and Valandro, Roberto",
    title = "{Global Orientifolded Quivers with Inflation}",
    eprint = "1706.06128",
    archivePrefix = "arXiv",
    primaryClass = "hep-th",
    doi = "10.1007/JHEP11(2017)134",
    journal = "JHEP",
    volume = "11",
    pages = "134",
    year = "2017"
}

@article{Cicoli:2013cha,
    author = "Cicoli, Michele and Klevers, Denis and Krippendorf, Sven and Mayrhofer, Christoph and Quevedo, Fernando and Valandro, Roberto",
    title = "{Explicit de Sitter Flux Vacua for Global String Models with Chiral Matter}",
    eprint = "1312.0014",
    archivePrefix = "arXiv",
    primaryClass = "hep-th",
    doi = "10.1007/JHEP05(2014)001",
    journal = "JHEP",
    volume = "05",
    pages = "001",
    year = "2014"
}

@article{Cicoli:2013mpa,
    author = "Cicoli, Michele and Krippendorf, Sven and Mayrhofer, Christoph and Quevedo, Fernando and Valandro, Roberto",
    title = "{D3/D7 Branes at Singularities: Constraints from Global Embedding and Moduli Stabilisation}",
    eprint = "1304.0022",
    archivePrefix = "arXiv",
    primaryClass = "hep-th",
    doi = "10.1007/JHEP07(2013)150",
    journal = "JHEP",
    volume = "07",
    pages = "150",
    year = "2013"
}

@article{Cicoli:2012vw,
    author = "Cicoli, Michele and Krippendorf, Sven and Mayrhofer, Christoph and Quevedo, Fernando and Valandro, Roberto",
    title = "{D-Branes at del Pezzo Singularities: Global Embedding and Moduli Stabilisation}",
    eprint = "1206.5237",
    archivePrefix = "arXiv",
    primaryClass = "hep-th",
    reportNumber = "DAMTP-2012-47, ZMP-HH-12-10",
    doi = "10.1007/JHEP09(2012)019",
    journal = "JHEP",
    volume = "09",
    pages = "019",
    year = "2012"
}

@article{AbdusSalam:2022krp,
    author = "AbdusSalam, Shehu and Crin{\`o}, Chiara and Shukla, Pramod",
    title = "{On K3-fibred LARGE Volume Scenario with de Sitter vacua from anti-D3-branes}",
    eprint = "2206.12889",
    archivePrefix = "arXiv",
    primaryClass = "hep-th",
    doi = "10.1007/JHEP03(2023)132",
    journal = "JHEP",
    volume = "03",
    pages = "132",
    year = "2023"
}

@article{Garcia-Etxebarria:2015lif,
    author = "Garc{\'\i}a-Etxebarria, I{\~n}aki and Quevedo, Fernando and Valandro, Roberto",
    title = "{Global String Embeddings for the Nilpotent Goldstino}",
    eprint = "1512.06926",
    archivePrefix = "arXiv",
    primaryClass = "hep-th",
    reportNumber = "MPP-2015-311, DAMTP-2015-91",
    doi = "10.1007/JHEP02(2016)148",
    journal = "JHEP",
    volume = "02",
    pages = "148",
    year = "2016"
}

@article{Cicoli:2024bxw,
    author = "Cicoli, Michele and Grassi, Antonella and Lacombe, Osmin and Pedro, Francisco G.",
    title = "{Chiral global embedding of Fibre Inflation with $ \overline{\textrm{D}3} $ uplift}",
    eprint = "2412.08723",
    archivePrefix = "arXiv",
    primaryClass = "hep-th",
    doi = "10.1007/JHEP06(2025)090",
    journal = "JHEP",
    volume = "06",
    pages = "090",
    year = "2025"
}

@article{Leontaris:2022rzj,
    author = "Leontaris, George K. and Shukla, Pramod",
    title = {{Stabilising all K{\"a}hler moduli in perturbative LVS}},
    eprint = "2203.03362",
    archivePrefix = "arXiv",
    primaryClass = "hep-th",
    doi = "10.1007/JHEP07(2022)047",
    journal = "JHEP",
    volume = "07",
    pages = "047",
    year = "2022"
}

@article{Antoniadis:2019rkh,
    author = "Antoniadis, Ignatios and Chen, Yifan and Leontaris, George K.",
    title = "{Logarithmic loop corrections, moduli stabilisation and de Sitter vacua in string theory}",
    eprint = "1909.10525",
    archivePrefix = "arXiv",
    primaryClass = "hep-th",
    doi = "10.1007/JHEP01(2020)149",
    journal = "JHEP",
    volume = "01",
    pages = "149",
    year = "2020"
}

@article{Antoniadis:2018hqy,
    author = "Antoniadis, Ignatios and Chen, Yifan and Leontaris, George K.",
    title = "{Perturbative moduli stabilisation in type IIB/F-theory framework}",
    eprint = "1803.08941",
    archivePrefix = "arXiv",
    primaryClass = "hep-th",
    doi = "10.1140/epjc/s10052-018-6248-4",
    journal = "Eur. Phys. J. C",
    volume = "78",
    number = "9",
    pages = "766",
    year = "2018"
}

@article{Conlon:2009kt,
    author = "Conlon, Joseph P. and Palti, Eran",
    title = "{Gauge Threshold Corrections for Local Orientifolds}",
    eprint = "0906.1920",
    archivePrefix = "arXiv",
    primaryClass = "hep-th",
    reportNumber = "OUTP-09-13P",
    doi = "10.1088/1126-6708/2009/09/019",
    journal = "JHEP",
    volume = "09",
    pages = "019",
    year = "2009"
}

@article{Conlon:2009xf,
    author = "Conlon, Joseph P.",
    title = "{Gauge Threshold Corrections for Local String Models}",
    eprint = "0901.4350",
    archivePrefix = "arXiv",
    primaryClass = "hep-th",
    reportNumber = "OUTP-09-03P",
    doi = "10.1088/1126-6708/2009/04/059",
    journal = "JHEP",
    volume = "04",
    pages = "059",
    year = "2009"
}

@article{Klaewer:2020lfg,
    author = "Klaewer, Daniel and Lee, Seung-Joo and Weigand, Timo and Wiesner, Max",
    title = "{Quantum corrections in 4d $N$ = 1 infinite distance limits and the weak gravity conjecture}",
    eprint = "2011.00024",
    archivePrefix = "arXiv",
    primaryClass = "hep-th",
    reportNumber = "CTPU-PTC-20-24, IFT-UAM/CSIC-20-148, MITP/20-064, ZMP-HH/20-21",
    doi = "10.1007/JHEP03(2021)252",
    journal = "JHEP",
    volume = "03",
    pages = "252",
    year = "2021"
}

@article{Weissenbacher:2020cyf,
    author = "Weissenbacher, Matthias",
    title = "{On $\alpha'$-effects from $D$-branes in $4d \; \mathcal{N} = 1$}",
    eprint = "2006.15552",
    archivePrefix = "arXiv",
    primaryClass = "hep-th",
    doi = "10.1007/JHEP11(2020)076",
    journal = "JHEP",
    volume = "11",
    pages = "076",
    year = "2020"
}

@article{Weissenbacher:2019mef,
    author = "Weissenbacher, Matthias",
    title = "{F-theory vacua and $\alpha'$-corrections}",
    eprint = "1901.04758",
    archivePrefix = "arXiv",
    primaryClass = "hep-th",
    doi = "10.1007/JHEP04(2020)032",
    journal = "JHEP",
    volume = "04",
    pages = "032",
    year = "2020"
}

@article{Grimm:2013gma,
    author = "Grimm, Thomas W. and Savelli, Raffaele and Weissenbacher, Matthias",
    title = "{On {\textbackslash}alpha' corrections in N=1 F-theory compactifications}",
    eprint = "1303.3317",
    archivePrefix = "arXiv",
    primaryClass = "hep-th",
    doi = "10.1016/j.physletb.2013.07.024",
    journal = "Phys. Lett. B",
    volume = "725",
    pages = "431--436",
    year = "2013"
}

@article{Acharya:2018deu,
    author = "Acharya, Bobby Samir and Maharana, Anshuman and Muia, Francesco",
    title = "{Hidden Sectors in String Theory: Kinetic Mixings, Fifth Forces and Quintessence}",
    eprint = "1811.10633",
    archivePrefix = "arXiv",
    primaryClass = "hep-th",
    doi = "10.1007/JHEP03(2019)048",
    journal = "JHEP",
    volume = "03",
    pages = "048",
    year = "2019"
}

@article{Cicoli:2016olq,
    author = "Cicoli, Michele and Dutta, Koushik and Maharana, Anshuman and Quevedo, Fernando",
    title = "{Moduli Vacuum Misalignment and Precise Predictions in String Inflation}",
    eprint = "1604.08512",
    archivePrefix = "arXiv",
    primaryClass = "hep-th",
    doi = "10.1088/1475-7516/2016/08/006",
    journal = "JCAP",
    volume = "08",
    pages = "006",
    year = "2016"
}

@article{Cicoli:2021rub,
    author = "Cicoli, Michele and Quevedo, Fernando and Savelli, Raffaele and Schachner, Andreas and Valandro, Roberto",
    title = "{Systematics of the {\ensuremath{\alpha}}' expansion in F-theory}",
    eprint = "2106.04592",
    archivePrefix = "arXiv",
    primaryClass = "hep-th",
    doi = "10.1007/JHEP08(2021)099",
    journal = "JHEP",
    volume = "08",
    pages = "099",
    year = "2021"
}

@article{deCarlos:1993wie,
    author = "de Carlos, B. and Casas, J. A. and Quevedo, F. and Roulet, E.",
    title = "{Model independent properties and cosmological implications of the dilaton and moduli sectors of 4-d strings}",
    eprint = "hep-ph/9308325",
    archivePrefix = "arXiv",
    reportNumber = "CERN-TH-6958-93, NEIP-93-006, IEM-FT-75-93",
    doi = "10.1016/0370-2693(93)91538-X",
    journal = "Phys. Lett. B",
    volume = "318",
    pages = "447--456",
    year = "1993"
}

@article{Banks:1993en,
    author = "Banks, Tom and Kaplan, David B. and Nelson, Ann E.",
    title = "{Cosmological implications of dynamical supersymmetry breaking}",
    eprint = "hep-ph/9308292",
    archivePrefix = "arXiv",
    reportNumber = "UCSD-PTH-93-26, RU-37",
    doi = "10.1103/PhysRevD.49.779",
    journal = "Phys. Rev. D",
    volume = "49",
    pages = "779--787",
    year = "1994"
}

@article{Coughlan:1983ci,
    author = "Coughlan, G. D. and Fischler, W. and Kolb, Edward W. and Raby, S. and Ross, Graham G.",
    title = "{Cosmological Problems for the Polonyi Potential}",
    reportNumber = "LA-UR-83-1423",
    doi = "10.1016/0370-2693(83)91091-2",
    journal = "Phys. Lett. B",
    volume = "131",
    pages = "59--64",
    year = "1983"
}

@article{Cicoli:2024yqh,
    author = "Cicoli, Michele and Cunillera, Francesc and Padilla, Antonio and Pedro, Francisco G.",
    title = "{From inflation to quintessence: a history of the universe in string theory}",
    eprint = "2407.03405",
    archivePrefix = "arXiv",
    primaryClass = "hep-th",
    doi = "10.1007/JHEP10(2024)141",
    journal = "JHEP",
    volume = "10",
    pages = "141",
    year = "2024"
}

@article{Cicoli:2012tz,
    author = "Cicoli, Michele and Pedro, Francisco G. and Tasinato, Gianmassimo",
    title = "{Natural Quintessence in String Theory}",
    eprint = "1203.6655",
    archivePrefix = "arXiv",
    primaryClass = "hep-th",
    doi = "10.1088/1475-7516/2012/07/044",
    journal = "JCAP",
    volume = "07",
    pages = "044",
    year = "2012"
}

@article{Cicoli:2011qg,
    author = "Cicoli, Michele and Mayrhofer, Christoph and Valandro, Roberto",
    title = "{Moduli Stabilisation for Chiral Global Models}",
    eprint = "1110.3333",
    archivePrefix = "arXiv",
    primaryClass = "hep-th",
    reportNumber = "DESY-11-179, ZMP-HH-11-15",
    doi = "10.1007/JHEP02(2012)062",
    journal = "JHEP",
    volume = "02",
    pages = "062",
    year = "2012"
}

@article{Cicoli:2011it,
    author = "Cicoli, Michele and Kreuzer, Maximilian and Mayrhofer, Christoph",
    title = "{Toric K3-Fibred Calabi-Yau Manifolds with del Pezzo Divisors for String Compactifications}",
    eprint = "1107.0383",
    archivePrefix = "arXiv",
    primaryClass = "hep-th",
    reportNumber = "DESY-11-103",
    doi = "10.1007/JHEP02(2012)002",
    journal = "JHEP",
    volume = "02",
    pages = "002",
    year = "2012"
}

@article{Kachru:2003aw,
    author = "Kachru, Shamit and Kallosh, Renata and Linde, Andrei D. and Trivedi, Sandip P.",
    title = "{De Sitter vacua in string theory}",
    eprint = "hep-th/0301240",
    archivePrefix = "arXiv",
    reportNumber = "SLAC-PUB-9630, SU-ITP-03-01, TIFR-TH-03-03",
    doi = "10.1103/PhysRevD.68.046005",
    journal = "Phys. Rev. D",
    volume = "68",
    pages = "046005",
    year = "2003"
}

@article{Gallego:2017dvd,
    author = "Gallego, Diego and Marsh, M. C. David and Vercnocke, Bert and Wrase, Timm",
    title = "{A New Class of de Sitter Vacua in Type IIB Large Volume Compactifications}",
    eprint = "1707.01095",
    archivePrefix = "arXiv",
    primaryClass = "hep-th",
    doi = "10.1007/JHEP10(2017)193",
    journal = "JHEP",
    volume = "10",
    pages = "193",
    year = "2017"
}

@article{Cicoli:2015ylx,
    author = "Cicoli, Michele and Quevedo, Fernando and Valandro, Roberto",
    title = "{De Sitter from T-branes}",
    eprint = "1512.04558",
    archivePrefix = "arXiv",
    primaryClass = "hep-th",
    doi = "10.1007/JHEP03(2016)141",
    journal = "JHEP",
    volume = "03",
    pages = "141",
    year = "2016"
}

@article{Burgess:2020qsc,
    author = "Burgess, C. P. and Cicoli, Michele and Ciupke, David and Krippendorf, Sven and Quevedo, Fernando",
    title = "{UV Shadows in EFTs: Accidental Symmetries, Robustness and No-Scale Supergravity}",
    eprint = "2006.06694",
    archivePrefix = "arXiv",
    primaryClass = "hep-th",
    doi = "10.1002/prop.202000076",
    journal = "Fortsch. Phys.",
    volume = "68",
    number = "10",
    pages = "2000076",
    year = "2020"
}

@article{Aghababaie:2003wz,
    author = "Aghababaie, Y. and Burgess, C. P. and Parameswaran, S. L. and Quevedo, F.",
    title = "{Towards a naturally small cosmological constant from branes in 6-D supergravity}",
    eprint = "hep-th/0304256",
    archivePrefix = "arXiv",
    reportNumber = "MCGILL-03-08, DAMTP-2003-38",
    doi = "10.1016/j.nuclphysb.2003.12.015",
    journal = "Nucl. Phys. B",
    volume = "680",
    pages = "389--414",
    year = "2004"
}

@article{Chauhan:2026gid,
    author = "Chauhan, Aman and Cicoli, Michele and Krippendorf, Sven and Maharana, Anshuman and Piantadosi, Pellegrino and Schachner, Andreas",
    title = "{Parameter compression in the flux landscape}",
    eprint = "2603.04941",
    archivePrefix = "arXiv",
    primaryClass = "hep-th",
    reportNumber = "LITP-26-05",
    month = "3",
    year = "2026"
}

@article{Chauhan:2025rdj,
    author = "Chauhan, Aman and Cicoli, Michele and Krippendorf, Sven and Maharana, Anshuman and Piantadosi, Pellegrino and Schachner, Andreas",
    title = "{Deep observations of the Type IIB flux landscape}",
    eprint = "2501.03984",
    archivePrefix = "arXiv",
    primaryClass = "hep-th",
    doi = "10.1007/JHEP07(2025)271",
    journal = "JHEP",
    volume = "07",
    pages = "271",
    year = "2025"
}

@article{Jockers:2005zy,
    author = "Jockers, Hans and Louis, Jan",
    title = "{D-terms and F-terms from D7-brane fluxes}",
    eprint = "hep-th/0502059",
    archivePrefix = "arXiv",
    doi = "10.1016/j.nuclphysb.2005.04.011",
    journal = "Nucl. Phys. B",
    volume = "718",
    pages = "203--246",
    year = "2005"
}

@article{Freed:1999vc,
    author = "Freed, Daniel S. and Witten, Edward",
    title = "{Anomalies in string theory with D-branes}",
    eprint = "hep-th/9907189",
    archivePrefix = "arXiv",
    journal = "Asian J. Math.",
    volume = "3",
    pages = "819",
    year = "1999"
}

@article{oguiso:theorempaper,
  author = {Oguiso, Keiji},
  title = {On Algebraic Fiber Space Structures on a Calabi-Yau 3-fold},
  journal = {International Journal of Mathematics},
  volume = {4},
  year = {1993},
  pages = {439--465}
}

@article{Cicoli:2017axo,
  author = {Cicoli, Michele and Ciupke, David and Diaz, Victor A. and Guidetti, Veronica and Muia, Francesco and Shukla, Pramod},
  title = {Chiral Global Embedding of Fibre Inflation Models},
  journal = {JHEP},
  volume = {11},
  year = {2017},
  pages = {207},
  doi = {10.1007/JHEP11(2017)207},
  eprint = {1709.01518},
  archivePrefix = {arXiv},
  primaryClass = {hep-th}
}

@article{Altman:2014bfa,
  author = {Altman, Ross and Gray, James and He, Yang-Hui and Jejjala, Vishnu and Nelson, Brent D.},
  title = {A Calabi-Yau Database: Threefolds Constructed from the Kreuzer-Skarke List},
  journal = {JHEP},
  volume = {02},
  year = {2015},
  pages = {158},
  doi = {10.1007/JHEP02(2015)158},
  eprint = {1411.1418},
  archivePrefix = {arXiv},
  primaryClass = {hep-th}
}

@article{Becker:2002nn,
  author = {Becker, Katrin and Becker, Melanie and Haack, Michael and Louis, Jan},
  title = {Supersymmetry breaking and alpha-prime corrections to flux induced potentials},
  journal = {JHEP},
  volume = {06},
  year = {2002},
  pages = {060},
  doi = {10.1088/1126-6708/2002/06/060},
  eprint = {hep-th/0204254},
  archivePrefix = {arXiv},
  primaryClass = {hep-th}
}

@article{Balasubramanian:2005zx,
  author = {Balasubramanian, Vijay and Berglund, Per and Conlon, Joseph P. and Quevedo, Fernando},
  title = {Systematics of moduli stabilisation in Calabi-Yau flux compactifications},
  journal = {JHEP},
  volume = {03},
  year = {2005},
  pages = {007},
  doi = {10.1088/1126-6708/2005/03/007},
  eprint = {hep-th/0502058},
  archivePrefix = {arXiv},
  primaryClass = {hep-th}
}

@article{Cicoli:2021skd,
  author = {Cicoli, Michele and Cunillera, Ferran and Padilla, Antonio and Pedro, Francisco G.},
  title = {Quintessence and the Swampland: The Numerically Controlled Regime of Moduli Space},
  journal = {Fortschritte der Physik},
  volume = {70},
  number = {4},
  year = {2022},
  pages = {2200008},
  doi = {10.1002/prop.202200008},
  eprint = {2112.10783},
  archivePrefix = {arXiv},
  primaryClass = {hep-th}
}

@article{Berg:2005ja,
  author = {Berg, Marcus and Haack, Michael and Kors, Boris},
  title = {String loop corrections to Kahler potentials in orientifolds},
  journal = {JHEP},
  volume = {11},
  year = {2005},
  pages = {030},
  doi = {10.1088/1126-6708/2005/11/030},
  eprint = {hep-th/0508043},
  archivePrefix = {arXiv},
  primaryClass = {hep-th}
}

@article{Berg:2007wt,
  author = {Berg, Marcus and Haack, Michael and Pajer, Enrico},
  title = {Jumping Through Loops: On Soft Terms from Large Volume Compactifications},
  journal = {JHEP},
  volume = {09},
  year = {2007},
  pages = {031},
  doi = {10.1088/1126-6708/2007/09/031},
  eprint = {0704.0737},
  archivePrefix = {arXiv},
  primaryClass = {hep-th}
}

@article{Ciupke:2015msa,
  author = {Ciupke, David and Louis, Jan and Westphal, Alexander},
  title = {Higher-Derivative Supergravity and Moduli Stabilization},
  journal = {JHEP},
  volume = {10},
  year = {2015},
  pages = {094},
  doi = {10.1007/JHEP10(2015)094},
  eprint = {1505.03092},
  archivePrefix = {arXiv},
  primaryClass = {hep-th}
}

@article{Cicoli:2023njy,
  author = {Cicoli, Michele and Licheri, Michele and Piantadosi, Pierfrancesco and Quevedo, Fernando and Shukla, Pramod},
  title = {Higher derivative corrections to string inflation},
  journal = {JHEP},
  volume = {02},
  year = {2024},
  pages = {115},
  doi = {10.1007/JHEP02(2024)115},
  eprint = {2309.11697},
  archivePrefix = {arXiv},
  primaryClass = {hep-th}
}

@article{Anderson:2017aux,
    author = "Anderson, Lara B. and Gao, Xin and Gray, James and Lee, Seung-Joo",
    title = "{Fibrations in CICY Threefolds}",
    eprint = "1708.07907",
    archivePrefix = "arXiv",
    primaryClass = "hep-th",
    doi = "10.1007/JHEP10(2017)077",
    journal = "JHEP",
    volume = "10",
    pages = "077",
    year = "2017"
}

@article{ElkiesKumar2014,
  title={K3 surfaces and equations for {Hilbert} modular surfaces},
  author={Elkies, Noam and Kumar, Abhinav},
  journal={Algebra \& Number Theory},
  volume={8},
  number={10},
  pages={2297--2342},
  year={2014},
  publisher={Mathematical Sciences Publishers},
  doi={10.2140/ant.2014.8.2297},
  url={https://projecteuclid.org/journals/algebra-and-number-theory/volume-8/issue-10/K3-surfaces-and-equations-for-Hilbert-modular-surfaces/10.2140/ant.2014.8.2297.full}
}

@article{LopesCardoso:1996hq,
    author = "Lopes Cardoso, Gabriel and Curio, Gottfried and Lust, Dieter and Mohaupt, Thomas",
    title = "{On the duality between the heterotic string and F theory in eight-dimensions}",
    eprint = "hep-th/9609111",
    archivePrefix = "arXiv",
    reportNumber = "CERN-TH-96-232, HUB-EP-96-49",
    doi = "10.1016/S0370-2693(96)01303-2",
    journal = "Phys. Lett. B",
    volume = "389",
    pages = "479--484",
    year = "1996"
}

@article{Morrison:1996pp,
    author = "Morrison, David R. and Vafa, Cumrun",
    title = "{Compactifications of F theory on Calabi-Yau threefolds. 2.}",
    eprint = "hep-th/9603161",
    archivePrefix = "arXiv",
    reportNumber = "DUKE-TH-96-107, HUTP-96-A012",
    doi = "10.1016/0550-3213(96)00369-0",
    journal = "Nucl. Phys. B",
    volume = "476",
    pages = "437--469",
    year = "1996"
}

@article{Greene:1989ya,
    author = "Greene, Brian R. and Shapere, Alfred D. and Vafa, Cumrun and Yau, Shing-Tung",
    title = "{Stringy Cosmic Strings and Noncompact Calabi-Yau Manifolds}",
    reportNumber = "HUTP-89-A047, IASSNS-HEP-89-47",
    doi = "10.1016/0550-3213(90)90248-C",
    journal = "Nucl. Phys. B",
    volume = "337",
    pages = "1--36",
    year = "1990"
}

@article{Candelas:2012uu,
    author = "Candelas, Philip and Constantin, Andrei and Skarke, Harald",
    title = "{An Abundance of K3 Fibrations from Polyhedra with Interchangeable Parts}",
    eprint = "1207.4792",
    archivePrefix = "arXiv",
    primaryClass = "hep-th",
    doi = "10.1007/s00220-013-1802-2",
    journal = "Commun. Math. Phys.",
    volume = "324",
    pages = "937--959",
    year = "2013"
}

@article{doi_yotsutani_2014,
  title={Doubling construction of Calabi-Yau threefolds},
  author={Doi, Mamoru and Yotsutani, Naoto},
  journal={New York Journal of Mathematics},
  volume={20},
  pages={1--33},
  year={2014},
  url={http://nyjm.albany.edu/j/2014/20-56p.pdf}
}

@article{KawamataNamikawa1994,
  author  = {Kawamata, Yujiro and Namikawa, Yoshinori},
  title   = {Logarithmic deformations of normal crossing varieties and smoothing of degenerate {Calabi-Yau} varieties},
  journal = {Inventiones mathematicae},
  year    = {1994},
  volume  = {118},
  number  = {3},
  pages   = {395--410},
  doi     = {10.1007/BF01231533},
  url     = {https://doi.org}
}

@article{talbot2017gluing,
  title={Gluing and deformation of asymptotically cylindrical Calabi-Yau manifolds in complex dimension three},
  author={Talbot, Tim},
  journal={arXiv preprint arXiv:1703.09201},
  year={2017}
}

@article{Shiu:2024sbe,
    author = "Shiu, Gary and Tonioni, Flavio and Tran, Hung V.",
    title = "{Analytic bounds on late-time axion-scalar cosmologies}",
    eprint = "2406.17030",
    archivePrefix = "arXiv",
    primaryClass = "hep-th",
    doi = "10.1007/JHEP09(2024)158",
    journal = "JHEP",
    volume = "09",
    pages = "158",
    year = "2024"
}

@article{Shiu:2023fhb,
    author = "Shiu, Gary and Tonioni, Flavio and Tran, Hung V.",
    title = "{Late-time attractors and cosmic acceleration}",
    eprint = "2306.07327",
    archivePrefix = "arXiv",
    primaryClass = "hep-th",
    doi = "10.1103/PhysRevD.108.063528",
    journal = "Phys. Rev. D",
    volume = "108",
    number = "6",
    pages = "063528",
    year = "2023"
}

@article{Shiu:2023nph,
    author = "Shiu, Gary and Tonioni, Flavio and Tran, Hung V.",
    title = "{Accelerating universe at the end of time}",
    eprint = "2303.03418",
    archivePrefix = "arXiv",
    primaryClass = "hep-th",
    doi = "10.1103/PhysRevD.108.063527",
    journal = "Phys. Rev. D",
    volume = "108",
    number = "6",
    pages = "063527",
    year = "2023"
}

@article{Arkani-Hamed:1998sfv,
    author = "Arkani-Hamed, Nima and Dimopoulos, Savas and Dvali, G. R.",
    title = "{Phenomenology, astrophysics and cosmology of theories with submillimeter dimensions and TeV scale quantum gravity}",
    eprint = "hep-ph/9807344",
    archivePrefix = "arXiv",
    reportNumber = "SLAC-PUB-7864, SU-ITP-98-142, IC-98-44",
    doi = "10.1103/PhysRevD.59.086004",
    journal = "Phys. Rev. D",
    volume = "59",
    pages = "086004",
    year = "1999"
}

@article{Corti:2013ers,
    author = "Corti, Alessio and Haskins, Mark and Nordstr{\~A}{\textparagraph}m, Johannes and Pacini, Tommaso",
    title = "{Asymptotically cylindrical Calabi{\textendash}Yau 3{\textendash}folds from weak Fano 3{\textendash}folds}",
    eprint = "1206.2277",
    archivePrefix = "arXiv",
    primaryClass = "math.AG",
    doi = "10.2140/gt.2013.17.1955",
    journal = "Geom. Topol.",
    volume = "17",
    number = "4",
    pages = "1955--2059",
    year = "2013"
}

@article{Cicoli:2011yy,
    author = "Cicoli, M. and Burgess, C. P. and Quevedo, F.",
    title = "{Anisotropic Modulus Stabilisation: Strings at LHC Scales with Micron-sized Extra Dimensions}",
    eprint = "1105.2107",
    archivePrefix = "arXiv",
    primaryClass = "hep-th",
    reportNumber = "DESY-11-073",
    doi = "10.1007/JHEP10(2011)119",
    journal = "JHEP",
    volume = "10",
    pages = "119",
    year = "2011"
}

@article{Davis:2014:STS,
  author = {Davis, Ryan and Doran, Charles and Gewiss, Adam and Novoseltsev, Andrey and Skjorshammer, Dmitri and Syryczuk, Alexa and Whitcher, Ursula},
  title = {Short Tops and Semistable Degenerations},
  journal = {Experimental Mathematics},
  volume = {23},
  number = {4},
  pages = {351--362},
  year = {2014},
  doi = {10.1080/10586458.2014.910850},
  url = {https://doi.org/10.1080/10586458.2014.910850}
}

@article{Braun:2017ryx,
    author = "Braun, Andreas P. and Del Zotto, Michele",
    title = "{Mirror Symmetry for $G_2$-Manifolds: Twisted Connected Sums and Dual Tops}",
    eprint = "1701.05202",
    archivePrefix = "arXiv",
    primaryClass = "hep-th",
    doi = "10.1007/JHEP05(2017)080",
    journal = "JHEP",
    volume = "05",
    pages = "080",
    year = "2017"
}

@article{Cvetic:2015uwu,
    author = "Cvetic, Mirjam and Grassi, Antonella and Klevers, Denis and Poretschkin, Maximilian and Song, Peng",
    title = "{Origin of Abelian Gauge Symmetries in Heterotic/F-theory Duality}",
    eprint = "1511.08208",
    archivePrefix = "arXiv",
    primaryClass = "hep-th",
    reportNumber = "UPR-1275-T, CERN-PH-TH-2015-273",
    doi = "10.1007/JHEP04(2016)041",
    journal = "JHEP",
    volume = "04",
    pages = "041",
    year = "2016"
}

@article{Braun:2016igl,
    author = "Braun, Andreas P.",
    title = "{Tops as building blocks for G$_{2}$ manifolds}",
    eprint = "1602.03521",
    archivePrefix = "arXiv",
    primaryClass = "hep-th",
    doi = "10.1007/JHEP10(2017)083",
    journal = "JHEP",
    volume = "10",
    pages = "083",
    year = "2017"
}

@article{Grimm:2017okk,
  author = {Grimm, Thomas W. and Mayer, Kilian and Weissenbacher, Matthias},
  title = {Higher derivatives in Type II and M-theory on Calabi-Yau threefolds},
  journal = {JHEP},
  volume = {02},
  year = {2018},
  pages = {127},
  doi = {10.1007/JHEP02(2018)127},
  eprint = {1702.08404},
  archivePrefix = {arXiv},
  primaryClass = {hep-th}
}

@article{Bansal:2024uzr,
    author = "Bansal, Suk{\c{r}}ti and Brunelli, Luca and Cicoli, Michele and Hebecker, Arthur and Kuespert, Ruben",
    title = "{Loop blow-up inflation}",
    eprint = "2403.04831",
    archivePrefix = "arXiv",
    primaryClass = "hep-th",
    doi = "10.1007/JHEP07(2024)289",
    journal = "JHEP",
    volume = "07",
    pages = "289",
    year = "2024"
}

@article{Cicoli:2013swa,
    author = "Cicoli, Michele and Conlon, Joseph P. and Maharana, Anshuman and Quevedo, Fernando",
    title = "{A Note on the Magnitude of the Flux Superpotential}",
    eprint = "1310.6694",
    archivePrefix = "arXiv",
    primaryClass = "hep-th",
    reportNumber = "DAMTP-2013-48, HRI-ST-1305",
    doi = "10.1007/JHEP01(2014)027",
    journal = "JHEP",
    volume = "01",
    pages = "027",
    year = "2014"
}

@article{vonGersdorff:2005bf,
    author = "von Gersdorff, Gero and Hebecker, Arthur",
    title = "{Kahler corrections for the volume modulus of flux compactifications}",
    eprint = "hep-th/0507131",
    archivePrefix = "arXiv",
    doi = "10.1016/j.physletb.2005.08.024",
    journal = "Phys. Lett. B",
    volume = "624",
    pages = "270--274",
    year = "2005"
}

@article{Cicoli:2007xp,
    author = "Cicoli, Michele and Conlon, Joseph P. and Quevedo, Fernando",
    title = "{Systematics of String Loop Corrections in Type IIB Calabi-Yau Flux Compactifications}",
    eprint = "0708.1873",
    archivePrefix = "arXiv",
    primaryClass = "hep-th",
    reportNumber = "DAMTP-2007-75",
    doi = "10.1088/1126-6708/2008/01/052",
    journal = "JHEP",
    volume = "01",
    pages = "052",
    year = "2008"
}

@article{Gao:2022uop,
    author = "Gao, Xin and Hebecker, Arthur and Schreyer, Simon and Venken, Victoria",
    title = "{Loops, local corrections and warping in the LVS and other type IIB models}",
    eprint = "2204.06009",
    archivePrefix = "arXiv",
    primaryClass = "hep-th",
    doi = "10.1007/JHEP09(2022)091",
    journal = "JHEP",
    volume = "09",
    pages = "091",
    year = "2022"
}

@article{Cicoli:2026bqo,
    author = "Cicoli, Michele",
    title = "{Recent progress on inflation and dark energy from string theory}",
    eprint = "2604.16281",
    archivePrefix = "arXiv",
    primaryClass = "hep-th",
    doi = "10.1007/s10714-026-03538-x",
    journal = "Gen. Rel. Grav.",
    volume = "58",
    number = "4",
    pages = "33",
    year = "2026"
}

@article{Cicoli:2012fh,
    author = "Cicoli, Michele and Maharana, Anshuman and Quevedo, F. and Burgess, C. P.",
    title = "{De Sitter String Vacua from Dilaton-dependent Non-perturbative Effects}",
    eprint = "1203.1750",
    archivePrefix = "arXiv",
    primaryClass = "hep-th",
    reportNumber = "DAMTP-2012-7",
    doi = "10.1007/JHEP06(2012)011",
    journal = "JHEP",
    volume = "06",
    pages = "011",
    year = "2012"
}

@article{Dudas:2025yqm,
    author = "Dudas, Emilian and Parameswaran, Susha and Serra, Marco",
    title = "{The Cosmological Constant and Dark Dimensions from Non-Supersymmetric Strings}",
    eprint = "2512.20570",
    archivePrefix = "arXiv",
    primaryClass = "hep-th",
    month = "12",
    year = "2025"
}

@article{Anchordoqui:2025nmb,
    author = "Anchordoqui, Luis A. and Antoniadis, Ignatios and Lust, Dieter",
    title = "{Two Micron-Size Dark Dimensions}",
    eprint = "2501.11690",
    archivePrefix = "arXiv",
    primaryClass = "hep-th",
    reportNumber = "MPP-2025-5, LMU-ASC 02/25",
    doi = "10.1002/prop.70015",
    journal = "Fortsch. Phys.",
    volume = "73",
    number = "8",
    pages = "e70015",
    year = "2025"
}

@article{Ebelt:2023clh,
  author = {Ebelt, Jan and Krippendorf, Sven and Schachner, Andreas},
  title = {W0\_sample = np.random.normal(0,1)?},
  journal = {Phys. Lett. B},
  volume = {855},
  year = {2024},
  pages = {138786},
  doi = {10.1016/j.physletb.2024.138786},
  eprint = {2307.15749},
  archivePrefix = {arXiv},
  primaryClass = {hep-th}
}

@article{Grimm:2018ohb,
  author = {Grimm, Thomas W. and Palti, Eran and Valenzuela, Irene},
  title = {Infinite Distances in Field Space and Massless Towers of States},
  journal = {JHEP},
  volume = {08},
  year = {2018},
  pages = {143},
  doi = {10.1007/JHEP08(2018)143},
  eprint = {1802.08264},
  archivePrefix = {arXiv},
  primaryClass = {hep-th}
}

@article{Monnee:2025ynn,
  author = {Monnee, Jules and Weigand, Timo and Wiesner, Max},
  title = {Physics and geometry of complex structure limits in type IIB Calabi-Yau compactifications},
  journal = {JHEP},
  volume = {03},
  year = {2026},
  pages = {063},
  doi = {10.1007/JHEP03(2026)063},
  eprint = {2509.07056},
  archivePrefix = {arXiv},
  primaryClass = {hep-th}
}

@inproceedings{Tyurin2004,
  author = {Tyurin, Andrey N.},
  title = {Fano versus Calabi-Yau},
  booktitle = {The Fano Conference},
  year = {2004},
  pages = {701--734},
  publisher = {Univ. Torino},
  address = {Turin}
}

@inproceedings{Doran_Harder_Thompson_2017,
  author = {Doran, Charles F. and Harder, Andrew and Thompson, Alan},
  title = {Mirror symmetry, Tyurin degenerations and fibrations on Calabi-Yau manifolds},
  booktitle = {String-Math 2015},
  series = {Proceedings of Symposia in Pure Mathematics},
  volume = {96},
  year = {2017},
  pages = {93--131},
  publisher = {American Mathematical Society}
}

@article{Hassfeld:2025uoy,
  author = {Hassfeld, Bastian and Monnee, Jules and Weigand, Timo and Wiesner, Max},
  title = {Emergent strings in Type IIB Calabi-Yau compactifications},
  journal = {JHEP},
  volume = {01},
  year = {2026},
  pages = {140},
  doi = {10.1007/JHEP01(2026)140},
  eprint = {2504.01066},
  archivePrefix = {arXiv},
  primaryClass = {hep-th}
}

@article{Arkani-Hamed:1998jmv,
  author = {Arkani-Hamed, Nima and Dimopoulos, Savas and Dvali, Gia},
  title = {The Hierarchy problem and new dimensions at a millimeter},
  journal = {Phys. Lett. B},
  volume = {429},
  year = {1998},
  pages = {263--272},
  doi = {10.1016/S0370-2693(98)00466-3},
  eprint = {hep-ph/9803315},
  archivePrefix = {arXiv},
  primaryClass = {hep-ph}
}

@article{Montero:2022prj,
  author = {Montero, Miguel and Vafa, Cumrun and Valenzuela, Irene},
  title = {The dark dimension and the Swampland},
  journal = {JHEP},
  volume = {02},
  year = {2023},
  pages = {022},
  doi = {10.1007/JHEP02(2023)022},
  eprint = {2205.12293},
  archivePrefix = {arXiv},
  primaryClass = {hep-th}
}

@article{Gonzalo:2022jac,
  author = {Gonzalo, Eduardo and Montero, Miguel and Obied, Georges and Vafa, Cumrun},
  title = {Dark dimension gravitons as dark matter},
  journal = {JHEP},
  volume = {11},
  year = {2023},
  pages = {109},
  doi = {10.1007/JHEP11(2023)109},
  eprint = {2209.09249},
  archivePrefix = {arXiv},
  primaryClass = {hep-ph}
}

@article{Vafa:2005ui,
  author = {Vafa, Cumrun},
  title = {The String landscape and the swampland},
  year = {2005},
  eprint = {hep-th/0509212},
  archivePrefix = {arXiv},
  primaryClass = {hep-th}
}

@article{Ooguri:2006in,
  author = {Ooguri, Hirosi and Vafa, Cumrun},
  title = {On the Geometry of the String Landscape and the Swampland},
  journal = {Nucl. Phys. B},
  volume = {766},
  year = {2007},
  pages = {21--33},
  doi = {10.1016/j.nuclphysb.2006.10.033},
  eprint = {hep-th/0605264},
  archivePrefix = {arXiv},
  primaryClass = {hep-th}
}

@article{Lust:2019zwm,
  author = {L{\"u}st, Dieter and Palti, Eran and Vafa, Cumrun},
  title = {AdS and the Swampland},
  journal = {Phys. Lett. B},
  volume = {797},
  year = {2019},
  pages = {134867},
  doi = {10.1016/j.physletb.2019.134867},
  eprint = {1906.05225},
  archivePrefix = {arXiv},
  primaryClass = {hep-th}
}

@article{Lee:2019wij,
  author = {Lee, Seung-Joo and Lerche, Wolfgang and Weigand, Timo},
  title = {Emergent strings from infinite distance limits},
  journal = {JHEP},
  volume = {02},
  year = {2022},
  pages = {190},
  doi = {10.1007/JHEP02(2022)190},
  eprint = {1910.01135},
  archivePrefix = {arXiv},
  primaryClass = {hep-th}
}

@article{Lee:2020zjt,
  author = {Lee, J. G. and Adelberger, E. G. and Cook, T. S. and Fleischer, S. M. and Heckel, B. R.},
  title = {New Test of the Gravitational $1/r^2$ Law at Separations down to 52 $\mu$m},
  journal = {Phys. Rev. Lett.},
  volume = {124},
  number = {10},
  year = {2020},
  pages = {101101},
  doi = {10.1103/PhysRevLett.124.101101},
  eprint = {2002.11761},
  archivePrefix = {arXiv},
  primaryClass = {hep-ex}
}

@article{Andriot:2026lac,
    author = "Andriot, David",
    title = "{Dark energy from string theory: an introductory review}",
    eprint = "2603.25797",
    archivePrefix = "arXiv",
    primaryClass = "hep-th",
    month = "3",
    year = "2026"
}

@article{Cicoli:2021fsd,
    author = "Cicoli, Michele and Cunillera, Francesc and Padilla, Antonio and Pedro, Francisco G.",
    title = "{Quintessence and the Swampland: The Parametrically Controlled Regime of Moduli Space}",
    eprint = "2112.10779",
    archivePrefix = "arXiv",
    primaryClass = "hep-th",
    doi = "10.1002/prop.202200009",
    journal = "Fortsch. Phys.",
    volume = "70",
    number = "4",
    pages = "2200009",
    year = "2022"
}

@article{Brinkmann:2022oxy,
    author = "Brinkmann, Max and Cicoli, Michele and Dibitetto, Giuseppe and Pedro, Francisco G.",
    title = "{Stringy multifield quintessence and the Swampland}",
    eprint = "2206.10649",
    archivePrefix = "arXiv",
    primaryClass = "hep-th",
    doi = "10.1007/JHEP11(2022)044",
    journal = "JHEP",
    volume = "11",
    pages = "044",
    year = "2022"
}

@article{Cicoli:2020noz,
    author = "Cicoli, Michele and Dibitetto, Giuseppe and Pedro, Francisco G.",
    title = "{Out of the Swampland with Multifield Quintessence?}",
    eprint = "2007.11011",
    archivePrefix = "arXiv",
    primaryClass = "hep-th",
    doi = "10.1007/JHEP10(2020)035",
    journal = "JHEP",
    volume = "10",
    pages = "035",
    year = "2020"
}

@article{Crino:2020qwk,
    author = "Crin{\`o}, Chiara and Quevedo, Fernando and Valandro, Roberto",
    title = "{On de Sitter String Vacua from Anti-D3-Branes in the Large Volume Scenario}",
    eprint = "2010.15903",
    archivePrefix = "arXiv",
    primaryClass = "hep-th",
    doi = "10.1007/JHEP03(2021)258",
    journal = "JHEP",
    volume = "03",
    pages = "258",
    year = "2021"
}

@article{Crino:2022zjk,
    author = "Crin{\`o}, Chiara and Quevedo, Fernando and Schachner, Andreas and Valandro, Roberto",
    title = "{A database of Calabi-Yau orientifolds and the size of D3-tadpoles}",
    eprint = "2204.13115",
    archivePrefix = "arXiv",
    primaryClass = "hep-th",
    doi = "10.1007/JHEP08(2022)050",
    journal = "JHEP",
    volume = "08",
    pages = "050",
    year = "2022"
}

@article{Collinucci:2008pf,
    author = "Collinucci, Andres and Denef, Frederik and Esole, Mboyo",
    title = "{D-brane Deconstructions in IIB Orientifolds}",
    eprint = "0805.1573",
    archivePrefix = "arXiv",
    primaryClass = "hep-th",
    doi = "10.1088/1126-6708/2009/02/005",
    journal = "JHEP",
    volume = "02",
    pages = "005",
    year = "2009"
}

@article{Collinucci:2008sq,
    author = "Collinucci, Andres and Kreuzer, Maximilian and Mayrhofer, Christoph and Walliser, Nils-Ole",
    title = "{Four-modulus 'Swiss Cheese' chiral models}",
    eprint = "0811.4599",
    archivePrefix = "arXiv",
    primaryClass = "hep-th",
    doi = "10.1088/1126-6708/2009/07/074",
    journal = "JHEP",
    volume = "07",
    pages = "074",
    year = "2009"
}

@article{Braun:2011zm,
    author = "Braun, Andreas P. and Collinucci, Andres and Valandro, Roberto",
    title = "{G-flux in F-theory and algebraic cycles}",
    eprint = "1107.5337",
    archivePrefix = "arXiv",
    primaryClass = "hep-th",
    reportNumber = "TUW-11-19, LMU-ASC-33-11, ZMP-HH-11-13",
    doi = "10.1016/j.nuclphysb.2011.10.034",
    journal = "Nucl. Phys. B",
    volume = "856",
    pages = "129--179",
    year = "2012"
}

@article{Braun:2020jrx,
    author = "Braun, Andreas P. and Valandro, Roberto",
    title = "{$G_{4}$ flux, algebraic cycles and complex structure moduli stabilization}",
    eprint = "2009.11873",
    archivePrefix = "arXiv",
    primaryClass = "hep-th",
    doi = "10.1007/JHEP01(2021)207",
    journal = "JHEP",
    volume = "01",
    pages = "207",
    year = "2021"
}

\end{document}